\documentclass[11pt,a4paper]{article}

\pdfoutput=1
\usepackage{jheppub}

\usepackage[english]{babel}
\usepackage{amsfonts,amsmath,amssymb,amsbsy,amsthm,bm,mathtools,fixmath}
\usepackage{comment,enumerate,footnote,graphicx,subfloat,relsize}
\usepackage{array,tabularx,tabu,multirow,framed}
\usepackage{mathabx}
\usepackage[usenames,dvipsnames]{xcolor}
\numberwithin{equation}{section}
\usepackage{cleveref}
\crefformat{pluralequation}{#2\black{eqs.~(}#1\black{)}#3}
\Crefformat{pluralequation}{#2\black{Equations~(}#1\black{)}#3}
\crefformat{pluralfigure}{#2\black{figs.~}#1#3}
\Crefformat{pluralfigure}{#2\black{Figures~}#1#3}

\usepackage{tikz}
\usetikzlibrary{arrows,shapes}
\usetikzlibrary{trees,patterns}
\usetikzlibrary{matrix,arrows} 				
\usetikzlibrary{positioning}				  
\usetikzlibrary{calc,through}				  
\usetikzlibrary{decorations.pathreplacing}  
\usepackage{pgffor}							

\usetikzlibrary{decorations.pathmorphing}	
\usetikzlibrary{decorations.markings}
\tikzset{
	>=stealth', 
    vector/.style={decorate, decoration={snake}, draw},
	provector/.style={decorate, decoration={snake,amplitude=2.5pt}, draw},
	antivector/.style={decorate, decoration={snake,amplitude=-2.5pt}, draw},
	bigvector/.style={decorate, decoration={snake,amplitude=4pt}, draw},
    fermion/.style={draw=black, postaction={decorate},
        decoration={markings,mark=at position .55 with {\arrow[draw=black]{>}}}},
    fermionbar/.style={draw=black, postaction={decorate},
        decoration={markings,mark=at position .55 with {\arrow[draw=black]{<}}}},
    fermionnoarrow/.style={draw=black},
    fermiongray/.style={draw=gray},
    anti/.style={double, draw=black, postaction={decorate},
        decoration={markings,mark=at position .55 with {\arrow[draw=black, scale=0.7]{>}}}},
    gluon/.style={decorate, draw=black,
        decoration={coil,amplitude=4pt, segment length=5pt}},
    scalar/.style={dashed,draw=black, postaction={decorate},
        decoration={markings,mark=at position .55 with {\arrow[draw=black]{>}}}},
    scalarbar/.style={dashed,draw=black, postaction={decorate},
        decoration={markings,mark=at position .55 with {\arrow[draw=black]{<}}}},
    scalarnoarrow/.style={dashed,draw=black},
    momentum/.style={draw=black, postaction={decorate},
        decoration={markings,mark=at position 1 with {\arrow[draw=black]{>}}}},
    antimomentum/.style={draw=black, postaction={decorate},
        decoration={markings,mark=at position 0.1 with {\arrow[draw=black]{<}}}}
}

\tikzstyle{block} = [draw, rectangle, 
    minimum height=3em, minimum width=6em]
    
\newcommand{\nc}{\newcommand}
\nc{\pd}{\partial}
\nc{\bea}{\begin{eqnarray}}
\nc{\eea}{\end{eqnarray}}
\nc{\bal}{\begin{alignedat}}
\nc{\eal}{\end{alignedat}}
\nc{\beq}{\begin{equation}}
\nc{\eeq}{\end{equation}}
\nc{\bit}{\begin{itemize}}
\nc{\eit}{\end{itemize}}
\nc{\benu}{\begin{enumerate}}
\nc{\eenu}{\end{enumerate}}
\nc{\bdes}{\begin{description}}
\nc{\edes}{\end{description}}
\nc{\bma}{\begin{pmatrix}}
\nc{\ema}{\end{pmatrix}}

\newcommand{\black}[1]	{{\color{black} #1}}

\nc{\nn}{\nonumber}
\nc{\hc}{\text{h.c.}}
\nc{\cc}{\text{c.c.}}

\nc{\slashed}[1]{{#1}\hspace{-2mm}/}

\nc{\abs}[1]{\left| #1 \right|}
\def\[{\left[}
\def\]{\right]}
\def\({\left(}
\def\){\right)}
\def\<{\langle}
\def\>{\rangle}

\def\g5{\gamma_{5}}

\def\GeV{{\rm GeV}}
\def\TeV{{\rm TeV}}

\def\a{\alpha}

\def\g{\gamma}
\def\d{\delta}

\def\k{\kappa}

\def\m{\mu}

\def\p{\pi}

\def\s{\sigma}

\def\f{\phi}
\def\x{\chi}

\def\W{\Omega}

\def\ag				{\alpha_g}
\def\as				{\alpha_s}

\def\asAnn			{\alpha_s^{\rm ann}}

\def\agScatt		{\alpha_g^{\mathsmaller{S}}}
\def\agScattRep		{\alpha_{g,\mathsmaller{\bf [\hat{R}]}}^{\mathsmaller{S}}}
\def\agScattSinglet	{\alpha_{g,\mathsmaller{\bf [1]}}^{\mathsmaller{S}}}
\def\agScattOctet	{\alpha_{g,\mathsmaller{\bf [8]}}^{\mathsmaller{S}}}
\def\asScatt		{\alpha_s^{\mathsmaller{S}}}

\def\agBound		{\alpha_g^{\mathsmaller{B}}}
\def\agBoundRep		{\alpha_{g,\mathsmaller{\bf [\hat{R}]}}^{\mathsmaller{B}}}
\def\agBoundSinglet	{\alpha_{g,\mathsmaller{\bf [1]}}^{\mathsmaller{B}}}

\def\asBound		{\alpha_s^{\mathsmaller{B}}}
\def\asBoundRep		{\alpha_{s,\mathsmaller{\bf [\hat{R}]}}^{\mathsmaller{B}}}
\def\asBoundSinglet	{\alpha_{s,\mathsmaller{\bf [1]}}^{\mathsmaller{B}}}

\def\asBSF			{\alpha_{s}^{\mathsmaller{\rm BSF}}}
\def\asBSFRep		{\alpha_{s,\mathsmaller{\bf [\hat{R}]}}^{\mathsmaller{\rm BSF}}}
\def\asBSFSinglet	{\alpha_{s,\mathsmaller{\bf [1]}}^{\mathsmaller{\rm BSF}}}

\def\asNA 			{\alpha_{s}^{\mathsmaller{\rm NA}}}
\def\asNARep		{\alpha_{s,\mathsmaller{\bf [\hat{R}]}}^{\mathsmaller{\rm NA}}}
\def\asNASinglet	{\alpha_{s,\mathsmaller{\bf [1]}}^{\mathsmaller{\rm NA}}}

\def\gs			{g_s}
\def\gsBSF		{g_{s}^{\mathsmaller{\rm BSF}}}

\def\ann		{{\rm ann}}
\def\BSF		{\mathsmaller{\rm BSF}}
\def\dec		{{\rm dec}}
\def\eff		{{\rm eff}}
\def\ion		{{\rm ion}}
\def\eq			{{\rm eq}}

\def\gstareffsqrt	{g_{*,\rm eff}^{1/2}}
\def\gstarS		{g_{*\mathsmaller{S}}}
\def\gx			{g_\x}
\def\gX			{g_{\mathsmaller{X}}}

\def\mpl		{m_{\mathsmaller{\rm Pl}}}
\def\mx			{m_\x}
\def\mX			{m_{\mathsmaller{X}}}

\def\sigmaeff	{\sigma_{\rm eff}}

\def\Xdagger	{X^\dagger}
\def\Yx			{Y_{\chi}}
\def\YX			{Y_{X}}
\def\YXdagger	{Y_{\Xdagger}}
\def\vrel		{v_{\rm rel}}

\def\zetaBound	{\zeta_{\mathsmaller{B}}}
\def\zetaScatt	{\zeta_{\mathsmaller{S}}}

\def\dG		{d_{\bf G}}
\def\dR		{d_{\bf R}}

\renewcommand{\vec}{\textbf}

\preprint{Nikhef-2018-023}

\title{
Radiative bound-state formation \\
in unbroken perturbative non-Abelian theories \\
and implications for dark matter
}

\author[1]{Julia Harz}
\author[1,2]{and Kalliopi Petraki}

\affiliation[1]{
Laboratoire de Physique Th\'eorique et Hautes Energies (LPTHE), 
UMR 7589 CNRS \& Sorbonne Universit\'e, 
4~Place Jussieu, F-75252, Paris, France
}
\affiliation[2]{Nikhef, Science Park 105, 1098 XG Amsterdam, The Netherlands}

\emailAdd{jharz@lpthe.jussieu.fr}
\emailAdd{kpetraki@lpthe.jussieu.fr}

\abstract{We compute the cross-sections for the radiative capture of non-relativistic particles into bound states, in unbroken perturbative non-Abelian theories. We find that the formation of bound states via emission of a gauge boson can be significant for a variety of dark matter models that feature non-Abelian long-range interactions, including multi-TeV scale WIMPs, dark matter co-annihilating with coloured partners and hidden-sector models. Our results disagree with previous computations, on the relative sign of the Abelian and non-Abelian contributions. In particular, in the case of capture of a particle-antiparticle pair into its tightest bound state, we find that these contributions add up, rather than partially canceling each other.  We apply our results to dark matter co-annihilating with particles transforming in the (anti)fundamental of $SU(3)_c$, as is the case in degenerate stop-neutralino scenarios in the MSSM. We show that the radiative formation and decay of particle-antiparticle bound states can deplete the dark matter density by $(40-240)\%$, for dark matter heavier than 500~GeV. This implies a larger mass difference between the co-annihilating particles, and allows for the dark matter to be as heavy as 3.3~TeV.}

\arxivnumber{1805.01200} 

\begin{document}
\maketitle


\section{Introduction \label{Sec:Intro}}

In theories with light force mediators, the formation of bound states may have significant implications. Early work pointed out aspects of the effect of bound states on the expected dark matter (DM) detection  signatures~\cite{Pospelov:2008jd,MarchRussell:2008tu,Shepherd:2009sa}. More recently, a variety of phenomenological implications have been identified or explored in greater depth. It has been shown that the formation and decay of unstable bound states can affect the density of thermal-relic DM~\cite{vonHarling:2014kha}, and is essential in determining the unitarity limit on the DM mass~\cite{Baldes:2017gzw}. Moreover, bound-state formation (BSF) processes enhance the expected annihilation signals of symmetric~\cite{Pospelov:2008jd,An:2016gad,An:2016kie,Asadi:2016ybp,Petraki:2016cnz,Cirelli:2016rnw,Kouvaris:2016ltf} and asymmetric DM~\cite{,Baldes:2017gzw,Baldes:2017gzu}. Since BSF cross-sections are typically dominated by different partial waves than the direct annihilation processes, they exhibit different velocity dependence and resonance structure, and can give rise to complementary signatures~\cite{An:2016kie,Petraki:2016cnz}. In the case of asymmetric DM~\cite{Petraki:2013wwa}, the formation of stable bound states may produce novel direct~\cite{Laha:2013gva,Butcher:2016hic} and indirect detection  signals~\cite{Pearce:2013ola,Cline:2014eaa,Detmold:2014qqa,Pearce:2015zca}, as well as affect the DM self-scattering inside haloes~\cite{Petraki:2014uza}. In confining theories, bound states may set the DM mass scale~\cite{Lonsdale:2014wwa,Lonsdale:2017mzg}, and relate it to that of ordinary matter~\cite{Lonsdale:2018xwd}. Finally, DM bound states may have implications for collider  experiments~\cite{Shepherd:2009sa,An:2015pva,Kang:2016wqi,Elor:2018xku}.\footnote{
Bound states may occur also in the spectrum of theories with contact interactions, in particular in the form of non-topological solitons~\cite{Coleman:1985ki,Kusenko:1997si,Kusenko:1997ad}, which have been considered in the context of DM~\cite{Kusenko:1997zq,Kusenko:2001vu}.}

Here, we consider BSF processes in non-Abelian theories. Non-Abelian interactions are particularly important in scenarios where DM is coannihilating with coloured partners, as well as in models where DM consists of TeV-scale Weakly interacting massive particles (WIMPs). Scenarios that feature DM coannihilation with coloured particles are encountered within the minimal supersymmetric standard model (MSSM)~\cite{
Harz:2012fz,
Harz:2014gaa,
Harz:2014tma,
Baker:2015qna,
Ibarra:2015nca,
Harz:2016dql,
Liew:2016hqo,
Pierce:2017suq}, 
and are in part motivated by the measurement of the Higgs mass~\cite{Haber:1996fp,Haber:1990aw}. These scenarios are being probed in high-precision collider experiments; the accurate prediction of the DM density within their parameter space is therefore necessary in order for the cosmological constraints to meaningfully complement those from colliders. On the other hand, TeV-scale WIMP models are being probed mostly via indirect searches. Their annihilation signals exhibit sharp resonances that depend on the DM mass~\cite{
Hryczuk:2014hpa,
Baumgart:2014saa,
Cirelli:2015bda,
Beneke:2016jpw,
Baumgart:2017nsr}. 
It follows that the precise value of the DM mass, which, under minimal assumptions, is predicted by the observed DM density, determines the viability of these models. Clearly, in both cases, an accurate computation of the DM freeze-out is necessary. It is then essential that the depletion of the DM abundance via BSF in the early universe -- which, as we show, can be a leading order effect -- is properly accounted for.

In this paper, we compute the cross-sections for the radiative capture into bound states of non-relativistic particles transforming under a non-Abelian gauge group. We consider, in particular, unbroken non-Abelian theories in the regime where the gauge coupling is perturbative. We do not specify the gauge group or the representation of the interacting particles, such that our results are applicable in a variety of models. 
Rather than an effective field theory approach~\cite{Beneke:1999zr} that has been employed in other studies~\cite{
An:2016gad,
Asadi:2016ybp,
Kim:2016zyy,
Kim:2016kxt,
Biondini:2017ufr,
Biondini:2018pwp,
Braaten:2017gpq,
Braaten:2017kci,
Braaten:2017dwq,
Mitridate:2017izz,
Keung:2017kot}, 
we use the method described in ref.~\cite{Petraki:2015hla}, where the non-relativistic approximation is carried out directly on the relativistic amplitude.

We apply our results to a simplified model where DM is co-annihilating with scalar particles transforming in the fundamental of $SU(3)_c$, and show that BSF has a very important effect on the DM relic density. In the MSSM incarnations of this scenario, the coloured particles typically possess also a sizeable coupling to the Higgs boson, which has been shown to  mediate a sufficiently long-range interaction that enhances the annihilation cross-section~\cite{Harz:2017dlj}. While we do not consider it here, the attractive force mediated by the Higgs is expected to make the BSF effect more pronounced~\cite{HiggsBoundStates}.

The formation of bound states in non-Abelian theories and their implications for DM have been considered in previous works. Reference~\cite{Asadi:2016ybp} considered DM transforming under the adjoint of $SU(2)_L$ (Wino-like DM), and performed computations in the broken electroweak phase, with the purpose of estimating the DM indirect detection signals. References~\cite{Kim:2016zyy,Kim:2016kxt,Biondini:2017ufr,Biondini:2018pwp} considered BSF via non-radiative scattering processes that can take place at a high rate in a thermal bath, and computed the effect of these processes on the DM density.  Various rearrangement processes in Abelian and non-Abelian theories have been discussed in Ref.~\cite{Geller:2018biy}. Finally, refs.~\cite{Mitridate:2017izz,Keung:2017kot} considered radiative BSF in unbroken non-Abelian theories. Our computations disagree with those of~\cite{Mitridate:2017izz,Keung:2017kot} on the relative sign of the Abelian and non-Abelian diagrams contributing to these processes, but are in agreement with the dissociation rate of heavy quarkonium via gluon absorption computed in earlier work~\cite{Brambilla:2011sg}. For the capture of a particle-antiparticle pair into its tightest bound state -- which is typically the most significant capture process --  our results imply that the leading order contributions add up, rather than partially canceling each other. This has very significant phenomenological implications, as we showcase in \cref{Sec:S3Model}.

The paper is organised as follows. In \cref{Sec:BSF}, we compute the radiative BSF cross-sections. In \cref{Sec:S3Model}, we calculate the DM freeze-out including BSF in the scenario of DM co-annihilation with coloured partners.  We conclude in \cref{Sec:Conclusions}. Several important calculations are included in the appendices. In particular, in \cref{App:OverlapIntegrals}, we compute the overlap integrals that enter the BSF cross-sections. In \cref{App:EFT}, we adopt an effective field theory standpoint, we derive the non-relativistic Hamiltonian of the interactions that determine the formation of bound states, and point out the disagreement with previous works~\cite{Mitridate:2017izz,Keung:2017kot}. This provides an independent check of the validity of our results.

\section{Radiative bound-state formation in non-Abelian theories \label{Sec:BSF} }

We consider two complex scalar fields $X_1$ and $X_2$, transforming in the representations ${\bf R_1}$ and ${\bf R_2}$ of a non-Abelian gauge group ${\bf G}$. The Lagrangian is

\beq
{\cal L} = 
(D_{\m} X_1)^\dagger \, (D^\m X_1) +
(D_{\m} X_2)^\dagger \, (D^\m X_2) 
- m_1^2 \, |X_1|^2 - m_2^2 \, |X_2|^2 \,,
\label{eq:Lagrangian_X1X2}
\eeq
where $D_{\m} = \partial_\mu + i \gs \, G_\m^a T^a$ is the covariant derivative, with 
$G_\m^a$ being the gluon fields (we shall denote the corresponding particles with $g$, as usual) and $T^a = T_1^a$ or $T_2^a$ being the generators of $X_1$ and $X_2$ respectively. The fine structure constant is
\beq
\as \equiv \gs^2/(4\pi) \,.
\label{eq:alpha_s}
\eeq
In the following, we shall compute  the radiative BSF processes
\beq
X_1 + X_2 \to {\cal B} (X_1 X_2) + g \,.
\label{eq:BSF_process}
\eeq
While we express our results in terms of the capture processes~\eqref{eq:BSF_process}, it is straightforward to generalise them to transitions between scattering states (bremsstrahlung) and bound-state excitation or de-excitation processes, by simply substituting the appropriate wavefunctions for the initial and final states~\cite{Petraki:2015hla}. 
Since we shall only compute the leading order terms to these transition processes, our results apply also to fermionic $X_1$ and/or $X_2$. Spin-orbit coupling arises only in higher orders in the non-relativistic regime.

We begin in \cref{sec:BSF_definitions}, with a summary of various formulae we use in our computations. In \cref{sec:BSF_Potential}, we give the interaction potential, and discuss the running of $\as$. In \cref{sec:BSF_Amplitude}, we compute the amplitude for the transitions \eqref{eq:BSF_process}, for general representations and masses of $X_1$ and $X_2$. In  \cref{sec:BSF_ColourDecomp} we apply the result to particles in conjugate representations, but with arbitrary masses. 
For transitions involving a colour-singlet scattering or bound state, we compute explicitly the projected amplitude, and in \cref{sec:BSF_CrossSections}, we calculate the corresponding cross-sections.

\subsection{Definitions and useful formulae \label{sec:BSF_definitions}}

For easy reference, we first summarise various formualae that will be used in the following sections.

We define the total and the reduced mass of the interacting particles,
\beq
\label{eq:Masses}
M   	\equiv m_1 + m_2 \,,
\qquad 
\mu 	\equiv \frac{m_1 m_2}{m_1+m_2} \,,
\eeq
and the dimensionless factors 
\beq
\eta_1 	\equiv \frac{m_1}{m_1+m_2} \,, \qquad
\eta_2 	\equiv \frac{m_2}{m_1+m_2} \,.
\eeq
For the momenta $p_1$ and $p_2$ of $X_1$ and $X_2$, we shall often use the following momentum transformation, which allows to separate the center-of-momentum (CM) from the relative motion~\cite{Itzykson:1980rh,Petraki:2015hla},
\begin{subequations}
\label{eq:MomentumTransformation}
\label[pluralequation]{eqs:MomentumTransformation}
\begin{align}
P &\equiv p_1 +p_2,&  
p &\equiv \eta_2 p_1 - \eta_1 p_2,& 
\label{eq:Pp}
\\
p_1 &= \eta_1 P + p,&  
p_2 &= \eta_2 P - p.&  
\label{eq:p1p2}
\end{align}
\end{subequations}

Let $S_1(p_1)$ and $S_2(p_2)$ be the propagators of the $X_1$ and $X_2$, 
\beq
S_j (p_j) \equiv \frac{i}{p_j^2 - m_j^2 + i \epsilon} \,, \qquad i=1,2 \,.
\label{eq:propagators_single}
\eeq
For convenience, we also define
\begin{align}
S(p;P) &\equiv S_1 (\eta_1 P + p) \ S_2 (\eta_2 P - p) \,,
\label{eq:propagators_product}
\\
{\cal S}_0(\vec{p};P) &\equiv \int \frac{dp^0}{2\pi} \ S (p;P) \,.
\label{eq:propagators_product_integral}
\end{align}
To leading order in the non-relativistic regime~\cite[appendix~C]{Petraki:2015hla},
\beq
{\cal S}_0 (\vec{p};P) \simeq  
\[-i 4 M \mu \(P^0 - M - \dfrac{\vec{P}^2}{2M} - \dfrac{\vec{p}^2}{2\mu}\) \]^{-1} \,, 
\label{eq:propagators_product_integral_NR}
\eeq
and~\cite[appendix~E]{Petraki:2015hla}
\begin{subequations}
\label{eq:DipoleIntegrals}
\label[pluralequation]{eqs:DipoleIntegrals}
\begin{align}
\int \dfrac{dq^0}{2\pi} \dfrac{dp^0}{2\pi}	
\ \dfrac{S(q;K) \, S_1(\eta_1 P + p)}{{\cal S}_0 (\vec{q};K) \, {\cal S}_0 (\vec{p};P)} 
\ (2\pi)^4 \, \delta^4(q-p-\eta_2 P_g)
&\simeq 
2m_2 \ (2\pi)^3 \, \delta^3(\vec{q}-\vec{p}-\eta_2 \vec{P}_g) ,
\label{eq:DipoleIntegrals_1}
\\
\int \dfrac{dq^0}{2\pi} \dfrac{dp^0}{2\pi}	
\ \frac{S(q;K) \, S_2(\eta_1 P - p)}{{\cal S}_0 (\vec{q};K) \, {\cal S}_0 (\vec{p};P)} 
\ (2\pi)^4 \, \delta^4(q-p+\eta_1 P_g)
&\simeq 
2m_1 \ (2\pi)^3 \, \delta^3(\vec{q}-\vec{p}+\eta_1 \vec{P}_g) .
\label{eq:DipoleIntegrals_2}
\end{align}
\end{subequations}
As we shall see in \cref{sec:BSF_Amplitude}, we use \cref{eq:propagators_product_integral_NR,eq:DipoleIntegrals} to integrate out the virtuality of $X_1, X_2$ in the radiative part of the BSF diagrams.

\subsection{Potential and the running of the coupling \label{sec:BSF_Potential}}

The interaction between $X_1$ and $X_2$ can be decomposed into irreducible representations, 
\beq 
{\bf R_1} \otimes {\bf R_2} = \sum_{\bf \hat{R}} {\bf \hat{R}} \,. 
\label{eq:R1crossR2}
\eeq 
For each ${\bf \hat{R}}$, the interaction is described in the non-relativistic regime by a static Coulomb potential~\cite{Kats:2009bv}
\beq
V(r) = - \ag / r \,,
\label{eq:V(r)}
\eeq
where the coupling $\ag$ is related to  $\as$ according to
\beq 
\ag = \as \times \frac12 [C_2({\bf R_1}) + C_2({\bf R_2}) -C_2({\bf \hat{R}})]  \,.  
\label{eq:alpha_g_def} \eeq
Here, $C_2 ({\bf R})$ is the quadratic Casimir invariant of the representation ${\bf R}$.
The Coulomb potential \eqref{eq:V(r)} distorts the scattering-state wavefunctions and, if attractive, gives rise to bound states. The scattering-state and bound-state wavefunctions are reviewed in \cref{App:Wavefunctions}.

In general, the coupling $\as$ depends on the momentum transfer $Q$, 
\beq \as = \as(Q) \label{eq:alphaofQ} \,, \eeq
which is different in the various interaction vertices that appear in the transitions we consider. In \cref{tab:MomentumTransfers}, we list the various vertices, specify the symbols we use, and give the average $Q$ in each case.
\begin{table}[h!]
\centering
\renewcommand{\arraystretch}{3}

\begin{tabular}{|c|c|c|c|} 
\hline
{\bf Vertices} 
&	\boldmath{$\as$}
&	\boldmath{$\ag$} 
&	\parbox[c]{21ex}{\centering 
	{\bf Average \\ momentum transfer}~\boldmath{$Q$}
	}
\\[1ex] \hline  \hline 
\parbox[c]{20.5ex}{\centering 
Wavefunction \\
(ladder diagrams) \\ 
of scattering state \\ 
in colour rep. $\hat{\bf R}$
} 
&	$\asScatt$
&	\parbox[c]{26ex}{\centering
	\begin{align*} 
	&\agScattRep = (\asScatt / 2) \times \\
	&\times [C_2({\bf R_1}) + C_2({\bf R_2}) -C_2({\bf \hat{R}})] 
	\end{align*}
	}
&	$ k \equiv \mu \, \vrel$
\\[2ex] \hline
\parbox[c]{20.5ex}{\centering
\medskip	
Wavefunction \\
(ladder diagrams)  \\ 
of bound state \\ 
in colour rep. $\hat{\bf R}$
}
&$\asBoundRep$ 
&\parbox[c]{26ex}{\centering
\begin{align*} 
&\agBoundRep = (\asBoundRep / 2) \times \\
&\times [C_2({\bf R_1}) + C_2({\bf R_2}) -C_2({\bf \hat{R}})] 
\end{align*}
}
&$\kappa_{\mathsmaller{\hat{\bf R}}} \equiv \mu \, \agBoundRep $
\\[3.5ex] \hline
\parbox[c]{20.5ex}{\centering 
\medskip
Formation of bound states 
of colour rep. $\hat{\bf R}$: 
gluon emission}
&$\asBSFRep$
&	
&	
\parbox[c]{21ex}{\centering
$ {\cal E}_{\vec{k}} - {\cal E}_{n\ell} 
=\dfrac{\mu}{2} \left[ \vrel^2 + (\agBoundRep/n)^2 \right]$
}
\\[2ex] \hline
\parbox[c]{20.5ex}{\centering 
\medskip
$g X_i^\dagger X_i$ vertices in 
non-Abelian diagram
for capture in
colour rep. $\hat{\bf R}$}
&$\asNARep$
&	
&	
$\mu \sqrt{ \vrel^2 + {\agBoundRep}^2 }$
\\[2.5ex] \hline
\end{tabular}
	
\caption[]{\label{tab:MomentumTransfers} 
The momentum transfer $Q$ at which the coupling $\as(Q)$ is evaluated. With the exception of $\asScatt$, the couplings depend on the representation ${\bf \hat{R}}$, as denoted; however, in our general computations, we shall often omit the representation index, for brevity. 
}
\end{table}

\subsection{Amplitude for radiative transitions \label{sec:BSF_Amplitude}}

Radiative transitions are represented by the diagram of \cref{fig:FeynmanDiagrams_BSF_Amplitude}, which can be separated into the wavefunctions of the asymptotic states and the radiative vertex. The wavefunctions resum the two-particle interactions at infinity. The long-range $X_1 - X_2$ interaction arises from the one-gluon exchange kernel, which gives rise to the static potential of \cref{eq:V(r)} in the non-relativistic regime. The low momentum transfer ($\sim \mu \vrel$ for the scattering states and $\sim \mu \alpha_s$ for the bound states) via the exchanged gluons is responsible for the appearance of non-perturbative phenomena, the Sommerfeld effect~\cite{SakharovEffect,Sommerfeld:1931} and the mere existence of bound states. The radiative vertex is computed perturbatively, with the leading order contributions shown in \cref{fig:FeynmanDiagrams_BSF_RadiativeVertices}. We discuss them further below.

In the instantaneous approximation, the amplitude for the radiative capture into a bound state is~\cite{Petraki:2015hla}
\beq
[{\cal M}^{\nu}_{\vec{k}\to \{n\ell m\}}]_{ii',jj'}^a 
= \frac{1}{\sqrt{2\mu}} \int \frac{d^3 q}{(2\pi)^3} \frac{d^3 p}{(2\pi)^3} 
\ \tilde{\psi}_{n\ell m}^* (\vec{p}) \ \tilde{\phi}_{\vec{k}}^{}(\vec{q})
\ [{\cal M}_{\rm trans}^{\nu} (\vec{q},\vec{p})]_{ii',jj'}^a \,,
\label{eq:MBSF_definition}
\eeq
where the Latin indices $i,i',j,j'$ and $a$ denote the colour of the initial and final state particles, as shown in \cref{fig:FeynmanDiagrams_BSF_RadiativeVertices}, and $\tilde{\phi}_{\vec{k}}^{}(\vec{q})$ and $\tilde{\psi}_{n\ell m} (\vec{p})$ are the scattering-state and bound-state wavefunctions in  momentum space that obey the Schr\"odinger equation. Here, ${\bf q}$ ($- {\bf q}$) and ${\bf p}$ (${\bf -p}$) are the 3-momenta of $X_1 (X_2)$ in the CM frame, in the scattering state and in the bound state, respectively. The scattering state wavefunction is characterised by the continuous quantum number ${\bf k}$, which specifies the expectation value of ${\bf q}$. In a central potential, such as \cref{eq:V(r)}, the bound state wavefunction is characterised by the standard discrete principal and angular-momentum quantum numbers $\{n, \ell, m\}$, which specify the expectation value of ${\bf p}$. The wavefunctions in a Coulomb potential are reviewed in~\cref{App:Wavefunctions}.\footnote{
We note that the integrand in \cref{eq:MBSF_definition} admits corrections of higher order in ${\bf q}$ and ${\bf p}$ that arise from the relativistic normalisation of states. Here we are interested only in the leading order terms, and we shall neglect these corrections. However, these corrections become important when there is a cancellation between the lowest order contributions to ${\cal M}_{\rm trans}$~\cite{Petraki:2015hla,Petraki:2016cnz}. 
}

${\cal M}_{\rm trans}^{\nu} (\vec{q},\vec{p})$ is the perturbative transition amplitude with the virtuality of the interacting particles integrated out, as follows~\cite[sec.~3.3]{Petraki:2015hla}
\beq
[{\cal M}_{\rm trans}^{\nu} (\vec{q},\vec{p})]_{ii',jj'}^a
= \frac{1}{ {\cal S}_0 (\vec{q};K)  \, {\cal S}_0 (\vec{p};P) }
\ \int \frac{dq^0}{2\pi} \frac{dp^0}{2\pi}
\ [{\cal C}^{\nu} (q,p;K,P)]_{ii',jj'}^a \,.
\label{eq:MBSF_trans_definition}
\eeq
Here, $[{\cal C}^{\nu} (q,p;K,P)]_{ii',jj'}^a$ is the sum of all connected diagrams contributing to the process
\beq
X_{1,i}  \,(\eta_1 K + q) + X_{2,j}  \,(\eta_2 K - q) \ \to \
X_{1,i'} \,(\eta_1 P + p) + X_{2,j'} \,(\eta_2 P - p) + g^a (P_g), 
\label{eq:X1X2_to_X1X2g}
\eeq
with the incoming and outgoing $X_1, X_2$ being off-shell and only the emitted gluon $g$ being on-shell and amputated. We emphasise that the $X_1, X_2$ incoming and outgoing legs should \emph{not} be amputated in the computation of ${\cal C}^\nu$; the proper amputation is done by the prefactor in \cref{eq:MBSF_trans_definition}. (We recall that ${\cal S}_0$ is defined in \cref{eq:propagators_product_integral}.) Note that the connected diagrams contributing to ${\cal C}^\nu$ may include not-fully-connected diagrams that are non-zero due to the off-shellness of the legs, such as the diagrams in which the radiation is emitted from one of the legs (cf.~\cref{fig:FeynmanDiagrams_BSF_RadiativeVertices}).\footnote{  
If in a certain theory, all diagrams contributing to \eqref{eq:X1X2_to_X1X2g} are fully connected, then ${\cal M}_{\rm trans}$ can be computed at leading order from the sum of the fully amputated diagrams by simply setting all incoming and outgoing particles on-shell~\cite{Petraki:2015hla}.}

In \cref{eq:X1X2_to_X1X2g}, the momenta of the particles are indicated inside the parentheses. While the 3-momenta ${\bf q}$ and ${\bf p}$ follow the probability distributions given by the wavefunctions $\tilde{\phi}_{\vec{k}}^{}(\vec{q})$ and $\tilde{\psi}_{n\ell m} (\vec{p})$ that appear in \cref{eq:MBSF_definition}, $q^0$ and $p^0$ are determined by the poles of ${\cal C}^\nu$, upon the integration denoted in \cref{eq:MBSF_trans_definition}.  The total 4-momenta of the scattering state, the bound state and the radiated gluon, $K$, $P$ and $P_g$ respectively, essentially contain all the (discrete and continuous) quantum numbers that fully specify the system. In the non-relativistic regime, they can be expressed as
\begin{subequations}
\label{eq:TotalMomenta_NR}
\label{eqs:TotalMomenta_NR}
\begin{align}
K &= \( M + \frac{{\bf K}^2}{2M} + {\cal E}_{\bf k}, \ {\bf K} \) \,,
\\
P &= \( M + \frac{{\bf P}^2}{2M} +{\cal E}_{n\ell}, \ {\bf P} \) \,, \\
P_g &= \(\omega, \ {\bf P}_g \) \,,
\end{align}
\end{subequations}
where ${\cal E}_{\bf k} = {\bf k}^2/(2\mu) = \mu \vrel^2/2$ is the kinetic energy of the scattering state in the CM frame, with $\vrel$ being the relative velocity of the interacting particles, and ${\cal E}_{n\ell} < 0$ is the binding energy of the bound state. Note that $M_{n\ell} \equiv M+{\cal E}_{n\ell}$ is the mass of the bound state. For a Coulomb potential, ${\cal E}_{n\ell} = -\kappa^2/(2n^2 \mu)$, with $\kappa \equiv \mu \asBound$ (cf.~\cref{App:Wavefunctions}). Energy-momentum conservation, $K=P+P_g$, implies
\beq
\omega = |{\bf P}_g| \simeq {\cal E}_{\bf k} - {\cal E}_{n\ell} \,.
\label{eq:omega}
\eeq
%

\begin{subfigures}
\label{fig:FeynmanDiagrams_BSF}
\label[pluralfigure]{figs:FeynmanDiagrams_BSF}	
\begin{figure}
\centering
\begin{tikzpicture}[line width=1.1pt, scale=1.2]

\begin{scope}[shift={(0,0)}]
\node at (-2.75, 1) {$X_1$};
\node at (-2.75, 0) {$X_2$};
\draw[gluon]    (-2.2,0) -- (-2.2,1);
\node at (-1.6,0.5) {$\cdots$};
\draw[gluon]    (-1,0) -- (-1,1);
\draw[fermion] (-2.5,1) -- (-0.5,1);
\draw[fermion] (-0.5,1) -- (-0.05,0.55);
\draw[fermion] (-0.5,0) -- (-0.05,0.45);
\draw[fermion] (-2.5,0) -- (-0.5,0);
\draw[gluon]    (2.2,0) -- (2.2,1);
\node at (1.6,0.5) {$\cdots$};
\draw[gluon]    (1,0) -- (1,1);
\draw[fermionbar] (2.5,1) -- (0.5,1);
\draw[fermionbar] (0.5,1) -- (0.1,0.6);
\draw[fermionbar] (0.5,0) -- (0.1,0.4);
\draw[fermionbar] (2.5,0) -- (0.5,0);
\draw[line width=0.7pt,gray] (1.6,0.5) ellipse (1cm and 0.9cm);
\node at (2.85,0.5) {${\cal B}$};
\draw[gluon] (0,1.5) -- (0,0.75);
\node at (0,1.7) {$g$};
\filldraw[lightgray]  (0,0.5) circle (7pt);
\node at (0,0.5) {${\cal C}^\nu$};
\end{scope}

\end{tikzpicture}
\caption[]{\label{fig:FeynmanDiagrams_BSF_Amplitude} 
The amplitude for the radiative capture consists of the (non-perturbative) initial and final state wavefunctions, and the perturbative 5-point function that includes the radiative vertices. 
}
\bigskip
\begin{tikzpicture}[line width=1.1pt, scale=1.6]
\begin{scope}[shift={(0,0)}]
\begin{scope}[shift={(-2.15,0)}]
\node at (-1.7, 1) {$i$};
\node at (-1.7, 0) {$j$};
\node at 	(-1, 1.3) {$\eta_1 K+q$};
\draw[fermion]    (-1.5,1) -- (0,1);
\draw[fermion] (-1.5,0) -- (0,0);
\node at 	(-1,-0.3) {$\eta_2 K-q$};
\draw[gluon] (0,0)   -- (0,0.5);
\draw[gluon] (0,1)   -- (0,0.5);
\draw[gluon] (0.8,0.5) -- (0,0.5);
\node at 	 ( 1.1 ,0.45) {$a, \nu$};
\node at 	 (-0.35,0.75) {$b, \rho$};
\node at 	 (-0.35,0.25) {$c, \mu$};
\draw[->]	(0.3,0.7) -- (0.7,0.7);
\node at 	(0.85,0.7) {$P_g$};
\node at 	(1, 1.3) {$\eta_1 P+p$};
\draw[fermion]    (0,1) -- (1.5,1);
\draw[fermion] (0,0) -- (1.5,0);
\node at 	(1,-0.3) {$\eta_2 P-p$};
\node at (1.7, 1) {$i'$};
\node at (1.7, 0) {$j'$};
\end{scope}
\node at (0, 0.5) {$+$};
\begin{scope}[shift={(2.4,0)}]
\begin{scope}[shift={(-1.2,0)}]
\draw[fermion] (-0.8,1) -- (0,1);
\draw[fermion] (-0.8,0) -- (0.8,0);
\draw[gluon] (0,1) -- (0,1.7);
\draw[fermion] (0,1) -- (0.8,1);
\end{scope}
\node at (0, 0.5) {$+$};
\begin{scope}[shift={(1.2,0)}]
\draw[fermion] (-0.8,1) -- (0.8,1);
\draw[fermion] (-0.8,0) -- (0,0);
\draw[gluon] (0,-0.7) -- (0,0);
\draw[fermion] (0,0) -- (0.8,0);
\end{scope}
\end{scope}
\end{scope}
\end{tikzpicture}
\caption[]{\label{fig:FeynmanDiagrams_BSF_RadiativeVertices} 
The leading order diagrams contributing to ${\cal C}^\nu$. The external-momentum, colour-index and space-time-index assignments are the same in all three diagrams.
}
\end{figure}
\end{subfigures}

\bigskip

The leading order contributions to $[{\cal C}^\nu]_{ii',jj'}^{a}$ are shown in \cref{fig:FeynmanDiagrams_BSF_RadiativeVertices}. We compute them next using the Feynman rules from~\cite{Sparticles}.
\begin{subequations}
\label{eq:BSF_offshellamplitudes_0}
\label[pluralequation]{eqs:BSF_offshellamplitudes_0}
\paragraph{Emission from the mediator}
\begin{align}
&i({\cal C}_{\rm med}^\nu)^{a}_{ii',jj'} =
\nn \\
&=
S_1(\eta_1 P + p) 
\[-i \gs (T_1^b)_{i'i} \ (\eta_1 K + \eta_1 P + q+p)_\rho \]
S_1(\eta_1 K + q)
\ \frac{-i}{(\eta_1 K + q -\eta_1 P - p)^2} 
\nn \\
&\times
S_2(\eta_2 P - p) 
\[-i \gs \, (T_2^c)_{j'j} \ (\eta_2 K + \eta_2 P - q - p)_\mu \]
S_2(\eta_2 K - q)
\ \frac{-i}{(\eta_2 K - q -\eta_2 P + p)^2}
\nn \\
&\times
(-\gsBSF f^{abc})  \  \left\{
g^{\rho \mu} [(\eta_1 K + q - \eta_1 P - p) - (\eta_2 K - q - \eta_2 P + p)]^\nu
\right. \nn \\ 
&\left.  
+ g^{\nu \rho} [-P_g- (\eta_1 K + q - \eta_1 P - p)]^\mu
+ g^{\mu \nu}  [(\eta_2 K - q - \eta_2 P + p) + P_g]^{\rho}
\right\}, 
\label{eq:BSF_iCmed}
\end{align}
\paragraph{Emission from \boldsymbol{$X_1$}}
\begin{align}
i({\cal C}_1^\nu)^{a}_{ii',jj'} &=
\delta_{j'j} \ S_2(\eta_2 K-q) \times 
S_1(\eta_1 P + p) \, S_1(\eta_1 K + q)
\nn \\
&\times
\[-i \gsBSF (T_1^a)_{i'i}  \ (\eta_1 K + \eta_1 P + q+p)^\nu
\, (2\pi)^4 \delta^4 (\eta_1 K + q - \eta_1 P - p - P_g) \],
\label{eq:BSF_iC1} 
\end{align}
\paragraph{Emission from \boldsymbol{$X_2$}}
\begin{align}
i({\cal C}_2^\nu)^{a}_{ii',jj'} &=
\delta_{i'i} \ S_1(\eta_1 K + q) 
\times S_2(\eta_2 P - p) \, S_2(\eta_2 K - q)
\nn \\
&\times
\[-i \gsBSF \, (T_2^a)_{j'j} \ (\eta_2 K + \eta_2 P - q - p)^\nu
\, (2\pi)^4 \delta^4 (\eta_2 K - q - \eta_2 P + p - P_g)  \] . 
\label{eq:BSF_iC2} 
\end{align}
\end{subequations}

\medskip

We are interested only in the spatial components of ${\cal C}^\nu$, $\nu = 1,2,3$,
\beq
\boldsymbol{{\cal C}}^a_{ii',jj'} = 
(\boldsymbol{{\cal C}}_{\rm med})^a_{ii',jj'}  +
(\boldsymbol{{\cal C}}_1)^a_{ii',jj'}  +
(\boldsymbol{{\cal C}}_2)^a_{ii',jj'} \,.
\label{eq:C_total}
\eeq
The wavefunctions in \cref{eq:MBSF_definition} impose $|{\bf q}| \sim k = \mu \vrel$ and for a Coulomb potential $|{\bf p}| \sim \kappa/n = \mu \asBound/n$. Noting the hierarchy of scales $|{\bf q}|, |{\bf p}|, k, \kappa \ll M,\mu$, and using \cref{eq:TotalMomenta_NR,eq:omega}, we find from \cref{eqs:BSF_offshellamplitudes_0} that to leading order in $\vrel$ and $\as$, in the CM frame,  
\begin{subequations}
\label{eq:BSF_offshellamplitudes}
\label[pluralequation]{eqs:BSF_offshellamplitudes}
\begin{align}
(\boldsymbol{{\cal C}}_{\rm med})^a_{ii',jj'} 
&\simeq +i  f^{abc} \, (T_1^b)_{i'i} (T_2^c)_{j'j} 
\times 8\gsBSF g_s^2 \, M \mu \ \frac{(\vec{q}-\vec{p})~}{(\vec{q}-\vec{p})^4} 
\times S(q;K) \, S(p;P),  
\label{eq:BSF_Cmed}
\\
(\boldsymbol{{\cal C}}_1)^a_{ii',jj'} 
&\simeq - \, (T_1^a)_{i'i} \, \delta_{j'j} 
\times \gsBSF \ (\vec{q}+\vec{p})  
\times S(q;K) \, S_1(\eta_1 P + p) \ (2\pi)^4 \delta^4 (q-p - \eta_2 P_g),
\label{eq:BSF_C1} 
\\
(\boldsymbol{{\cal C}}_2)^a_{ii',jj'} 
&\simeq + \delta_{i'i} \, (T_2^a)_{j'j} 
\times \gsBSF \ (\vec{q}+\vec{p})   
\times S(q;K) \, S_2(\eta_2 P - p) \ (2\pi)^4 \delta^4 (q-p + \eta_1 P_g).
\label{eq:BSF_C2} 
\end{align}
\end{subequations}
A few comments on $\boldsymbol{{\cal C}}_{\rm med}$ are in order.  
\begin{itemize}
\item
The leading order contribution arises from the contraction of the momenta of the two vertices on the scalar legs. 
\item
The factor $\gs^2$ arises from the same vertices, where the momentum transfer is $\sim |\vec{p}-\vec{q}|$, with $|\vec{p}| \sim \kappa = \mu \agBound$ and $|\vec{q}| \sim k = \mu \vrel$. This implies that $\gs$ should be evaluated at the scale $Q \approx \sqrt{\kappa^2 + k^2}$ (cf.~\cref{tab:MomentumTransfers}).\footnote{\label{foot:alphaNA}
Since BSF is insignificant when $\kappa \ll k$, in \cref{Sec:S3Model} we approximate the momentum transfer in these vertices with that inside the bound state, $Q \sim \kappa$, thus setting $\asNA = \asBound$.}
\item
In \cref{eq:BSF_Cmed}, we have neglected the energy transfer along the gluon propagators (as is also done in the one-boson exchange diagrams that are resummed into the non-relativistic potential). Upon the integration indicated in \cref{eq:MBSF_trans_definition}, the poles of the scalar propagators, $S(q;K) \, S(p;P)$, set $q^0 \sim \vec{q}^2/\mu \sim \vec{k}^2 / \mu$ and $p^0 \sim \vec{p}^2/\mu \sim \kappa^2/\mu$. In contrast, the poles of the gluon propagators set $q^0 - p^0 \sim |\vec{q} - \vec{p}| \sim \sqrt{\vec{k}^2 + \kappa^2}$. Since $q^0$ and $p^0$ indicate the off-shellness of the two scalar particles, the gluon poles --- on which the off-shellness of the scalars is greater for $\as,\vrel <1$ --- yield subdominant contributions to the transition amplitude.
\item 
Naively, $\boldsymbol{{\cal C}}_{\rm med}$ may be expected to be of higher order in $\as$ than $\boldsymbol{{\cal C}}_1$ and $\boldsymbol{{\cal C}}_2$. However, the scaling of $q^0, p^0, |{\bf q}|$ and $|{\bf p}|$ with $\as$ and $\vrel$ described above implies all three diagrams are of the same order. This will become apparent in the following.
\end{itemize}

Collecting \cref{eq:MBSF_trans_definition,eq:C_total,eq:BSF_offshellamplitudes}, and using \cref{eq:propagators_product_integral,eq:DipoleIntegrals}, we find
\begin{align}
[\boldsymbol{{\cal M}}_{\rm trans}]_{ii',jj'}^a \simeq
-\gsBSF 4M
&\left\{
-i f^{abc} (T_1^b)_{i'i} (T_2^c)_{j'j}
\times 8\pi \mu \asNA
\ \frac{\vec{q}-\vec{p}}{(\vec{q}-\vec{p})^4}
\right. 
\nn \\ 
&
+ \eta_2 \, (T_1^a)_{i'i} \, \delta_{j'j} 
\times \vec{p} \ (2\pi)^3 \, \delta^3(\vec{q}-\vec{p}-\eta_2 \vec{P}_g)
\nn \\ 
&\left.
- \eta_1 \, \delta_{i'i} \, (T_2^a)_{j'j} 
\times \vec{p} \ (2\pi)^3 \, \delta^3(\vec{q}-\vec{p}+\eta_1 \vec{P}_g)
\right\} \,. 
\label{eq:MBSF_trans_computed}
\end{align}
Plugging \cref{eq:MBSF_trans_computed} into \cref{eq:MBSF_definition}, we obtain 
\begin{multline}
[\boldsymbol{{\cal M}}_{\vec{k}\to \{n\ell m\}}]_{ii',jj'}^a =
-\(2^5\pi\asBSF \, M^2 / \mu \)^{1/2} \times 
\left\{
- i f^{abc} (T_1^b)_{i'i} (T_2^c)_{j'j}
\ \boldsymbol{{\cal Y}}_{\vec{k}, \{n\ell m\}}
\right. \\ \left.
+ \eta_2 \, (T_1^a)_{i'i} \, \delta_{j'j} 
\ \boldsymbol{{\cal J}}_{\vec{k}, \{n\ell m\}} (\eta_2 \vec{P}_g)
- \eta_1 \, \delta_{i'i} \, (T_2^a)_{j'j} 
\ \boldsymbol{{\cal J}}_{\vec{k}, \{n\ell m\}} (-\eta_1 \vec{P}_g)
\right\} 
\,, \label{eq:MBSF_general}
\end{multline}
where we have defined the overlap vector integrals (see also~\cite{Petraki:2015hla,Petraki:2016cnz})
\begin{subequations}
\label{eq:calIntegrals_definition}
\label[pluralequation]{eqs:calIntegrals_definition}
\begin{align}
\boldsymbol{{\cal J}}_{\vec{k}, \{n\ell m\}} (\vec{b}) &\equiv
\int \frac{d^3 p}{(2\pi)^3} \ \vec{p} 
\, \tilde{\psi}_{n\ell m}^*(\vec{p}) 
\, \tilde{\phi}_{\vec{k}}^{} (\vec{p}+\vec{b}) \,,
\label{eq:Jcal_definition}
\\
\boldsymbol{{\cal Y}}_{\vec{k}, \{n\ell m\}} &\equiv
8\pi\mu\asNA
\ \int \frac{d^3 p}{(2\pi)^3} \frac{d^3 q}{(2\pi)^3} 
\ \frac{\vec{q}-\vec{p}}{(\vec{q}-\vec{p})^4} 
\, \tilde{\psi}_{n\ell m}^*(\vec{p}) 
\, \tilde{\phi}_{\vec{k}}^{} (\vec{q}) \,.
\label{eq:Ycal_definition}
\end{align}
\end{subequations}
$\boldsymbol{{\cal J}}_{\vec{k}, \{n\ell m\}}$ has been computed in refs.~\cite{Petraki:2015hla,Petraki:2016cnz}. We review the result, and compute $\boldsymbol{{\cal Y}}_{\vec{k}, \{n\ell m\}}$ in \cref{App:OverlapIntegrals}.


\subsection{Colour decomposition for conjugate representations \label{sec:BSF_ColourDecomp}}

We now focus on particles transforming under conjugate representations ${\bf R}_1 = {\bf R}$ and ${\bf R}_2 = \bar{\bf R}$, such that 
\beq
T_1^a = T^a\,, \qquad T_2^a = -T^{a \, *} \,.
\label{eq:Generators_ParticleAntiparticle}
\eeq
We shall \emph{not} assume though that the masses of the interacting particles are equal, so that our results are more widely applicable. 
For the overlap integral $\boldsymbol{{\cal J}}_{\vec{k} \to \{n\ell m\}} (\vec{b})$, with $|\vec{b}| \propto |\vec{P}_g|$, the dominant contribution is independent of ${\vec{b}}$~\cite{Petraki:2015hla,Petraki:2016cnz}. In the following, we denote $\boldsymbol{{\cal J}}_{\vec{k}, \{n\ell m\}} = \boldsymbol{{\cal J}}_{\vec{k}, \{n\ell m\}} ({\bf b}=0)$, and the amplitude \eqref{eq:MBSF_general} becomes
\begin{multline}
[\boldsymbol{{\cal M}}_{\vec{k}\to \{n\ell m\}}]_{ii',jj'}^a =
-(2^5\pi\asBSF \, M^2 / \mu )^{1/2} \ \times 
\\
\times
\left\{
\( \eta_2 \, T^a_{i'i} \, \delta_{j'j}  +  \eta_1 \, \delta_{i'i} \, T^a_{jj'} \) 
\ \boldsymbol{{\cal J}}_{\vec{k}, \{n\ell m\}}
+ i f^{abc} T^b_{i'i} T^c_{jj'} \ \boldsymbol{{\cal Y}}_{\vec{k}, \{n\ell m\}}
\right\} 
\,. \label{eq:MBSF_general_LO}
\end{multline}
The tensor product of two conjugate representations contains always a singlet and an adjoint, and possibly other states,
\beq
{\bf R \otimes \bar{R}} = {\bf 1 \oplus adj \oplus \cdots} \,.
\label{eq:RxRbar}
\eeq
It is clear from \cref{eq:alpha_g_def} that, among all irreducible representations, the singlet configuration ($C_2({\bf 1})=0$) exhibits the most attractive potential, and thus accommodates the tightest bound state. \Cref{eq:RxRbar} implies that at least the following capture processes are allowed by the group algebra, provided that the potential in the final state is attractive, such that the bound state exists:
\begin{subequations}
\label{eq:Transitions}
\label[pluralequation]{eqs:Transitions}
\begin{align}
(X+ X^\dagger)_{\bf [adj]} 
&\ \to \ 
{\cal B} (XX^\dagger)_{\bf [1]} + g_{\bf [adj]} \,,
\label{eq:BSF_AdjointToSinglet_0}
\\
(X+ X^\dagger)_{\bf [1]}
&\ \to \ 
{\cal B} (XX^\dagger)_{\bf [adj]} + g_{\bf [adj]} \,,
\label{eq:BSF_SingletToAdjoint_0}
\\
(X+ X^\dagger)_{\bf [adj]} 
&\ \to \ 
{\cal B} (XX^\dagger)_{\bf [adj]} + g_{\bf [adj]} \,.
\label{eq:BSF_AdjointToAdjoint_0}
\end{align}
\end{subequations}
Depending on the group ${\bf G}$ and the representation ${\bf R}$, more transitions may be possible. The amplitudes for the various transitions may be computed from \cref{eq:MBSF_general_LO} by projecting onto the appropriate colour representations, using the Clebsch-Gordan coefficients. Below we compute explicitly the amplitudes for the transitions \eqref{eq:BSF_AdjointToSinglet_0} and \eqref{eq:BSF_SingletToAdjoint_0} only. 

In the following, $\dR$, $C({\bf R})$ and $C_2({\bf R})$ stand for the dimension and the Casimir invariants of the representation ${\bf R}$, and $\dG$, $C({\bf G})$ and $C_2({\bf G})$ are the corresponding quantities for the group ${\bf G}$.
Evidently, the wavefunctions and therefore $\boldsymbol{{\cal J}}_{\vec{k}, \{n\ell m\}}$ and $\boldsymbol{{\cal Y}}_{\vec{k}, \{n\ell m\}}$ depend on the colour representations of the scattering and bound states, ${\bf R}_S$ and ${\bf R}_B$. Whenever appropriate, we shall use the notation 
$\boldsymbol{{\cal J}}_{\vec{k}, \{n\ell m\}}^{[{\bf R}_S,{\bf R}_B]}$ and 
$\boldsymbol{{\cal Y}}_{\vec{k}, \{n\ell m\}}^{[{\bf R}_S,{\bf R}_B]}$.

\subsubsection{Adjoint scattering states to singlet bound states   \label{sec:BSF_AdjointToSinglet}}

The radiative capture into colour-singlet bound states can occur \emph{only} from adjoint scattering states,
\beq
(X + X^\dagger)_{\bf [adj]} \to {\cal B}(X X^\dagger)_{\bf [1]} + g_{\bf [adj]} \,.
\label{eq:BSF_AdjointToSinglet}
\eeq
It thus suffices to project only the final $X-X^\dagger$ state onto the singlet configuration; upon summing the squared amplitude over colours, the group algebra will project the initial state onto the adjoint. The amplitude for the process~\eqref{eq:BSF_AdjointToSinglet} is
\begin{align}
\(\boldsymbol{{\cal M}}_{\vec{k}\to \{n\ell m\}}^{\bf [adj] \to [1]} \)_{i,j}^a 
&= \frac{\delta_{i'j'}}{\sqrt{\dR}}
\( \boldsymbol{{\cal M}}_{\vec{k}\to \{n\ell m\}}
\)_{ii',jj'}^a 
\nn \\
&=
-\( \frac{2^5\pi\asBSF \, M^2}{\mu} \)^{1/2} \times 
\frac{1}{\sqrt{\dR}} 
\[
T^a_{ji} \ \boldsymbol{{\cal J}}_{\vec{k}, \{n\ell m\}}^{\bf [adj,1]}
+ i f^{abc} (T^c T^b)_{ji} \ \boldsymbol{{\cal Y}}_{\vec{k}, \{n\ell m\}}^{\bf [adj,1]} 
\]
\nn \\ 
&=
-\( \frac{2^5\pi\asBSF \, M^2}{\mu} \)^{1/2} \times 
\frac{1}{\sqrt{\dR}} 
\[
\boldsymbol{{\cal J}}_{\vec{k}, \{n\ell m\}}^{\bf [adj,1]}
+\frac{C_2({\bf G})}{2} \ \boldsymbol{{\cal Y}}_{\vec{k}, \{n\ell m\}}^{\bf [adj,1]}
\] T^a_{ji}
\,, \label{eq:MBSF_BoundStateSinglet}
\end{align}
where in the last step we used $f^{abc} T^b T^c = (i/2) \, C_2({\bf G}) \, T^a$. 
The amplitude squared, colour-summed and averaged over the colour of the initial particles is
\begin{align}
\frac{1}{\dR^2}
\abs{ \boldsymbol{{\cal M}}_{\vec{k}\to \{n\ell m\}}^{\bf [adj] \to [1]} }^2 =
\( \frac{2^5\pi\asBSF \, M^2}{\mu} \) \times 
\frac{C_2({\bf R})}{\dR^2} 
\abs{
	\boldsymbol{{\cal J}}_{\vec{k}, \{n\ell m\}}^{\bf [adj,1]}  +
	\frac{C_2({\bf G})}{2} \ \boldsymbol{{\cal Y}}_{\vec{k}, \{n\ell m\}}^{\bf [adj,1]}
}^2 \,.
\label{eq:MBSFsquared_BoundStateSinglet}
\end{align}

\subsubsection{Singlet scattering states to adjoint bound states  \label{sec:BSF_SingletToAdjoint}}

With the emission of a gluon, a colour-singlet scattering state can turn \emph{only} into an adjoint state,
\beq
(X + X^\dagger)_{\bf [1]} \to {\cal B}(X X^\dagger)_{\bf [adj]} + g_{\bf [adj]} \,.
\label{eq:BSF_SingletToAdjoint}
\eeq
Similarly to the above, the amplitude for the process \eqref{eq:BSF_SingletToAdjoint} is deduced from \cref{eq:MBSF_general_LO} by projecting the initial $X-X^\dagger$ state onto the singlet configuration
\begin{align}
\(\boldsymbol{{\cal M}}_{\vec{k}\to \{n\ell m\}}^{\bf [1] \to [adj]} \)_{i',j'}^a 
&= \frac{\delta_{ij}}{\sqrt{\dR}}
\( \boldsymbol{{\cal M}}_{\vec{k}\to \{n\ell m\}}
\)_{ii',jj'}^a 
\nn \\
&=
-\( \frac{2^5\pi\asBSF \, M^2}{\mu} \)^{1/2} \times 
\frac{1}{\sqrt{\dR}} 
\[
T^a_{i'j'} \ \boldsymbol{{\cal J}}_{\vec{k}, \{n\ell m\}}^{\bf [1,adj]}  +  
i f^{abc} (T^b T^c)_{i'j'} \ \boldsymbol{{\cal Y}}_{\vec{k}, \{n\ell m\}}^{\bf [1,adj]}
\]
\nn \\ 
&=
-\( \frac{2^5\pi\asBSF \, M^2}{\mu} \)^{1/2} \times 
\frac{1}{\sqrt{\dR}}
\[
\boldsymbol{{\cal J}}_{\vec{k}, \{n\ell m\}}^{\bf [1,adj]}
-\frac{C_2({\bf G})}{2} \ \boldsymbol{{\cal Y}}_{\vec{k}, \{n\ell m\}}^{\bf [1,adj]}
\] T^a_{i'j'} 
\,. \label{eq:MBSF_ScatteringStateSinglet}
\end{align}
Then,
\begin{align}
\frac{1}{\dR^2}
\abs{ \boldsymbol{{\cal M}}_{\vec{k}\to \{n\ell m\}}^{\bf [1] \to [adj]} }^2 =
\( \frac{2^5\pi\asBSF \, M^2}{\mu} \) \times 
\frac{C_2({\bf R})}{\dR^2} 
\abs{
\boldsymbol{{\cal J}}_{\vec{k}, \{n\ell m\}}^{\bf [1,adj]}
-\frac{C_2({\bf G})}{2} \ \boldsymbol{{\cal Y}}_{\vec{k}, \{n\ell m\}}^{\bf [1,adj]}
}^2  \,.
\label{eq:MBSFsquared_ScatteringStateSinglet}
\end{align}

\subsubsection{Remaining transitions  \label{sec:BSF_RemainingTransitions}}

Depending on ${\bf R}$, other transitions may be possible. In this case, appropriate projections of the amplitude \eqref{eq:MBSF_general_LO} have to be computed. It is sometimes possible to obtain the amplitude-squared for a transition of interest from the total amplitude-squared by subtracting the contributions of other known transitions. For this reason, we provide here the total colour-averaged squared amplitude,
\begin{align}
\frac{1}{\dR^2}
\abs{ \boldsymbol{{\cal M}}_{\vec{k}\to \{n\ell m\}}^{\bf total}  }^2 
&= 
\( \frac{2^5\pi\asBSF \, M^2}{\mu} \) \times
\nn \\
&\times 
C_2({\bf R})
\left\{
(\eta_1^2 + \eta_2^2)
\abs{ \boldsymbol{{\cal J}}_{\vec{k}, \{n\ell m\}} }^2 
+
\frac{ C_2({\bf R}) C_2({\bf G}) }{\dG}
\abs{ \boldsymbol{{\cal Y}}_{\vec{k}, \{n\ell m\}} }^2 
\right\} \,.
\label{eq:MBSFsquared_Total}
\end{align}
Note that $\boldsymbol{{\cal J}}_{\vec{k}, \{n\ell m\}}$ and $\boldsymbol{{\cal Y}}_{\vec{k}, \{n\ell m\}}$ depend on the representations of the scattering and the bound states. Thus, if more than one transitions are possible, in which either the scattering and/or the bound states belong to different representations, then their contributions to \cref{eq:MBSFsquared_Total} need to be separated, and $\boldsymbol{{\cal J}}_{\vec{k}, \{n\ell m\}}$, $\boldsymbol{{\cal Y}}_{\vec{k}, \{n\ell m\}}$ should be evaluated using the wavefunctions of the corresponding representations.

\subsection{Cross-sections for capture into the ground state \label{sec:BSF_CrossSections}}

The differential BSF cross-section for capture into the ground state $\{n\ell m\} = \{100\}$ is given by
\beq
\vrel \frac{ d\sigma_{\vec{k}\to \{100\}} }{d\Omega}  =  
\frac{|\vec{P}_{g}|}{64 \pi^2 M^2 \mu} 
\(
|\boldsymbol{{\cal M}}_{\vec{k}\to \{100\}}|^2
- |\hat{\vec{P}}_{g} \cdot \boldsymbol{{\cal M}}_{\vec{k}\to \{100\}}|^2 
\) \,,
\label{eq:dsigmadOmega_BSF}
\eeq
where energy-momentum conservation implies (cf.~\cref{eq:omega})
\beq
|\vec{P}_{g}| = {\cal E}_{\vec{k}} - {\cal E}_{10} 
=  \frac{\mu}{2} \, \[ \(\agBound\)^2 + \vrel^2 \] \,.
\label{eq:Pg}
\eeq
The leading-order contributions to the amplitude are 
$\boldsymbol{{\cal J}}_{\vec{k}, \{100\}} \propto \hat{\vec{k}}$ and $\boldsymbol{{\cal Y}}_{\vec{k}, \{100\}} \propto \hat{\vec{k}}$ (cf.~refs.~\cite{Petraki:2015hla,Petraki:2016cnz} and \cref{App:OverlapIntegrals}). Thus
\beq
|\boldsymbol{{\cal M}}_{\vec{k}\to \{100\}}|^2
- |\hat{\vec{P}}_{g} \cdot \boldsymbol{{\cal M}}_{\vec{k}\to \{100\}}|^2 =
|\boldsymbol{{\cal M}}_{\vec{k}\to \{100\}}|^2 \sin^2 \theta \,,
\nn
\eeq
where $\theta$ is the angle between $\vec{k}$ and $\vec{P}_g$, and $|\boldsymbol{{\cal M}}_{\vec{k}\to \{100\}}|^2$ is independent of $\theta$. Thus,
\beq
\sigma_{\vec{k}\to \{100\}} \vrel = 
\frac{(\agBound)^2+ \vrel^2}{48 \pi M^2}
\ |\boldsymbol{{\cal M}}_{\vec{k}\to \{100\}}|^2 \,.
\label{eq:sigmaBSF_general}
\eeq

For convenience, in the following we shall use the parameters [cf.~\cref{eqs:zetas_scattANDbound_App}]
\begin{subequations}
\label{eq:zetas_scattANDbound}
\label[pluralequation]{eqs:zetas_scattANDbound}
\begin{align}
\zetaScatt &\equiv \agScatt/\vrel \,, \label{eq:zeta_scatt} \\
\zetaBound &\equiv \agBound/\vrel \,. \label{eq:zeta_bound}
\end{align}
\end{subequations}
From the amplitudes of  \cref{eq:MBSFsquared_BoundStateSinglet,eq:MBSFsquared_ScatteringStateSinglet,eq:MBSFsquared_Total}, and the expressions \eqref{eqs:IntegralsCal_Coulomb} for the overlap integrals $\boldsymbol{{\cal J}}$ and $\boldsymbol{{\cal Y}}$, we find that the colour-averaged BSF cross-sections are
\begin{subequations}
\label{eq:sigmaBSF_GeneralGroup_Coulomb_All}
\label[pluralequation]{eqs:sigmaBSF_GeneralGroup_Coulomb_All}
\beq
\sigma_{\vec{k}\to \{100\}} \vrel = \frac{\pi \asBSF \agBound}{\mu^2}
\ \frac{2^7 C_2({\bf R})}{3 \dR^2}  
\ f_c \times S_{\BSF}  (\zetaScatt, \zetaBound) \,,
\tag{\ref{eq:sigmaBSF_GeneralGroup_Coulomb_All}}
\label{eq:sigmaBSF_GeneralGroup_Coulomb}
\eeq
where $f_c$ is a numerical factor that depends on the transition,
\beq
f_c = \left\{
\begin{alignedat}{100}
&\[1 + \frac{C_2({\bf G})}{2 \, C_2({\bf R})} \(\frac{\asNA}{\asBound} \) \]^2,&
\quad
&{\bf [adj] \to [1],}& 
\\
&\[1 - \frac{C_2({\bf G})}{2C_2({\bf R}) - C_2({\bf G})} \(\frac{\asNA}{\asBound} \) \]^2,&
\quad
&{\bf [1] \to [adj],}& 
\\
&\dR^2(\eta_1^2 + \eta_2^2)-2 + 
C_2({\bf G}) \[\dR C({\bf R}) - \frac{C_2({\bf G})}{2} \]  \(\frac{\asNA}{\agBound} \)^2,&
\quad
&{\bf rest,}& 
\end{alignedat}
\right.
\label{eq:fc}
\eeq
and
\beq
S_{\BSF} (\zetaScatt, \zetaBound) \equiv
\( \frac{2\pi \zetaScatt}{1-e^{-2\pi \zetaScatt}} \)
(1+\zetaScatt^2)
\[ \frac{\zetaBound^4 \ \exp \[ - 4 \, \zetaScatt \ {\rm arccot} (\zetaBound) \]}{(1+\zetaBound^2)^3}  \] .
\label{eq:SBSF_Coulomb}
\eeq
\end{subequations}
A few remarks are in order:
\bit
\item
Clearly, in the ${\bf [adj] \to [1]}$ transition, the contributions from all three diagrams of \cref{fig:FeynmanDiagrams_BSF_RadiativeVertices} add up. No (partial) cancellation occurs, for any group or representation, contrary to what was found in~\cite{Mitridate:2017izz,Keung:2017kot}. We note that our result reproduces the dissociation rate via gluon absorption of the colour-singlet bound state of a particle-antiparticle pair transforming in the (anti)fundamental of $SU(N)$, that was computed in ref.~\cite[eq.~(19)]{Brambilla:2011sg}.\footnote{The gluo-dissociation (\cite[eq.~(19)]{Brambilla:2011sg}) and  the radiative capture cross-sections [\cref{eqs:sigmaBSF_GeneralGroup_Coulomb_All}] are related via the Milne relation, which we review in \cref{App:MilneRelation} and use in \cref{Sec:S3Model}. Note that the gluo-dissociation cross-section of~\cite[eq.~(19)]{Brambilla:2011sg} is not averaged over the gluon degrees of freedom, while $\sigma_\ion$ in \cref{eq:MilneRelation} is.}

The potential in the singlet state is always attractive and gives rise to the tightest bound state. Thus, our results quite generally suggest that BSF can be very significant for phenomenology.\footnote{
For particle-antiparticle pairs transforming in the (anti)fundamental $SU(N)$, the opposite relative sign between the Abelian and non-Abelian contributions leads to an accidental near cancellation and thus a suppression of the adjoint-to-singlet capture cross-section by a factor of $(2N^2-1)^2$.

}

Moreover, the radiative transitions contribute to the self-energy of the initial state. From the optical theorem and \cref{eq:fc}, it follows that the forward scattering amplitude (or equivalently, the index of refraction) of the adjoint state is enhanced by the non-Abelian contribution, as is reasonable to expect.

\item
Since $\as$ runs only logarithmically, to a good approximation we may set
$\asNA\simeq\asBound$, at least in the parameter space where BSF is significant (cf.~\cref{foot:alphaNA}). Then, for the ${\bf [adj] \to [1]}$ and ${\bf [1] \to [adj]}$ transitions, the $f_c$  factors simplify.
\item
We recall that the couplings $\asBSF$, $\asNA$, $\asBound$, $\agBound$, $\agScatt$, and thus $\zetaScatt$ and $\zetaBound$, depend on the colour representations of the initial and final states (cf.~\cref{tab:MomentumTransfers}), and are different for every transition. 
\item
As noted in \cref{sec:BSF_RemainingTransitions}, if the group algebra allows for more than one transitions in the category ``rest'', then their contributions to $f_c$ have to be disentangled, in order for $S_\BSF$ to be computed. Note that a transition 
$(XX^\dagger)_{\bf \hat{R}} \to (XX^\dagger)_{\bf \hat{R}'} + g$ 
allowed by the group algebra contributes to $f_c$ even if ${\bf \hat{R}'}$  has a repulsive potential and cannot accommodate a bound state. 
\eit

\bigskip

The function $S_{\BSF}  (\zetaScatt, \zetaBound)$ encapsulates all the velocity dependence of $\sigma_{\BSF} \vrel$. The first two factors in \cref{eq:SBSF_Coulomb} arise solely from the scattering-state wavefunction and coincide with the Coulomb Sommerfeld enhancement of $p$-wave annihilation processes 
$S_1 (\zetaScatt) = [2\pi \zetaScatt/ (1-e^{-2\pi \zetaScatt})] (1+\zetaScatt^2)$. The factors inside the square brackets in \cref{eq:SBSF_Coulomb} arise from the convolution of the scattering-state and bound-state wavefunctions with the radiative vertices. 

Let us now discuss the asymptotic behaviour of $S_{\BSF}$ in various cases. 
\begin{subequations}
\label{eq:SBSF_AsymptoticScalings}
\label[pluralequation]{eqs:SBSF_AsymptoticScalings}
\bit
\item
At large velocities, $|\zetaScatt|, \zetaBound \ll 1$, BSF is very suppressed,
\beq
S_{\BSF} \simeq \zetaBound^4 \ll 1 .
\label{eq:SBSF_LargeVelocities}
\eeq
\item
For an attractive interaction in the scattering state ($\zetaScatt > 0$), and at low enough $\vrel$ such that $\zetaScatt \gtrsim 1$ and $\zetaBound \gtrsim 1$, 
\beq
S_{\BSF} \simeq 2\pi \zetaScatt \times
\(\frac{\zetaScatt}{\zetaBound} \)^2 
\exp\(-\frac{4 \zetaScatt}{\zetaBound} \) .
\label{eq:SBSF_AttractiveSS}
\eeq  
Since $\zetaScatt / \zetaBound = \agScatt/\agBound$ is constant, $S_{\BSF}$ exhibits the characteristic scaling $S_{\BSF} \propto 1/\vrel$.\footnote{This scaling appears also in the upper limit on inelastic cross-sections imposed by unitarity. This implies that the unitarity limit may be approached or realised only by Sommerfeld enhanced processes~\cite{vonHarling:2014kha,Baldes:2017gzw}.
}
We observe that $S_{\BSF}$ becomes maximal for $\agScatt/\agBound =0.5$. That is, the transition probability decreases for transitions between states governed by very different potentials ($\agScatt/\agBound \gg 1~\text{or}~\ll 1$). 
\item
For a repulsive interaction in the scattering state ($\zetaScatt < 0$), and at low enough $\vrel$ such that $\zetaScatt \lesssim -1$ and $\zetaBound \gtrsim 1$,
\beq 
S_{\BSF} \simeq 2\pi |\zetaScatt| 
\(\frac{\zetaScatt}{\zetaBound} \)^2 
\ \exp\[\(\frac{4}{\zetaBound} - 2\pi\) |\zetaScatt| \] .
\label{eq:SBSF_RepulsiveSS}
\eeq 
At low $\vrel$, $S_{\BSF}$ becomes exponential suppressed. However, the exponential suppression sets at $\zetaBound > 1$, by when BSF may already have an important effect on the DM density. It is interesting that the exponential suppression is more severe for tighter bound states (larger $\zetaBound$), i.e.~two particles that repel each other are less likely to be captured into a very deep bound state. This is consistent with the behaviour exhibited by \cref{eq:SBSF_AttractiveSS}.
\eit

\end{subequations}

\section{Dark matter co-annihilating with coloured partners \label{Sec:S3Model}}

\subsection{Simplified model and Boltzmann equation \label{sec:S3Model_ModelDetails}}

We assume that DM is a Majorana fermion $\x$ of mass $\mx$, that co-annihilates with a complex scalar triplet under $SU(3)_c$, denoted by $X$. The gauge interactions of $X$ are specified by the Lagrangian
\beq
{\d \cal L} = 
(D_{\m,ij} X_j)^\dagger \, (D_{ij'}^\m X_{j'}) 
- \mX^2 \, \Xdagger_j X_j^{} \,,
\label{eq:Lagrangian}
\eeq
where $D_{\m,ij} = \d_{ij} \partial_\mu + i \gs \, G_\m^a T^a_{ij}$ is the covariant derivative, with $G_\m^a$ being the gluon fields and $T^a$ are the generators.
$\x$ and $X$ are the lightest and next-to-lightest particles that are odd under a $Z_2$ symmetry which prevents $\x$ from decaying. We also assume that the interactions between $\x$ and $X$ -- which we shall leave unspecified -- keep them in chemical equilibrium throughout the freeze-out of their annihilation processes into other species.

As long as the relative mass splitting between DM and its coannihilating partner,
\beq
\d \equiv (\mX - \mx)/\mx \,,
\label{eq:Delta}
\eeq
is small, $\d \ll 1$, the DM density is determined by the $\chi-\chi$, $\chi-X$, $\chi-X^\dagger$ and $X-X^\dagger$ (co-)annhilation processes. It can be tracked by considering the sum of densities of all co-annihilating species,
\beq 
\tilde{Y} \equiv \Yx + \YX^{} + \YXdagger  =   \Yx + 2\YX^{} \,,
\label{eq:Ytilde} 
\eeq
where $Y_j \equiv n_j / s$, with $n_j$ being the number density of the species $j$ and $s \equiv (2\pi^2/45) \, \gstarS^{} \, T^3$ being the entropy density of the universe. Using the time parameter 
\beq  x \equiv \mx / T,  \label{eq:x}  \eeq 
the evolution of $\tilde{Y}$ is governed by the Boltzmann equation~\cite{Edsjo:1997bg}
\begin{subequations}
\label{eq:BoltzmannEq_All}
\label[pluralequation]{eqs:BoltzmannEq_All}	
\beq
\frac{d\tilde{Y}}{dx} = 
- \frac{c \, \gstareffsqrt \ \<\sigmaeff \, \vrel \>}{x^2} \ 
\ (\tilde{Y}^2 - \tilde{Y}_\eq^2) \,,
\tag{\ref{eq:BoltzmannEq_All}}
\label{eq:BoltzmannEq}
\eeq
where
\begin{align}
c &\equiv \sqrt{\pi / 45} \ \mpl \mx \,, 
\label{eq:c}
\\
\gstareffsqrt &\equiv \frac{\gstarS}{\sqrt{g_*}}
\(1 + \frac{T}{3g_{*\mathsmaller{S}}} \frac{dg_{*\mathsmaller{S}}}{dT} \) \,,
\label{eq:gstareff}
\\
\Yx^\eq &= \frac{90}{(2\pi)^{7/2}} \ \frac{\gx}{\gstarS} \ x^{3/2} \ e^{-x} \,, 
\label{eq:Yx equil} 
\\
\YX^\eq = \YXdagger^\eq &= 
\frac{90}{(2\pi)^{7/2}} \ \frac{\gX}{\gstarS} \ [(1+\d)x]^{3/2} \ e^{-(1+\d)x} \,, 
\label{eq:YX equil}
\end{align}
\end{subequations}
with $\gx = 2$ and $\gX^{} = 3$ being the $\chi$ and $X$ degrees of freedom.

The \emph{effective} cross-section $\<\sigmaeff \, \vrel \>$ in \cref{eq:BoltzmannEq} includes all annihilation and co-annihilation processes weighted by the densities of the participating species. We shall assume that the dominant contribution arises from the processes that annihilate $XX^\dagger$, with total cross-section $\sigma_{XX^\dagger}$, such that
\beq
\<\sigmaeff \, \vrel \> = 
\frac{2\YX^\eq \YXdagger^\eq \, \<\s_{\mathsmaller{X\Xdagger}} \, \vrel\> }{\tilde{Y}_\eq^2} = 
\< \s_{\mathsmaller{X\Xdagger}} \, \vrel\> 
\( \frac{2 \gX^2 (1+\d)^3 \, e^{-2 x \, \d}}{\[\gx + 2\gX (1+\d)^{3/2} \, e^{- x \, \d }\]^2} \) \,.
\label{eq:sigma_eff}
\eeq
Both the direct annihilation and the BSF processes contribute to $\sigma_{XX^\dagger}$, as we discuss in the following.

In this work, we shall neglect thermal effects. The thermal bath may affect the DM freeze-out in a variety of ways, including, on one hand, screening of the long-range interactions and, on the other hand, frequent (non-radiative) scattering processes that precipitate DM depletion via BSF \cite{Kim:2016kxt}. In the context of DM coannihilation with coloured partners, the latter have been considered in Ref.~\cite{Biondini:2018pwp}. The inclusion of thermal corrections for the radiative BSF processes considered here requires a comprehensive study that we leave for future work.

\subsection{Colour states and the running of the coupling}

The $X - X^\dagger$ colour interaction may be decomposed as
\begin{align}
{\bf 3 \otimes  \bar{3}} &= {\bf 1 \oplus 8 } \,.
\label{eq:3x3bar}
\end{align}
In each irreducible representation ${\bf \hat{R}}$, the gluon exchange gives rise to the Coulomb potential of  \cref{eq:V(r)} with the coupling $\ag$ given by \cref{eq:alpha_g_def}. The quadratic Casimir invariants for the $SU(3)$ representations of interest are
$C_2({\bf 1})  = 0$, 
$C_2({\bf 3})  = C_2({\bf \bar{3}}) = 4/3$, 
$C_2({\bf 8})  = 3$, 
therefore
\beq
\ag \equiv \as \times \left\{
\begin{aligned}
4/3,  &\qquad {\bf \hat{R} = 1},
\\
-1/6, &\qquad {\bf \hat{R} = 8}.
\end{aligned}
\right.
\label{eq:alpha_g}
\eeq

As discussed in \cref{sec:BSF_Potential}, the strong coupling $\as$ depends on the momentum transfer $Q$. In \cref{tab:MomentumTransfers_SU3}, we list the average $Q$ for the various vertices appearing in the annihilation and BSF processes, in this model. For the bound states, the momentum transfer depends itself on the strong coupling, $Q= Q (\as)$. In this case, we determine $\as$ by solving the numerically the equation 
\beq 
\as ( Q(\tilde{\a}) ) = \tilde{\a} \,,
\label{eq:alpha_s_determination} 
\eeq 
for $\tilde{a}$. We discuss further the effect of the $\as$ running in the following.

\begin{table}[ht!]
\centering
\renewcommand{\arraystretch}{3}

\begin{tabular}{|c|c|c|c|} 
\hline
	{\bf Vertices} 
&	\boldmath{$\as$}
&	\boldmath{$\ag$} 
&	\parbox[c]{24ex}{\centering {\bf Average momentum transfer}~\boldmath{$Q$}}
\\[1ex] \hline  \hline 
	\parbox[c]{18ex}{\centering 
	Annihilation: \\ 
	gluon emission}
&	$\asAnn$
&	
&	$\mX$
\\[0.5ex] \hline
	\multirow{2}{*}{	
	\parbox[c]{18ex}{\centering 
	Scattering-state \\ 
	wavefunctions}
	} 
&	\multirow{2}{*}{$\asScatt$} 
&	\parbox[c]{18ex}{\centering 
	Colour singlet \\
	\medskip	
	$\agScattSinglet = 4\asScatt / 3 $}
&	\multirow{2}{*}{$ \dfrac{\mX \, \vrel}{2}$}
\\[1ex] \cline{3-3}

&
&	\parbox[c]{18ex}{\centering 
	Colour octet \\
	\medskip
	$\agScattOctet = -\asScatt/6$}
&	
\\[1ex] \hline
	\parbox[c]{27ex}{\centering
	\medskip	
	Colour-singlet
	bound-state  
	wavefunction} 
&	$\asBoundSinglet$ 
&	$\agBoundSinglet = \dfrac{4\asBoundSinglet}{3}$
&	$\dfrac{\mX}{2} \, \(\dfrac{4\asBoundSinglet}{3} \) $
\\[1.5ex] \hline
	\parbox[c]{27ex}{\centering 
	\medskip
	Colour-singlet	bound-state formation:	gluon emission}
&	$\asBSFSinglet$
&	
&	$ \dfrac{\mX}{4} \left[ \vrel^2 + \(\dfrac{4\asBoundSinglet}{3} \)^2 \right]$
	\\[1.5ex] \hline
\parbox[c]{27ex}{\centering 
	\medskip
	$g X^\dagger X$ vertices in \\ 
	non-Abelian diagram for \\ 
	colour-singlet BSF}
&$\asNASinglet \approx\asBoundSinglet$
&	
&	
\parbox[c]{24ex}{\centering 
$(\mX/2) \, \sqrt{ \vrel^2 + {\agBoundSinglet}^2 }$ \\[1ex]
approximated with \\
$(\mX/2) \agBoundSinglet$
}
\\[2ex] \hline
\end{tabular}

\caption[]{\label{tab:MomentumTransfers_SU3} 
The momentum transfer $Q$ at which the strong coupling $\as(Q)$ is evaluated, for the various processes and the states participating in these processes, in the model of \cref{Sec:S3Model}.}
\end{table}


\begin{subfigures}
\label{fig:FeynmanDiagrams_ANN} 
\label{figs:FeynmanDiagrams_ANN} 
\begin{figure}[t!]
\centering
\begin{tikzpicture}[line width=1.1pt, scale=1.2]

\begin{scope}[shift={(2.9,0)}]
\node at (-2.25, 1) {$X^{}$};
\node at (-2.25, 0) {$X^\dagger$};
\draw[gluon]    (-1.6,0) -- (-1.6,1);
\node at (-1.1,0.5) {$\cdots$};
\draw[gluon]    (-0.55,0) -- (-0.55,1);
\draw[fermion]    (-2,1) -- (  0,1);
\draw[fermion]    ( 0,1) -- (0.5,0.5);
\draw[fermionbar] ( 0,0) -- (0.5,0.5);
\draw[fermionbar] (-2,0) -- (  0,0);
\filldraw (0.5,0.5) circle (5pt);
\draw[fermionnoarrow] (0.5,0.5) -- (1,0);
\draw[fermionnoarrow] (0.5,0.5) -- (1,1);
\end{scope}
\end{tikzpicture}
\caption{\label{fig:FeynmanDiagrams_ANN_GE} 
The $XX^\dagger$ annihilation is influenced by the Sommerfeld effect due to gluon exchange. The black blob represents the perturbative part of the annihilation processes (hard scattering).}

\bigskip

\begin{tikzpicture}[line width=1.1pt, scale=1.2]

\begin{scope}[shift={(0,0)}]
\begin{scope}[shift={(-2.5,0)}]
\node at (-1.25, 1) {$X^{}$};
\node at (-1.25, 0) {$X^\dagger$};
\draw[fermion]    (-1,1) -- (0,1);
\draw[fermion]    ( 0,1) -- (0,0);
\draw[fermionbar] (-1,0) -- (0,0);
\draw[gluon] (0,0) -- (1,0);
\draw[gluon] (0,1) -- (1,1);
\node at (1.2,1) {$g$};
\node at (1.2,0) {$g$};
\end{scope}
\begin{scope}[shift={(0,0)}]
\node at (-0.25, 1) {$X^{}$};
\node at (-0.25, 0) {$X^\dagger$};
\draw[fermion]    ( 0,1) -- (0.5,0.5);
\draw[fermionbar] ( 0,0) -- (0.5,0.5);
\draw[gluon] (0.5,0.5) -- (1,0);
\draw[gluon] (0.5,0.5) -- (1,1);
\node at (1.2,1) {$g$};
\node at (1.2,0) {$g$};
\end{scope}
\begin{scope}[shift={(2.5,0)}]
\node at (-0.25, 1) {$X^{}$};
\node at (-0.25, 0) {$X^\dagger$};
\draw[fermion]    ( 0,1) -- (0.5,0.5);
\draw[fermionbar] ( 0,0) -- (0.5,0.5);
\draw[gluon] (0.5,0.5) -- (1,0.5);
\draw[gluon] (1,0.5)   -- (1.5,0);
\draw[gluon] (1,0.5)   -- (1.5,1);
\node at (1.7,1) {$g$};
\node at (1.7,0) {$g$};
\end{scope}
\end{scope}
\end{tikzpicture}
\caption[]{\label{fig:FeynmanDiagrams_ANN_perturb} 
Tree-level diagrams contributing to the hard process $XX^\dagger \to gg$. Besides the $t$-channel diagram, there is the corresponding $u$-channel (not shown). The $s$-channel yields a $p$-wave contribution, which we neglect. 
}
\end{figure}
\end{subfigures}

\smallskip

\begin{figure}[t]
\centering
\includegraphics[height=0.45\textwidth]{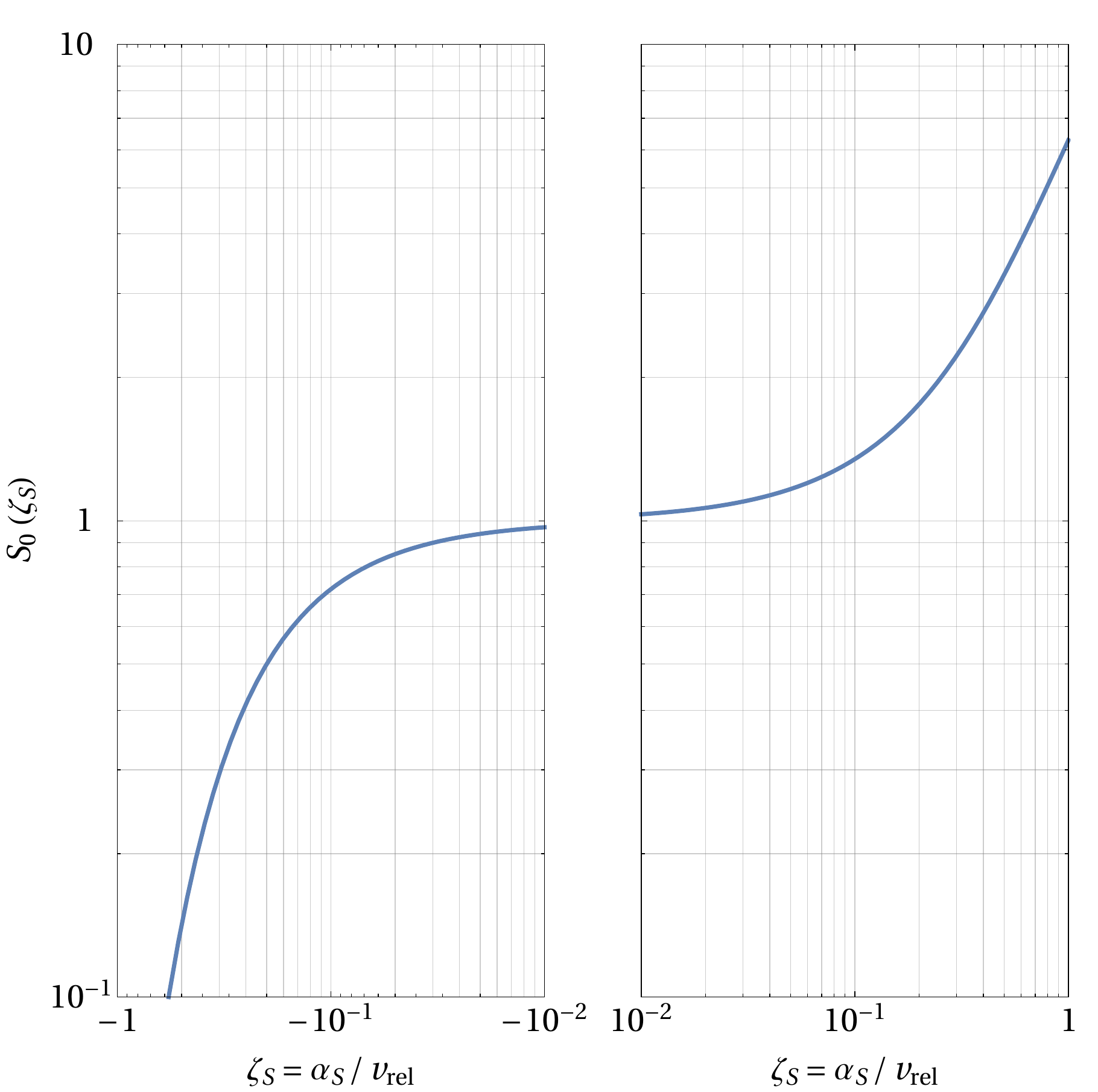}
~~~
\includegraphics[height=0.45\textwidth]{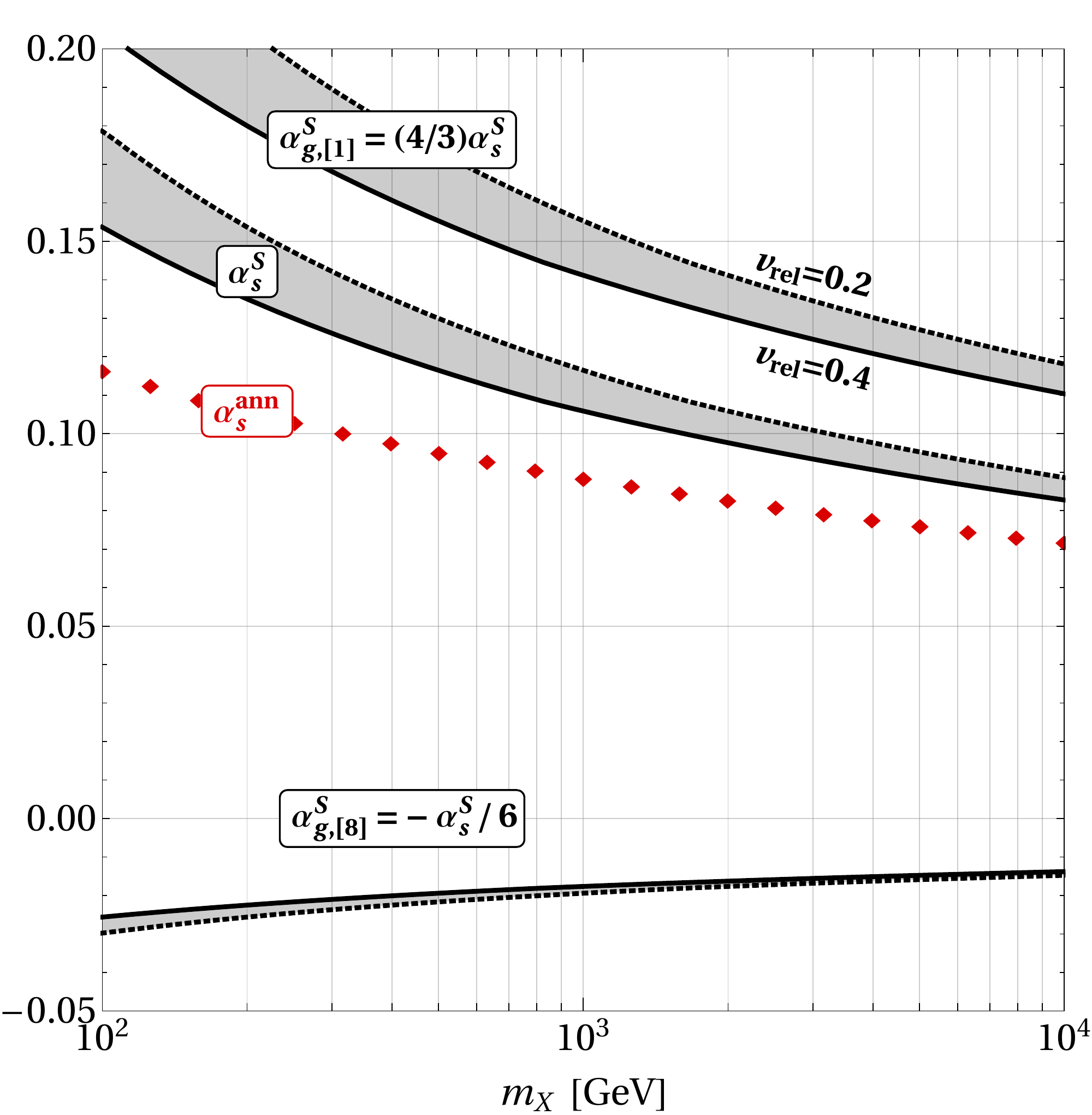}
\\ \smallskip
\includegraphics[height=0.45\textwidth]{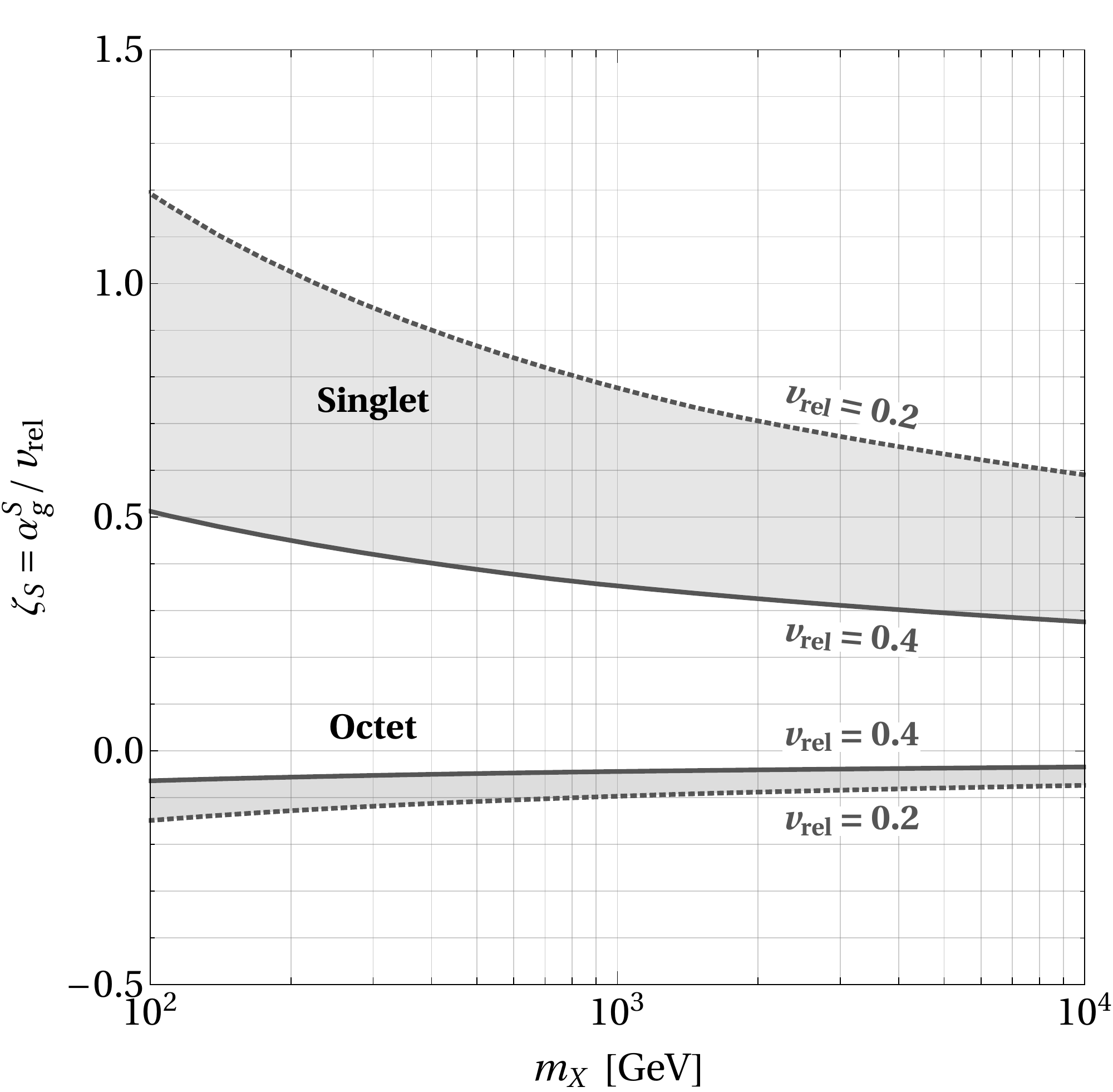}
~~~
\includegraphics[height=0.45\textwidth]{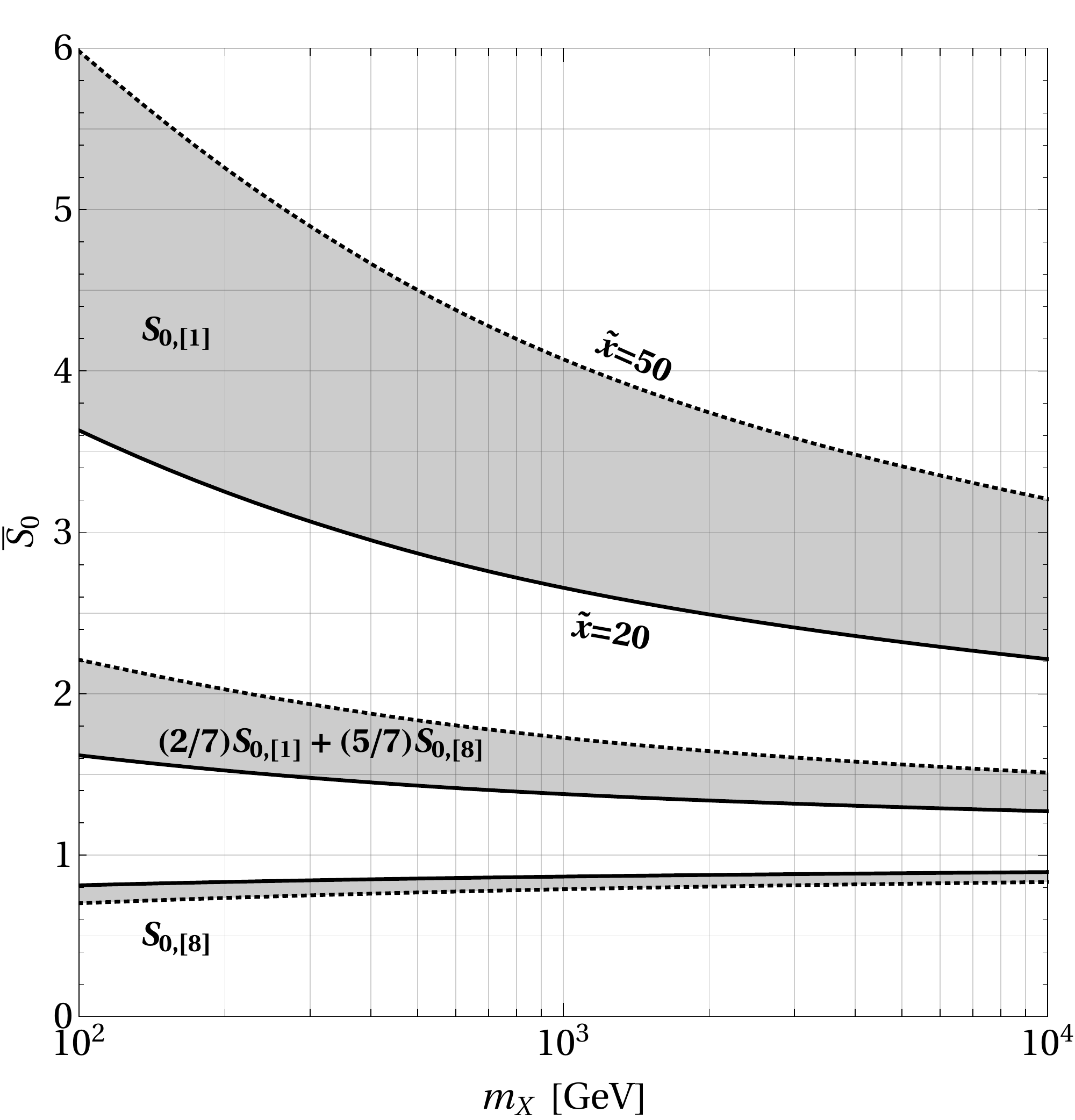}

\caption[]{\label{fig:S0}
\emph{Top left:} 
The Sommerfeld factor for $s$-wave annihilation, $S_0(\zetaScatt)$, vs $\zetaScatt \equiv \agScatt/v_{\rm rel}$, for both attractive ($\zetaScatt >0$) and repulsive ($\zetaScatt<0$) Coulomb interaction.
\emph{Top right:} 
The running of $\alpha_s$ in the scattering states, where the average momentum exchange is $Q = (\mX/2) v_{\rm rel}$. For the colour-singlet and the colour-octet states, $\agScattSinglet = (4/3)\asScatt$ and $\agScattOctet = -\asScatt/6$. We show $\asScatt$, $\agScattSinglet$ and $\agScattOctet$ in the velocity range $0.2 < v_{\rm rel} < 0.4$, that is typical during the DM freeze-out. For comparison, we also show the strong coupling at the gluon emission vertices of the annihilation processes, $\asAnn$, which corresponds to $Q = \mX$ (\emph{red diamonds}).  
\emph{Bottom left:}  The parameter $\zetaScatt = \agScatt / v_{\rm rel}$ that determines the Sommerfeld effect.  
\emph{Bottom right:} The thermally-averaged $s$-wave Sommerfeld factor, $\bar{S}_0$, for $\tilde{x} \equiv \mX / T$ within the indicative range $20 < \tilde{x} < 50$ during which the DM abundance freezes-out. 
}
\end{figure}

\subsection{Direct annihilation \label{sec:SU3model_Annihilation}}

$X X^\dagger$ pairs annihilate dominantly into gluons (cf.~\cref{fig:FeynmanDiagrams_ANN}), with cross-section~\cite{ElHedri:2016onc}
\beq
\sigma_{X \Xdagger \to gg} \vrel = 
\frac{14}{27} \frac{\pi (\asAnn)^2}{\mX^2} \times 
\( \frac27 S_{0, {\bf [1]}} + \frac57 S_{0, {\bf [8]}} \) \,,
\label{eq:sigma_XXbarTogg}
\eeq
where $S_{0, {\bf [1]}}$ and $S_{0, {\bf [8]}}$ are the $s$-wave Sommerfeld factors of the colour-singlet and colour-octet states, 
\begin{align}
S_{0, {\bf [1]}} \equiv S_{0} 
\(\frac{4\asScatt}{3\vrel} \) 
\qquad \text{and} \qquad
S_{0,{\bf [8]}} \equiv S_{0} 
\(-\frac{\asScatt}{6\vrel} \) \,.
\label{eq:SommerfeldFactors}
\end{align}
The function $S_0 (\zetaScatt)$ is the $s$-wave Sommerfeld enhancement factor (cf.~ref.~\cite{Cassel:2009wt} and \cref{App:Wavefunctions}),
\beq
S_0 (\zetaScatt) \equiv \frac{2\pi \zetaScatt}{1-e^{-2\pi \zetaScatt}} \,.
\label{eq:S0}
\eeq
The annihilation $X\Xdagger \to q \bar{q}$ is $p$-wave suppressed and we neglect it for simplicity.
In \cref{fig:S0}, we show $S_0(\zetaScatt)$, for both attractive and repulsive interactions, and depict the effect of the $\as$ running on the Sommerfeld factors. Because the momentum exchange in the scattering state is much smaller than on the gluon-emission vertices (cf.~\cref{tab:MomentumTransfers_SU3}), $\asScatt$ is considerably larger than $\asAnn$, as seen in the top right panel of \cref{fig:S0}.

\subsection{Bound-state formation, ionisation and decay \label{sec:SU3model_BoundStates}}

\subsubsection*{Formation \label{sec:SU3model_BSF}}

As seen from \cref{eq:alpha_g}, only the colour-singlet $XX^\dagger$ state interacts via an attractive potential and can form bound states. The only capture process via one-gluon emission is from the octet state,
\beq
(X + \Xdagger)_{\bf [8]} \ \to \ 
{\cal B} (X\Xdagger)_{\bf [1]} + g_{\bf [8]}  \,.
\label{eq:BSF 8to1}  
\eeq
Using $\dR = 3$,  $C({\bf 3}) = 1/2$,  $C_2({\bf 3}) = 4/3$ and $C_2({\bf G}) = 3$, and setting $\eta_1 = \eta_2=1/2$ and $\mu = \mX/2$, we find from \cref{eqs:sigmaBSF_GeneralGroup_Coulomb_All} the colour-averaged BSF cross-section,
\beq
\sigma_{\BSF}^{\bf [8] \to [1]} \vrel = 
\ \frac{2^7 17^2}{3^5} 
\ \frac{\pi \asBSF \asBoundSinglet}{\mX^2}
\times S_{\BSF}  (\zetaScatt, \zetaBound) \,,
\label{eq:sigmaBSF_SU3fundamental}
\eeq
where $S_{\BSF}  (\zetaScatt, \zetaBound)$ is given by \cref{eq:SBSF_Coulomb}, and here $\zetaScatt = \agScattOctet / \vrel$, $\zetaBound = \agBoundSinglet/\vrel$ [cf.~\cref{eqs:zetas_scattANDbound}].

\begin{figure}[t!]
\centering
\includegraphics[height=0.41\textwidth]{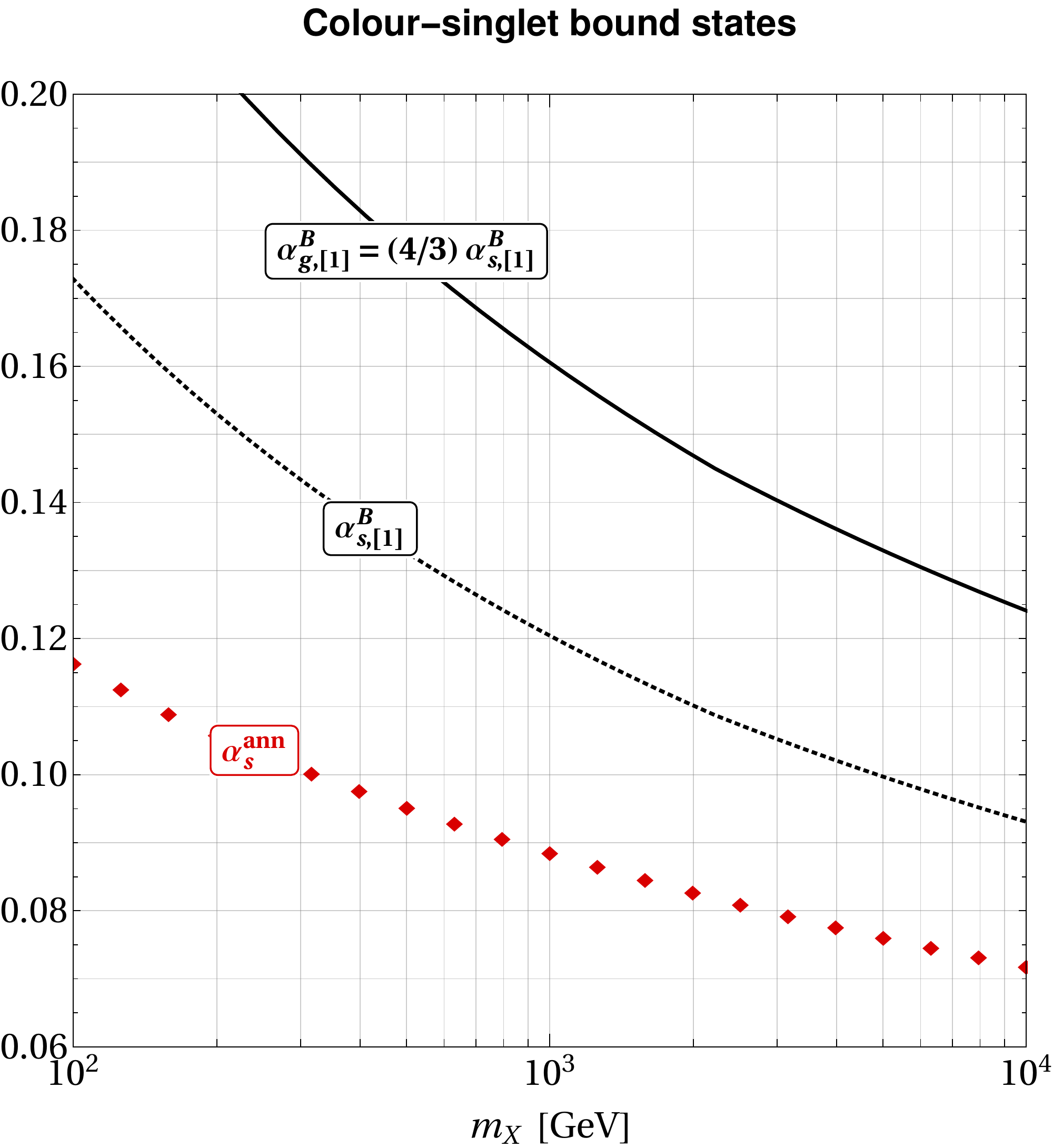}~~~~
\includegraphics[height=0.41\textwidth]{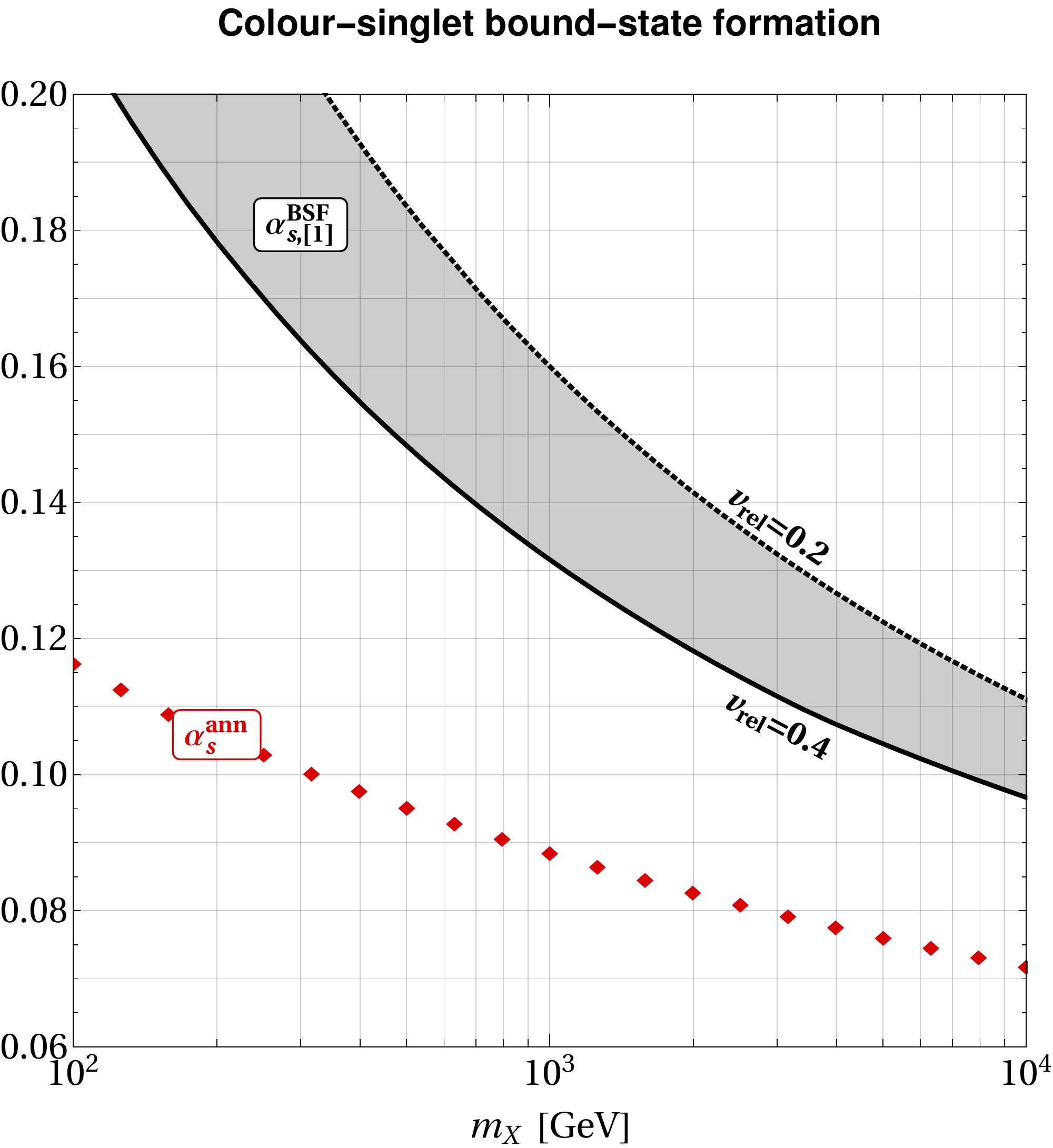}
\caption[]{\label{fig:AlphaRunning_BoundStates}
\emph{Left panel}: The strong coupling $\asBoundSinglet$ and the corresponding  $\agBoundSinglet$, that determine the colour-singlet bound-state wavefunction.  
\emph{Right panel}: The strong coupling at the gluon emission vertex during the formation of colour-singlet bound states, $\asBSFSinglet$. The emitted gluon carries away the binding energy of the bound state plus the kinetic energy of the scattering state; we show $\asBSFSinglet$ in the range $0.2 \leqslant v_{\rm rel} \leqslant 0.4$ that is indicative of the relative velocities during DM freeze-out. 
}
\end{figure}

\begin{figure}[t!]
\centering
\includegraphics[height=0.43\textwidth]{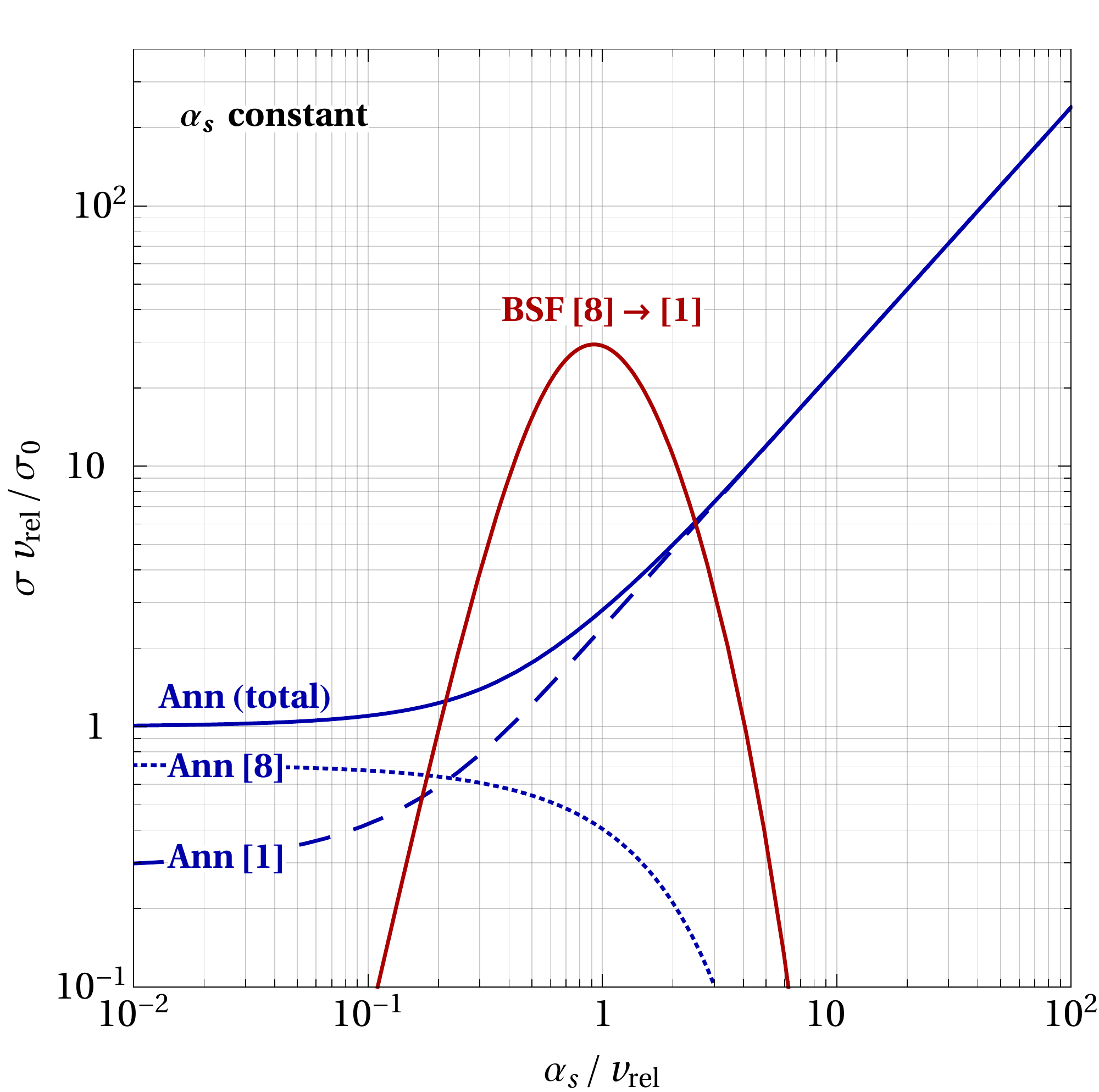}~~~~
\includegraphics[height=0.43\textwidth]{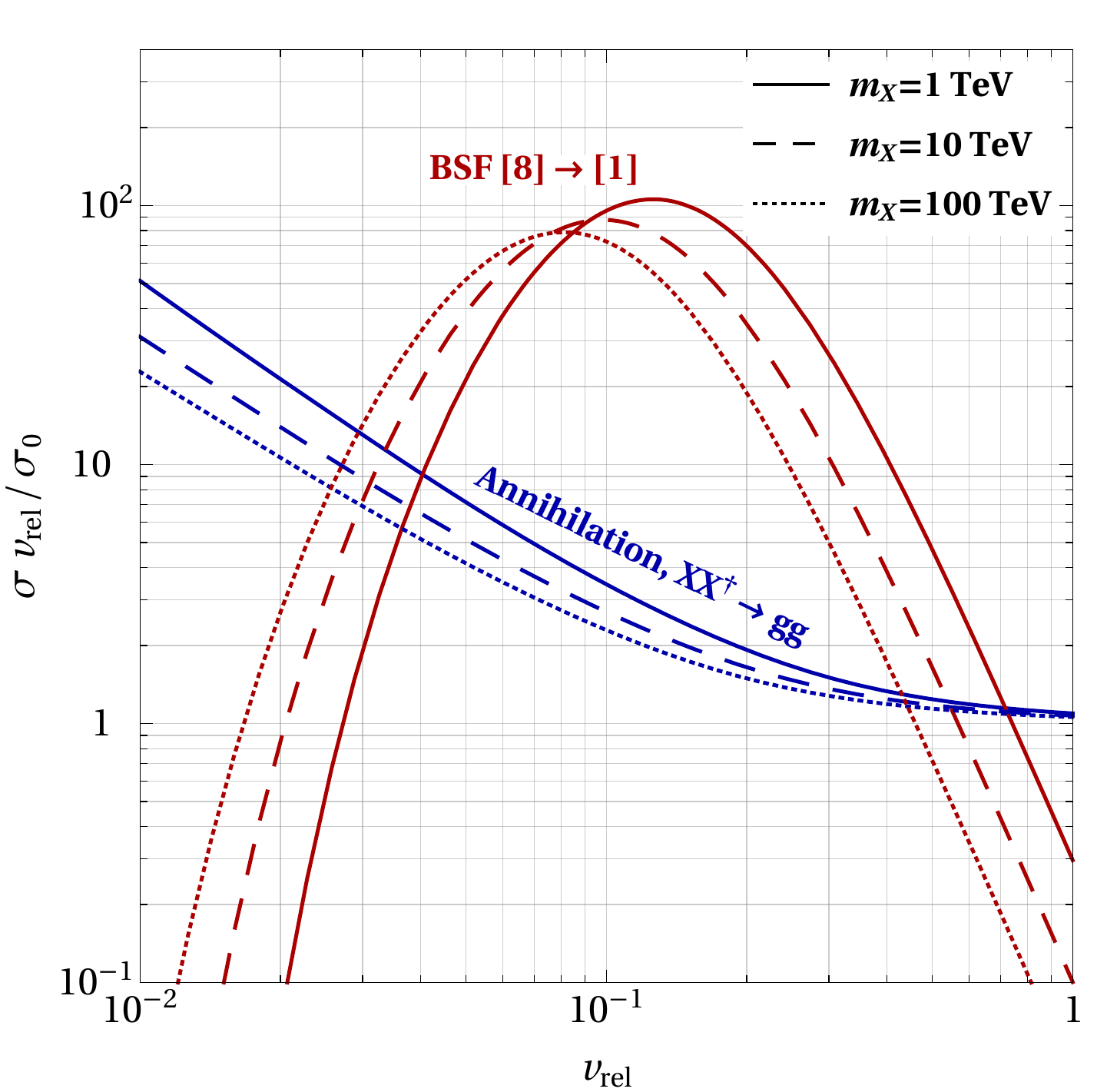}
\caption[]{
\label{fig:BSFvsANN}
\emph{Left}:
The cross-sections times relative velocity for annihilation into gluons, $XX^\dagger \to gg$ (\emph{blue lines}), and radiative capture into the ground state (\emph{red line}), normalised to the perturbative annihilation cross-section times relative velocity $\sigma_0 \equiv 14\pi\as^2/(27\mX^2)$. 
The \emph{blue dashed} and \emph{blue dotted} lines denote the contributions of the colour-singlet and the colour-octet scattering states in the total annihilation cross-section. We have ignored here the running of the strong coupling, which implies that $\sigma \vrel / \sigma_0$ depend only on $\alpha_s/\vrel$.
\emph{Right}: The cross-sections times relative velocity for annihilation into gluons, $XX^\dagger \to gg$ (\emph{blue lines}), and radiative capture into the ground state (\emph{red line}), normalised to $\sigma_0$. The lines corresponding to different values of $\mX$ differ due to the running of the strong coupling.}
\end{figure}

The coupling $\asBoundSinglet$ that determines the bound-state wavefunction, and the coupling $\asBSFSinglet$ that corresponds to the gluon radiation vertex in the capture process, are shown in \cref{fig:AlphaRunning_BoundStates}. Due to the small momentum transfer (cf.~\cref{tab:MomentumTransfers_SU3}), they are considerably larger than $\asAnn$ that corresponds to the gluon vertices in the $XX^\dagger \to gg$ annihilation.
This enhances further the BSF cross-section with respect to the annihilation cross-section, as seen by comparing the two panels in  \cref{fig:BSFvsANN}.

In \cref{fig:BSFvsANN}, we compare \cref{eq:sigmaBSF_SU3fundamental} to the cross-section for $XX^\dagger \to gg$ (cf.~\cref{eq:sigma_XXbarTogg}). At large velocities, $\as/\vrel \ll 1$, the BSF cross-section scales as $\sigma_{\BSF}^{\bf [8] \to [1]} \vrel \propto (\as/\vrel)^4$ and is subdominant to annihilation. At low velocities, $\as/\vrel \gg 1$, it becomes exponentially suppressed due to the Coulomb repulsion in the scattering state. 
$\sigma_{\BSF}^{\bf [8] \to [1]} \vrel$ peaks at $\as/\vrel \approx 1$, where it exceeds the annihilation cross-section by more than one order of magnitude.

The thermally-averaged BSF cross-section is 
\beq
\< \sigma_{\BSF}^{\bf [8] \to [1]} \vrel \> = 
\(\frac{\mu}{2\pi T}\)^{3/2}
\int d^3 \vrel 
\ \exp \(-\frac{\mu \vrel^2}{2T}\) 
[1+f_g(\omega)]
\ \sigma_{\BSF}^{\bf [8] \to [1]} \vrel \,,
\label{eq:sigmaBSF_averaged}
\eeq
where $\mu = \mX/2$ is the $X-X^\dagger$ reduced mass. Here, $f_g(\omega) = 1/(e^{\omega/T} - 1)$ is the gluon occupation number, with $\omega$ being the energy of the emitted gluon,
\beq
\omega = \frac{\mu}{2} [ (\agBound)^2 + \vrel^2] \,.
\label{eq:omega_SU3}
\eeq 
The factor $1+f_g(\omega)$ accounts for the Bose enhancement due to the final-state gluon, and is necessary to ensure the detailed balance between the bound-state formation and ionisation processes at $T \gtrsim \omega$~\cite{vonHarling:2014kha}, which encompasses a significant temperature range that is relevant to the DM freeze-out.

\subsubsection*{Ionisation \label{sec:SU3model_BSIonisation}}

The ionisation  and BSF cross-sections are related via the Milne relation, which we review in \cref{App:MilneRelation}. From \cref{eq:MilneRelation}, we find
\beq
\sigma_{\ion} =  \frac{\gX^2}{g_g g_{\cal B}^{}} 
\( \frac{\mu^2\vrel^2}{\omega^2}\)  \sigma_{\BSF} \,,
\label{eq:sigma_ion}
\eeq
where $g_g$ and $g_{\cal B}^{}$ are the gluon and bound-state degrees of freedom. The ionisation rate is
\beq
\Gamma_{\rm ion} = 
g_g \ \int_{\omega_{\min}}^\infty \dfrac{d\omega}{2\pi^2} \dfrac{\omega^2}{e^{\omega/T}-1} 
\ \sigma_{\ion} \,.
\nn
\eeq
Using \cref{eq:sigma_ion,eq:omega_SU3}, we obtain the ionisation rate of the colour-singlet bound states,
\beq
\Gamma_{\rm ion, \mathsmaller{\bf [1]}} = 
\dfrac{\gX^2 \mu^3}{2\pi^2 g_{\cal B, \bf [1]}^{}}
\int_0^\infty d\vrel \ 
\dfrac{\vrel^2}
{\exp \left\{ \dfrac{\mu [(\agBoundSinglet)^2 + \vrel^2] }{2T} \right\} - 1} 
\ \sigma_{\BSF}^{\bf [8] \to [1]} \vrel
\,.
\label{eq:GammaIon_def}
\eeq
We compute \eqref{eq:GammaIon_def} using $g_g=8$, $\gX^{} = 3$, $g_{{\cal B}, \bf [1]}^{} = 1$, $\mu = \mX/2$ and \cref{eq:sigmaBSF_SU3fundamental}.

\subsubsection*{Decay \label{sec:SU3model_BSDecay}}

The decay rate of $\ell=0$ bound states is related to the perturbative $s$-wave annihilation cross-section times relative velocity of the corresponding scattering states (see e.g.~\cite{Petraki:2015hla}),
\beq
\Gamma_{\dec} =
(\sigma_{\ann, \mathsmaller{\bf [\hat{R}]}}^{s-\rm wave} \vrel )
\ |\psi_{n \ell m}^{\mathsmaller{\bf [\hat{R}]}}(0)|^2 \,.
\label{eq:GammaDecaySinglet_def}  
\eeq
Note that $\sigma_{\ann, \mathsmaller{\bf [\hat{R}]}}^{s-\rm wave} \vrel$ corresponds to the colour configuration of the bound state and should be averaged over the bound-state colour degrees of freedom, rather than those of an unbound $XX^\dagger$ pair.

For the colour-singlet states (cf.~\cref{eq:sigma_XXbarTogg}, noting that $\gX^2 =9$ and $g_{\mathsmaller{\bf [1]}}^{} = 1$) 
\beq
\sigma_{\ann, \mathsmaller{\bf [1]}}^{s-\rm wave} \vrel =
\frac{4\pi (\asAnn)^2}{3\mX^2} \,.
\eeq
The colour-singlet ground-state wavefunction at the origin is [cf.~\cref{eq:psi_ground}]
\beq
|\psi_{1,0,0}^{\mathsmaller{\bf [1]}} (0)|^2 
=  \frac{\mu^3 (\agBoundSinglet)^3 }{\pi} 
=  \frac{2^3\mX^3 (\asBoundSinglet)^3 }{3^3 \pi} \,.
\label{eq:psi0squared_Coulomb}
\eeq
Thus, the decay rate of the colour-singlet ground state is
\beq
\Gamma_{\dec, \mathsmaller{\bf [1]}} = \frac{32}{81}  \, \mX \, (\asAnn)^2  (\asBoundSinglet)^3  \,.
\label{eq:GammaDecaySinglet}  
\eeq

\subsubsection*{Effective bound-state formation cross-section \label{sec:SU3model_EffectiveBSF}}

The effect of unstable bound states on the DM relic density is governed by a system of coupled Boltzmann equations for the unbound and bound particles that describe the interplay between bound-state formation, ionisation and decay processes~\cite{vonHarling:2014kha}. However, it is possible to incorporate the effect of bound states in a single Boltzmann equation for the unbound particles, using an effective BSF cross-section that is weighted by the fraction of bound states that decay (rather than getting ionised),
\beq
\< \sigma_{\BSF} \vrel\>_{\eff}  \equiv  
\< \sigma_{\BSF}^{\bf [8] \to [1]} \vrel \>
\times \(\frac{\Gamma_{\dec, \mathsmaller{\bf [1]}} }{\Gamma_{\dec, \mathsmaller{\bf [1]}} + \Gamma_{\ion, \mathsmaller{\bf [1]}}} \) \,.
\label{eq:sigmaBSFeff}
\eeq

\begin{figure}[t]
\centering
\includegraphics[height=0.43\textwidth]{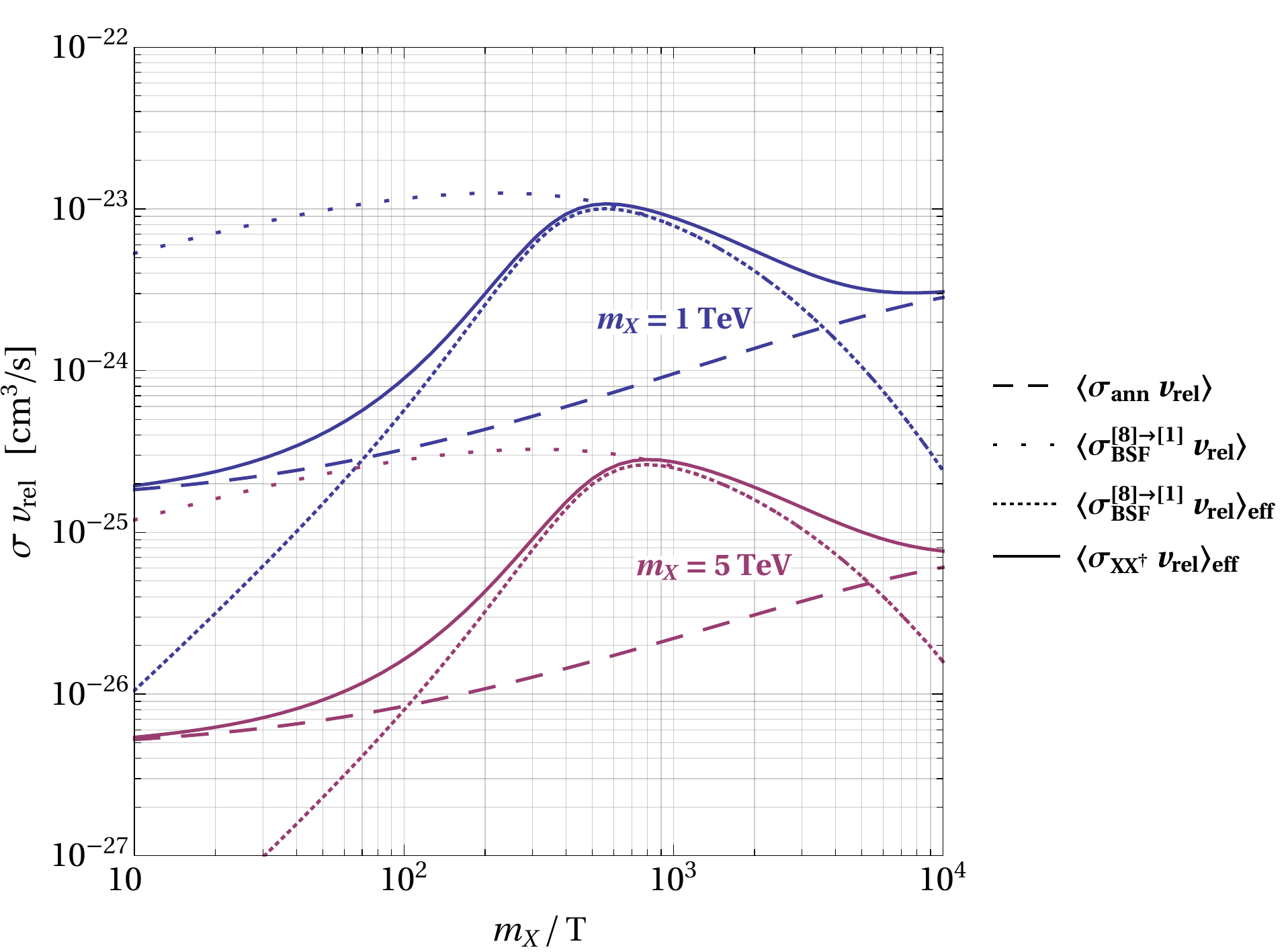}
\caption[]{\label{fig:BSFvsANN_ThermAver}
\emph{Dashed lines:} The thermally averaged cross-section for annihilation into gluons, $XX^\dagger \to gg$ [cf.~\cref{eq:sigma_XXbarTogg}].
\emph{Wide-spaced dotted lines:} The thermally averaged cross-section for the radiative capture into the colour-singlet bound state [cf.~\cref{eq:sigmaBSF_averaged}].
\emph{Densely spaced dotted lines:} The effective bound-state formation cross-section, which is weighted by the fraction of $XX^\dagger$ pairs that decay into radiation, rather than getting ionised [cf.~\cref{eq:sigmaBSFeff}].
\emph{Solid lines:} The total effective cross-section of processes that deplete $XX^\dagger$ pairs [cf.~\cref{eq:sigma_XXdagger}]. This determines the DM effective annihilation cross-section via \cref{eq:sigma_eff}.
}
\end{figure}

\subsection{Relic density \label{sec:Results}}

The total cross-section of the processes that deplete $XX^\dagger$ is
\beq
\< \sigma_{XX^\dagger} \,\vrel \> = 
\< \sigma_{XX^\dagger \to gg}  \,\vrel \> +
\< \sigma_{\BSF} \,\vrel \>_{\eff} \,,
\label{eq:sigma_XXdagger}
\eeq
where the individual cross-sections are given in \cref{{eq:sigma_XXbarTogg},eq:sigmaBSFeff}. From \cref{eq:sigma_eff,eq:sigma_XXdagger}, we obtain $\< \sigma_{\eff} \,\vrel \>$ which enters the Boltzmann \cref{eq:BoltzmannEq}. In \cref{fig:BSFvsANN_ThermAver}, we show $\< \sigma_{\eff} \,\vrel \>$ as a function of the time parameter $\mX/T$, together with the contributions it receives from direct annihilation and from BSF. At early times, the depletion of DM via BSF is impeded by the large ionisation rate of the bound states. However, BSF becomes more efficient than direct annihilation in depleting DM  at $\mX / T \gtrsim 70$ for $\mX \sim \TeV$, suggesting that a sizeable effect on the DM density should be expected.

\begin{figure}[t]
\centering
\includegraphics[height=0.45\textwidth]{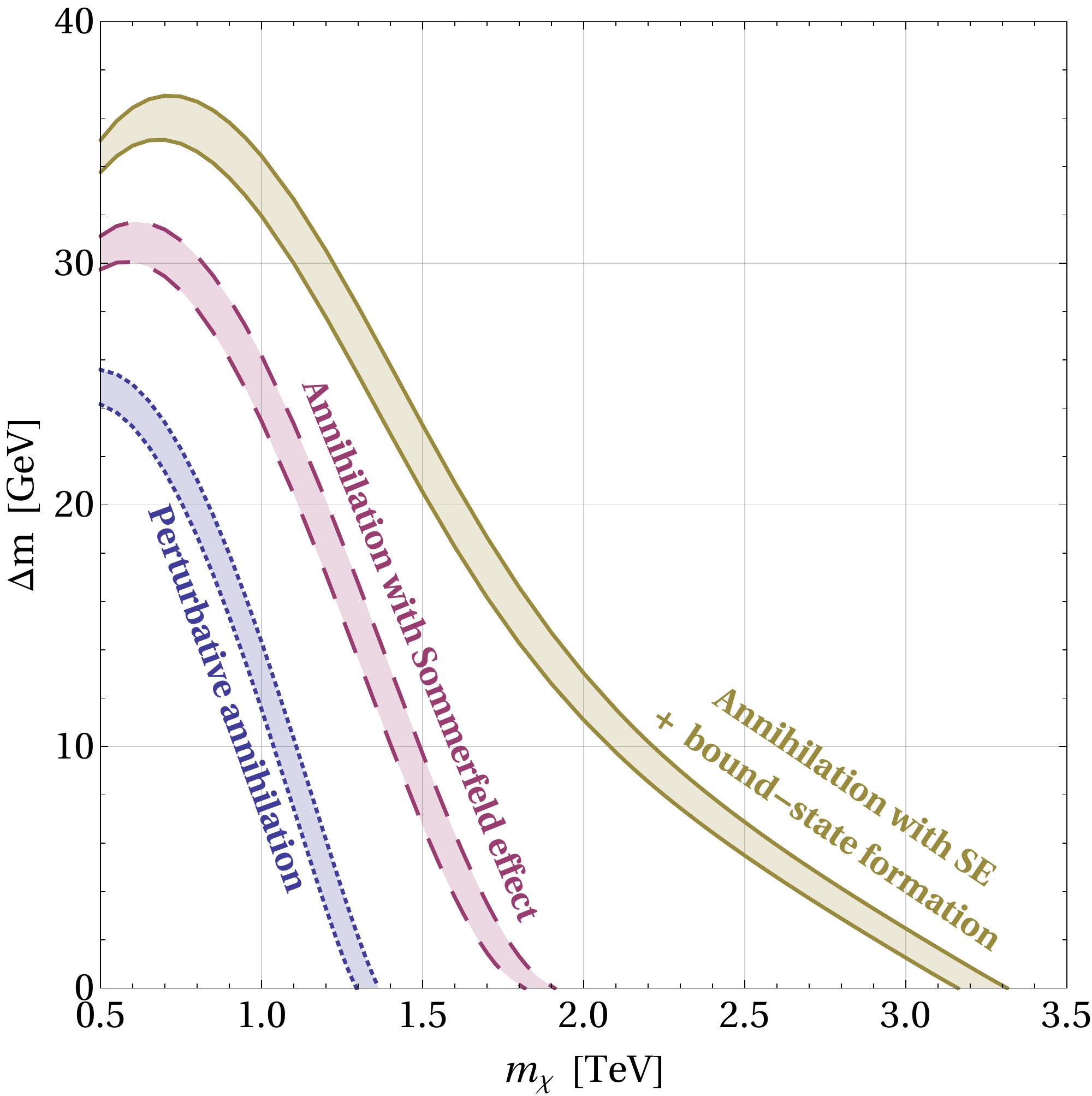}~~~
\includegraphics[height=0.45\textwidth]{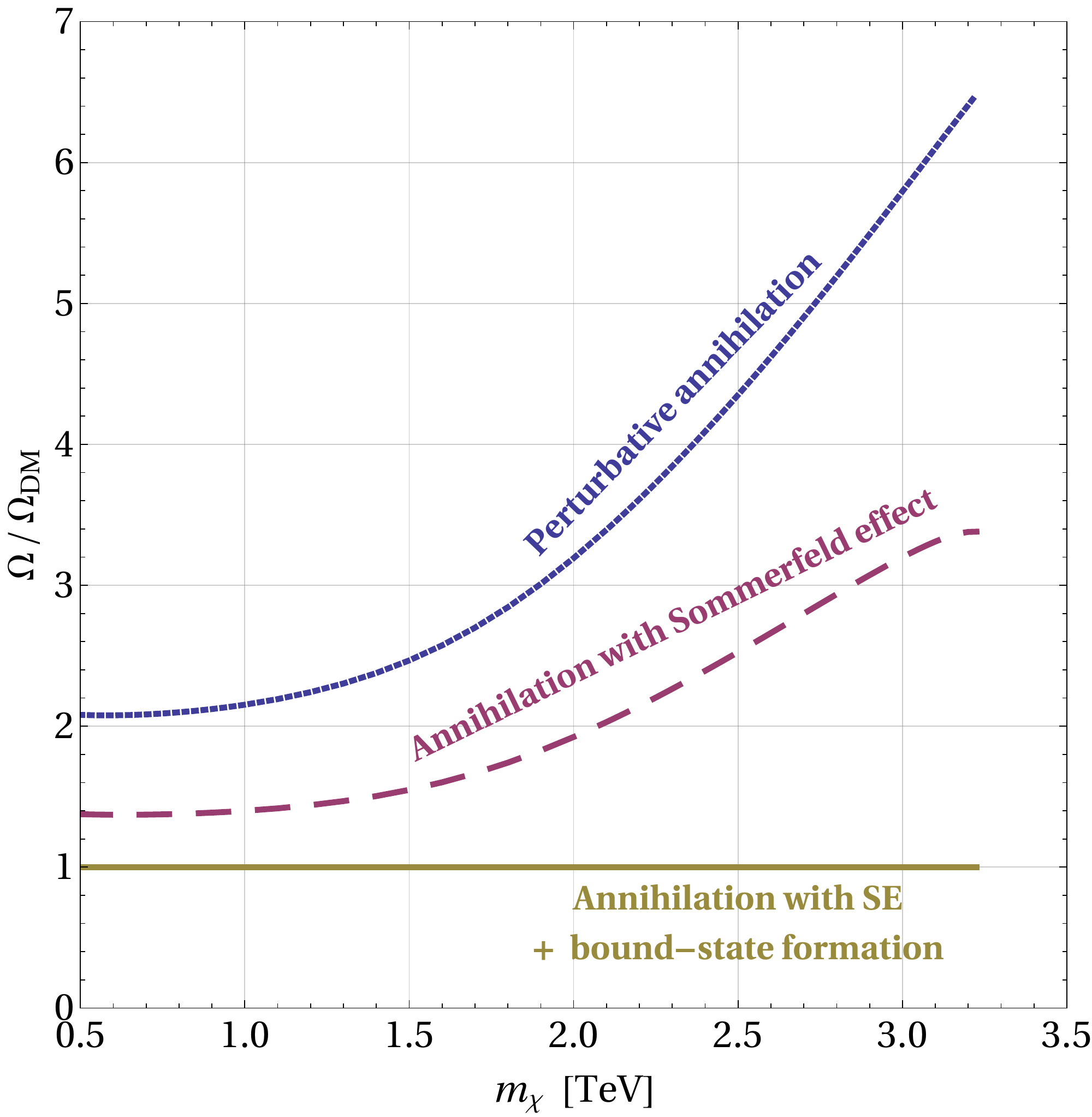}
\caption[]{\label{fig:m-Deltam_AnnBSF}	
\emph{Left panel:} The mass splitting $\Delta m$ between DM and its coloured co-annihilating partner, that is required to obtain the observed DM density. 
The \emph{blue dotted band} takes into account perturbative annihilation only, 
the \emph{purple dashed band} incorporates the Sommerfeld effect on the direct annihilation, and 
the \emph{yellow solid band} includes also the effect of bound-state formation and decay.
The width of the bands arises from the 3$\sigma$ uncertainty on the DM density.
\emph{Right panel:} The impact of the Sommerfeld effect and bound-state formation on the DM density. $\Delta m$ is fixed with respect to $m_\chi$ along the yellow solid band of the left panel.
We present the ratios of the relic densities predicted by perturbative annihilation only (blue dotted line) and by Sommerfeld-enhanced annihilation (purple dashed line), to the relic density predicted by the full computation that includes the effect of bound states. 
}
\end{figure}

In \cref{fig:m-Deltam_AnnBSF} we present the results of the relic density computation. 
In the left panel, we show the mass splitting $\Delta m \equiv \mX - \mx$ vs. $\mx$, in three different cases, (i) considering perturbative annihilation only, (ii) taking into account the Sommerfeld effect on the direct annihilation processes, and (iii) including the formation and decay of unstable bound states.  In agreement with previous works~\cite{Harz:2014gaa,Liew:2016hqo}, we confirm that the Sommerfeld effect has a considerable impact on the predicted $\Delta m$ and $m_\x$. In addition, we find that BSF has a significant effect. It implies that the mass splitting can be as high as $\sim 38~\GeV$, and DM can be as heavy as $3.3~\TeV$.
For the viable $\mx$ and $\Delta m$ values determined by the full computation, we show in the right panel of \cref{fig:m-Deltam_AnnBSF}, the depletion of DM due to the Sommerfeld enhancement of the direct annihilation and due to BSF. We find that
BSF depletes DM by $(40 - 240)\%$. Clearly, this far exceeds the experimental uncertainty on the DM density.

\section{Conclusion \label{Sec:Conclusions}}

Long-range interactions imply that non-perturbative effects and a variety of radiative processes come into play. Here, we have considered the radiative capture of non-relativistic particles into bound states, in unbroken non-Abelian gauge theories, in the regime where the gauge coupling is perturbative. This can be important in multi-TeV WIMP DM scenarios, 
in scenarios where DM co-annihilates with coloured particles, as well as in hidden sector models.

Our main results include the amplitude for the radiative formation of bound states via one-gluon emission, for arbitrary representations and masses of the interacting particles [cf.~\cref{eq:MBSF_general}], and the complete BSF cross-sections for particles transforming in conjugate representations [cf.~\cref{eqs:sigmaBSF_GeneralGroup_Coulomb_All}], but still for arbitrary masses.

As a first application of our results, we considered a simplified model where DM coannihilates with particles transforming in the fundamental of $SU(3)_c$, and showed that the formation and decay of particle-antiparticle bound states can affect the DM relic density very significantly.  This implies larger DM mass and/or mass splitting between DM and its coannihilating particles, thereby altering the interpretation of the experimental results, and affecting the detection prospects. In particular, larger mass splittings imply the production of harder jets that can be more easily probed in collider experiments. Moreover, larger DM masses motivate indirect searches in the multi-TeV regime. 

While the analytical formulae \eqref{eqs:sigmaBSF_GeneralGroup_Coulomb_All} assume a Coulomb potential, it is straightforward to generalise our results to other potentials, by computing the overlap integrals \eqref{eqs:calIntegrals_definition} using the wavefunctions arising from those potentials. This allows to include, for example, thermal masses for the gauge bosons, as well as the effect of multiple force mediators. The latter has been shown to be important in models where the (co-)annihilating particles possess a significant coupling to the Higgs~\cite{Harz:2017dlj,Biondini:2018xor}. We leave these extensions for future work.

\clearpage
\appendix

\section*{Appendices}

\section{Scattering-state and bound-state wavefunctions \label{App:Wavefunctions}}

The non-relativistic potentials due to gluon exchange, for the scattering and bound states are
\begin{subequations}
\label{eq:Potential}
\label[pluralequation]{eqs:Potential}
\begin{align}
V_{\rm scatt}(r) &= - \agScatt / r \,,
\label{eq:V(r) scatt}
\\
V_{\rm bound}(r) &= - \agBound / r \,.
\label{eq:V(r) bound}
\end{align}
\end{subequations}
where $\agScatt$ may be either positive or negative, but $\agBound > 0$. 
The scattering and bound states are characterised by the momenta
\begin{subequations}
\label{eq:momenta}
\label[pluralequation]{eqs:momenta}
\begin{align}
\vec{k} &\equiv \mu \vec{v}_{\rm rel} \,,
\label{eq:k}
\\
\kappa &\equiv  \mu \agBound \,,
\label{eq:kappa}
\end{align}
\end{subequations}
where $\vec{v}_{\rm rel}$ is the expectation value of the relative velocity in the scattering state and $\kappa$ is the Bohr momentum of the bound state. The corresponding wavefunctions, $\phi_{\vec k}(\vec r)$ and $\psi_{n\ell m}(\vec r)$, with $\{n,\ell,m\}$ being the principal and angular-momentum quantum numbers, obey the Schr\"odinger equations 
\begin{subequations}
\label{eq:SchrEq}
\label[pluralequation]{eqs:SchrEq}
\begin{align} 
\[-\frac{\nabla^2}{2\mu} + V_{\rm scatt} (\vec r)\] \f_{\vec k}(\vec r) 
&= {\cal E}_{\vec k} \, \f_{\vec k}(\vec r) \,,
\label{eq:SchrEq Scatt}
\\
\[-\frac{\nabla^2}{2\mu} + V_{\rm bound}(\vec r)\] \psi_{n\ell m}(\vec r) 
&= {\cal E}_{n\ell} \, \psi_{n\ell m}(\vec r) \, ,
\label{eq:SchrEq Bound}
\end{align}
\end{subequations}
where 
\begin{subequations}
\label{eq:Energies}
\label[pluralequation]{eqs:Energies}
\begin{align}
{\cal E}_{\vec{k}} &\equiv \frac{\vec{k}^2}{2\mu}  
= \frac{1}{2} \mu \vrel^2 \,,
\label{eq:Ek}
\\
{\cal E}_{n\ell}   &\equiv -\frac{\kappa^2}{2n^2\mu}   
= -\frac{1}{2n^2} \, \mu (\agBound)^2 \,,
\label{eq:Enl}
\end{align}
\end{subequations}
The wavefunctions are normalised according to
\begin{subequations}
\label{eq:WF norm}
\label[pluralequation]{eqs:WF norm}
\begin{align}
\int d^3r \: \f_{\vec k}^*(\vec r) \, \f_{\vec k'}(\vec r) 
&= (2\p)^3 \d^3(\vec k - \vec k') \, . \label{eq:phi norm}
\\
\int d^3r \: \psi_{n\ell m}^*(\vec r) \, \psi_{n'\ell' m'}^{} (\vec r) 
&= \d_{nn'} \d_{\ell\ell'} \d_{m m'} \,. \label{eq:psi norm}
\end{align}
\end{subequations}

\medskip

For convenience, we shall define
\begin{subequations}
\label{eq:zetas_scattANDbound_App}
\label[pluralequation]{eqs:zetas_scattANDbound_App}
\begin{align}
\zetaScatt &\equiv \agScatt/\vrel \,, \label{eq:zeta_scatt_App} \\
\zetaBound &\equiv \agBound/\vrel \,. \label{eq:zeta_bound_App}
\end{align}
\end{subequations}
and
\beq
S_0 (\zetaScatt) \equiv \frac{2\pi\zetaScatt}{1-e^{-2\pi\zetaScatt}} \,.
\label{eq:S0Coul}
\eeq
The solutions to \cref{eqs:SchrEq} with the potentials \eqref{eq:V(r)}, are (see e.g.~\cite{Messiah:1962})
\begin{subequations}
\begin{align}
\phi_{\vec k} (\vec r) 
&= 
\ \sqrt{S_0 (\zetaScatt)} 
\ {}_1F_1 [i\zetaScatt;\ 1; \ i(kr - \vec k \cdot \vec r)] 
\ e^{i \vec k \cdot \vec r} \, .
\label{eq:phi}
\\
\psi_{n\ell m}(\vec r) 
&= \k^{3/2} 
\ \[\frac{4(n-\ell-1)!}{n^4(n+\ell)!}\]^{1/2}
\ \(\frac{2\kappa r}{n}\)^{\ell} 
\ L_{n-\ell-1}^{(2\ell+1)} \( \frac{2\kappa r}{n} \)
\ e^{-\kappa r/n} 
\ Y_{\ell m} (\W_{\vec r}) \, ,
\label{eq:psi}
\end{align}
\end{subequations}
where ${}_1F_1$ is the confluent hypergeometric function of the first kind, and $L_n^a$ are the generalised Laguerre polynomials of degree $n$. (We assume the normalisation condition 
$\int_0^\infty z^a e^{-z} L_n^{(a)}(z) L_m^{(a)}(z) dz = [\Gamma(n+a+1)/n!] \, \d_{n,m}$.) For the ground state, $\{n,\ell,m\} =\{1,0,0\}$,
\beq
\psi_{100}^{} (\vec r) =  \sqrt{\kappa^3 / \pi} \ e^{-\kappa r} \,.
\label{eq:psi_ground}
\eeq
Note that $S_0$ is the Sommerfeld factor for $s$-wave annihilation (see e.g.~\cite{Cassel:2009wt})
\beq
S_0 (\zetaScatt) = |\f_{\vec{k}} (r=0)|^2 \,.
\label{eq:S0 def}
\eeq
In \cref{Sec:BSF}, we also need the Fourier transforms of the wavefunctions, defined as
\begin{subequations}
\label{eq:FourierTransformsWavefunctions}
\label[pluralequation]{eqs:FourierTransformsWavefunctions}
\begin{align}
\tilde{\phi}_{\vec{k}}^{}(\vec{q}) &= \int d^3 r \ \phi_{\vec{k}}^{}(\vec{r}) \ e^{-i\vec{q} \, \vec{r}} \,,&
\phi_{\vec{k}}^{}(\vec{r}) &= \int \frac{d^3 q}{(2\pi)^3} \ \tilde{\phi}_{\vec{k}}^{}(\vec{q}) \ e^{i\vec{q} \, \vec{r}} \,,
\\ 
\tilde{\psi}_{n\ell m}^{}(\vec{p}) &= \int d^3 r \ \psi_{n\ell m}^{}(\vec{r}) \ e^{-i\vec{p} \, \vec{r}} \,,&
\psi_{n\ell m}^{}(\vec{r}) &= \int \frac{d^3 p}{(2\pi)^3} \ \tilde{\psi}_{n\ell m}^{}(\vec{p}) \ e^{i\vec{p} \, \vec{r}} \,.
\end{align}
\end{subequations}

\section{Overlap integrals for capture into the ground state \label{App:OverlapIntegrals}}

For the BSF cross-sections of interest, we need to compute  the overlap integrals defined in~\cref{eqs:calIntegrals_definition}. In coordinate space, they become
\begin{subequations}
\label{eq:calIntegrals_CoordSpace}
\label[pluralequation]{eqs:calIntegrals_CoordSpace}
\begin{align}
\boldsymbol{{\cal J}}_{\vec{k}, \{n\ell m\}} (\vec{b}) 
&\equiv
\int \frac{d^3 p}{(2\pi)^3} \ \vec{p} 
\, \tilde{\psi}_{n\ell m}^*(\vec{p}) 
\, \tilde{\phi}_{\vec{k}}^{} (\vec{p}+\vec{b})
=
i \int d^3 r \ [\nabla \psi_{n\ell m}^* (\vec r)] \, \phi_{\vec{k}} (\vec{r}) \ e^{-i \vec{b} \vec{r}} \,,
\label{eq:Jcal_CoordSpace}
\\
\boldsymbol{{\cal Y}}_{\vec{k}, \{n\ell m\}}  
&\equiv
8\pi\mu\asNA
\int \frac{d^3 p}{(2\pi)^3} \frac{d^3 q}{(2\pi)^3} 
\frac{\vec{q}-\vec{p}}{(\vec{q}-\vec{p})^4} 
\, \tilde{\psi}_{n\ell m}^*(\vec{p}) 
\, \tilde{\phi}_{\vec{k}}^{} (\vec{q})
\nn \\
&=
-i \mu \asNA \int d^3 r \, \psi_{n\ell m}^* (\vec r) \, \phi_{\vec{k}} (\vec{r}) \, \hat{\vec{r}} \,.
\label{eq:Ycal_CoordSpace}
\end{align}
\end{subequations}
In deriving \cref{eq:Ycal_CoordSpace}, we Fourier-transformed $\tilde{\psi}_{n\ell m}^*(\vec{p})$ and $\tilde{\phi}_{\vec{k}}^{} (\vec{q})$, and used the following integral in the limit $m_g \to 0$,  
\begin{align}
\int \frac{d^3 q}{(2\pi)^3} \ \frac{ \vec{q} \, e^{-i \vec{q} \, \vec{r}} }{(\vec{q}^2+m_g^2)^2} 
&= i \, \nabla_\vec{r} 
\int \frac{d^3 q}{(2\pi)^3} \ \frac{e^{-i \vec{q} \, \vec{r}} }{(\vec{q}^2+m_g^2)^2} 
\nn \\
&= \frac{i}{4\pi^2} \nabla_\vec{r} \int_0^\infty dq \ q^2 
\int_{-1}^{1} d\cos\theta \ \frac{e^{-i q r \cos\theta}}{(q^2+m_g^2)^2} 
\nn \\
&= \frac{i}{2\pi^2} \nabla_\vec{r} \[ \frac{1}{r} \, \int_0^\infty dq \ \frac{q \sin (qr)}{(q^2+m_g^2)^2} \]
= \frac{i}{2\pi^2} \nabla_\vec{r} \[ r \, \int_0^\infty dq \ \frac{q \sin q}{(q^2+m_g^2 r^2)^2} \]
\nn \\
&= \frac{i}{2\pi^2} \nabla_\vec{r} \[ \frac{\pi e^{-m_g r}}{4m_g} \]
=-\frac{i \, e^{-m_g r}}{8\pi} \, \hat{\vec{r}} \,.  
\label{eq:Integral_qOVERq4}
\end{align}
In the following, we consider capture into the ground state only, $\{n\ell m\} = \{100\}$.

Following refs.~\cite{Pearce:2013ola,Petraki:2015hla,Petraki:2016cnz}, we compute the overlap integrals~\eqref{eq:calIntegrals_CoordSpace} using the identity~\cite{AkhiezerMerenkov_sigmaHydrogen}
\beq
\int d^3r 
\ \frac{ e^{ i(\vec{k}-\vec{b}) \cdot \vec{r} - \kappa r }  }{4\pi r}
\ {}_1F_1 [ i \zetaScatt, 1, i (kr - \vec{k} \cdot \vec{r}) ] 
= \dfrac{[\vec{b}^2 + (\kappa - i k)^2]^{-i \zetaScatt}}
{[(\vec{k}-\vec{b})^2 +\kappa^2]^{1-i\zetaScatt}}
\equiv f_{\vec{k},\vec{b}} (\kappa) \,.
\label{eq:Identity_HyperGeo}
\eeq
\Cref{eqs:calIntegrals_CoordSpace} become
\begin{subequations}
\label{eq:IntegralsCal_Coulomb}
\label[pluralequation]{eqs:IntegralsCal_Coulomb}
\begin{align}
\boldsymbol{{\cal J}}_{\vec{k}, \{100\}} (\vec{b})
&= \kappa
\sqrt{16 \pi \kappa^3 \, S_0 (\zetaScatt)} 
\ [ \nabla_{\vec{b}} f_{\vec{k},\vec{b}}(\kappa) ]
\label{eq:Jcal_Coulomb} \,,
\\
\boldsymbol{{\cal Y}}_{\vec{k}, \{100\}}
&= \mu \asNA 
\sqrt{16 \pi \kappa^3 \, S_0 (\zetaScatt)} 
\ [ \nabla_{\vec{b}} f_{\vec{k},\vec{b}}(\kappa) ]_{\vec{b}=0} 
\label{eq:Ycal_Coulomb} \,,
\end{align}
where
\beq
\[\nabla_{\vec b} f_{\vec k, \vec b}(\kappa)\]_{\vec b = \vec 0} 
= \hat{\vec{k}} \ \frac{2(1-i\,\zetaScatt)}{k^3} \ 
\frac{\exp\[-2 \zetaScatt \, {\rm arccot} (\zetaBound) \]}{(1+\zetaBound^2)^2} \,.
\label{eq:df/db@b=0}
\eeq
For the cross-sections of interest, we only need $\boldsymbol{{\cal J}}_{\vec{k}, \{100\}} (\vec{b})$ evaluated at $\vec{b}=\vec{0}$~\cite{Petraki:2015hla,Petraki:2016cnz}. Evidently,
\beq
\dfrac{ \boldsymbol{{\cal Y}}_{\vec{k}, \{100\}} }
{ \boldsymbol{{\cal J}}_{\vec{k}, \{100\}} (\vec{0}) }
= \dfrac{\asNA}{\agBound} \,,
\label{eq:YtoJratio}
\eeq
and
\beq
|\boldsymbol{{\cal J}}_{\vec{k}, \{100\}} (\vec{0})|^2 = 
\frac{2^6 \pi}{k}
\ S_0 (\zetaScatt) \, (1+\zetaScatt^2)
\ \dfrac{\zetaBound^5 \ \exp[-4\zetaScatt \, {\rm arccot}(\zetaBound)]}
{(1+\zetaBound^2)^4} \,.
\label{eq:Jcal_Coulomb_squared}
\eeq
\end{subequations}
We recall that $\zetaScatt$ and $\zetaBound$ are defined in \cref{eqs:zetas_scattANDbound_App}. In \cref{sec:BSF_CrossSections}, we use \cref{eqs:IntegralsCal_Coulomb} to obtain analytical expressions for the BSF cross-sections.


\section{The non-relativistic Hamiltonian from effective field theory\label{App:EFT}}

Our results in \cref{Sec:BSF} differ from previous computations~\cite{Mitridate:2017izz,Keung:2017kot} in the relative sign of the Abelian and non-Abelian contributions to the radiative transition amplitude. 
In this appendix, we use the non-relativistic QCD (NRQCD) approach of ref.~\cite{Manohar:2000kr,Manohar:1999xd,Beneke:1999zr} to derive the effective Hamiltonian for our system, and compare it with  refs.~\cite{Mitridate:2017izz,Keung:2017kot}, whose computations are based on effective field theory. This offers an independent check of our computations.

Before moving to NRQCD, we want to display that the discrepancy  between our computations and ref.~\cite{Mitridate:2017izz} originates from the expression for the transition amplitude. Indeed, starting from \cref{eq:MBSF_general} with $\eta_1=\eta_2=1/2$,
and using the coordinate-space expressions for the overlap integrals \eqref{eq:calIntegrals_CoordSpace} where we integrate \cref{eq:Jcal_CoordSpace} by parts, we arrive at
\begin{align}
i [{\cal M}_{\vec{k}\to \{n\ell m\}}]_{ii',jj'} 
&=
- i [\boldsymbol{{\cal M}}_{\vec{k}\to \{n\ell m\}}]_{ii',jj'}^a
\cdot \boldsymbol{\epsilon}^a
\nn \\
&= 
\sqrt{2^8 \pi \asBSF \, \mX} \times 
\left\{ 
\frac{1}{ 2} \left[ (T_1^a)_{i'i} \, \delta_{j'j}  - \delta_{i'i} \, (T_2^a)_{j'j} \right]
\int d^3 r \ [\psi_{n\ell m}^* (\vec r)] \, \nabla  \phi_{\vec{k}} (\vec{r}) \right. 
\nn \\
&
\left. 
- i f^{abc} (T_1^b)_{i'i} (T_2^c)_{j'j} \  \frac{\mX \asNA}{2} \int d^3 r \, \psi_{n\ell m}^* (\vec r) \, \phi_{\vec{k}} (\vec{r}) \, \hat{\vec{r}}
\right\} \cdot \boldsymbol{\epsilon}^a
\,, 
\label{eq:MBSF_comparison}
\end{align}
where $\boldsymbol{\epsilon}^a$ is the gluon polarisation vector. 
\Cref{eq:MBSF_comparison} may be directly compared to eqs.~(41)-(43) of  ref.~\cite{Mitridate:2017izz}. We observe that the relative sign between the two terms is different.

This discrepancy arises from the non-relativistic Hamiltonian assumed in  ref.~\cite{Mitridate:2017izz}. The interactions that give rise to the radiative BSF amplitude correspond to a non-relativistic potential $\tilde{V}_{\BSF} (\vec{q} , \vec{p})$ that can be deduced from the transition amplitude  
$\langle \vec{p} | i \mathcal{T} | \vec{q} \rangle$ in ordinary quantum mechanics,
\begin{align}
\langle \vec{p} | i \mathcal{T} | \vec{q} \rangle = 
- i \tilde{V}_{\BSF} (\vec{q},\vec{p})  \,,
\label{eq:potdef}
\end{align}
where $|\vec{q}\rangle$ denotes a state where the two interacting particles have momentum $\vec{q}$ and $-\vec{q}$ in the CM frame. The quantum mechanical transition amplitude is related to the matrix element 
${\cal M} (\vec{q},\vec{p}) = 
- \boldsymbol{{\cal M}}_{\rm trans}^a  (\vec{q},\vec{p}) 
\cdot \boldsymbol{\epsilon}^a$ 
via
\beq
({4\mX^2/A_0})  \ 
\langle \vec{p} | i \mathcal{T} | \vec{q} \rangle = 
- i \boldsymbol{{\cal M}}_{\rm trans}^a(\vec{q},\vec{p})  
\cdot  \boldsymbol{\epsilon}^a
\ (2 \pi) \delta(E_{\vec{q}} - E_{\vec{p}} - \omega) \,,
\label{eq:TransAmpl_QM}
\eeq
where the factor $4\mX^2/A_0$ accounts for the different normalization of fields in quantum field theory and quantum mechanics, with $A_0$ being the non-relativistic  normalisation of the gauge field. $E_{\vec{q}}$, $E_{\vec{p}}$ are the energies of the $|\vec{q}\>$, $|\vec{p}\>$ states and $\omega$ is the energy of the radiated gauge boson. From \cref{eq:potdef,eq:TransAmpl_QM}, we identify the non-relativistic potential in momentum space as
\beq
\tilde{V}_{\BSF} (\vec{q},\vec{p}) = 
\( \frac{A_0}{4\mX^2} \) \,  
\boldsymbol{{\cal M}}_{\rm trans}^a (\vec{q},\vec{p}) 
\cdot  \boldsymbol{\epsilon}^a
\ (2 \pi) \delta(E_{\vec{q}} - E_{\vec{p}} - \omega) \,.
\label{eq:VBSF(q-p)_def}
\eeq
The potential in coordinate space is\footnote{
Note that we Fourier transform only with respect to the energy and momentum differences between the initial and final $X_1 X_2$ states, thus the coordinate-space potential may still depend on $\vec{q}$ [cf.~\cref{eq:ourpotential}].}
\beq 
V_{\BSF} (t, \vec{r}) = \int  
\frac{d (E_{\vec q}-E_{\vec p})}{(2\pi)} 
\frac{d^3 (\vec{q}-\vec{p})}{(2\pi)^3} 
\ e^{ i [(E_{\vec q}-E_{\vec p}) \cdot t - (\vec{q}-\vec{p}) \cdot \vec{r}] }
\ \tilde{V}_{\BSF} (\vec{q},\vec{p})  \,.
\label{eq:VBSF(t-r)_def}
\eeq 
Using \cref{eq:MBSF_trans_computed} for 
$\boldsymbol{{\cal M}}_{\rm trans}^a (\vec{q},\vec{p})$ 
and the identity \eqref{eq:Integral_qOVERq4}, we obtain 
\begin{align}
V_{\BSF} (t,\vec{r}) = - A_0 \boldsymbol{\epsilon}^a \cdot
&\left\{ 
\frac{\gsBSF}{\mX}  \left[ 
(T_1^a)_{i'i} \, \delta_{j'j}  e^{i(\omega t - \vec{P}_g \cdot \vec{r}/2)} - 
\delta_{i'i} \, (T_2^a)_{j'j}  e^{i(\omega t + \vec{P}_g \cdot \vec{r}/2)} 
\right] \vec{q} 
\right.  \nn \\  
&\left.
- \gsBSF \asNA f^{abc} (T_1^b)_{i'i} (T_2^c)_{j'j} \, e^{i\omega t} \ \hat{\vec{r}}
\right\} 
\,.
\label{eq:VBSF(t-r)}
\end{align}
Identifying 
$\vec{A}^a (t,\vec{x}) = A_0 \exp [i(\omega t - \vec{P}_g \cdot \vec{x})] \, \boldsymbol{\epsilon}^a$ 
as the background field that induces the transition (see e.g.~\cite[sec.~5.7]{Sakurai_QMbook}), we rewrite the above as follows
\begin{align}
V_{\BSF} (t,\vec{r}) = -&
\left\{ 
\frac{\gsBSF}{\mX} \left[ 
  (T_1^a)_{i'i} \, \delta_{j'j}  \,\vec{q}  \cdot \vec{A}^a (t, \vec{r}/2) 
- \delta_{i'i} \, (T_2^a)_{j'j}  \,\vec{q}  \cdot \vec{A}^a (t,-\vec{r}/2)
\right]
\right.  \nn \\  &\left.
- \gsBSF \asNA f^{abc} (T_1^b)_{i'i} (T_2^c)_{j'j} 
\ \hat{\vec{r}} \cdot \vec{A}^a (t, \vec{0})
\right\}\,.
\label{eq:ourpotential}
\end{align}
Comparing this with eq.~(36) of ref.~\cite{Mitridate:2017izz}, we note the difference in the relative sign of the Abelian and non-Abelian contributions.

\begin{figure}
\centering
\begin{tikzpicture}[line width=1.1pt, scale=1.1]

\begin{scope}[shift={(0,0)}]
\begin{scope}[shift={(-5,2)}]
\node at (-1.3, 0) {$j$};
\draw[fermion] (-1,0) -- (0,0);
\draw[fermion] ( 0,0) -- (1,0);
\node at (0,-0.5) {$\psi$};
\node at (1.3,0) {$i$};
\filldraw (0.0,0.0) circle (3.0pt);
\draw[gluon]    ( 0,1) -- (0,0);
\node at (0,1.2) {$A^{a0}$};
\node at (0,-1.3) {$-ig_s T_{ij}^a$};
\end{scope}
\begin{scope}[shift={(-5,-2)}]
\draw[fermion] (-1,0) -- (0,0);
\draw[fermion] (0,0) -- (1,0);
\node at ( 0,-0.75) {$\psi$};
\node at (-1.25, 0) {$j$};
\node at ( 1.2,  0) {$i$};
\filldraw (-0.1,-0.1) rectangle (0.1,0.1);
\draw[gluon]    ( 0,1) -- (0,0);
\node at (0,1.2) {$\vec{A}^{a}$};
\draw[->] (-0.75,-0.25) -- (-0.25,-0.25);
\draw[->] ( 0.25,-0.25) -- ( 0.75,-0.25);
\node at (-0.5,-0.5) {$p_1$};
\node at ( 0.5,-0.5) {$p_2$};
\node at (0,-1.5) {$\dfrac{i g_s}{2 m_Q}(p_1 + p_2)^k \, T^a_{ij}$};
\end{scope}
\begin{scope}[shift={(0,2)}]
\node at (  0,1.2) {$\vec{A}^{ak}$};
\node at ( 1.35,0)   {$A^{b 0}$};
\node at (-1.35,0)   {$A^{c 0}$};
\draw[gluon] ( 1,0) -- (0,0);
\draw[gluon] ( 0,1) -- (0,0);
\draw[gluon] (-1,0) -- (0,0);
\draw[->] (-0.75,-0.25) -- (-0.25,-0.25);
\node at (-0.5,-0.5) {$r$};
\draw[<-] (0.25,-0.25) -- (0.75,-0.25);
\node at (0.5,-0.5) {$s$};
\draw[<-] (0.25,0.25) -- (0.25,0.75);
\node at (0.5,0.5) {$k$};
\node at (0,-1.3) {$ g_s f^{abc} (s-r)^k$};
\end{scope}
\begin{scope}[shift={(0,-2)}]
\node at (  0,1.2) {$\vec{A}^{ak}$};
\node at ( 1.35,0)   {$\vec{A}^{bm}$};
\node at (-1.35,0)   {$A^{c 0}$};
\draw[gluon] ( 1,0) -- (0,0);
\draw[gluon] ( 0,1) -- (0,0);
\draw[gluon] (-1,0) -- (0,0);
\draw[->] (-0.75,-0.25) -- (-0.25,-0.25);
\node at (-0.5,-0.5) {$r$};
\draw[<-] (0.25,-0.25) -- (0.75,-0.25);
\node at (0.5,-0.5) {$s$};
\draw[<-] (0.25,0.25) -- (0.25,0.75);
\node at (0.5,0.5) {$k$};
\node at (0,-1.3) {$g_s f^{abc} (k-s)_0 \delta_{km}$};
\end{scope}
\begin{scope}[shift={(0,-6)}]
\node at (  0,1.2) {$\vec{A}^{ak}$};
\node at ( 1.35,0)   {$\vec{A}^{bm}$};
\node at (-1.35,0)   {$\vec{A}^{cn}$};
\draw[gluon] ( 1,0) -- (0,0);
\draw[gluon] ( 0,1) -- (0,0);
\draw[gluon] (-1,0) -- (0,0);
\draw[->] (-0.75,-0.25) -- (-0.25,-0.25);
\node at (-0.5,-0.5) {$r$};
\draw[<-] (0.25,-0.25) -- (0.75,-0.25);
\node at (0.5,-0.5) {$s$};
\draw[<-] (0.25,0.25) -- (0.25,0.75);
\node at (0.5,0.5) {$k$};
\node at (0,-1.3) {$- g_s f^{abc} [(k-s)^n \delta_{km} + (s-r)^k  \delta_{mn} + (r-k)^m  \delta_{nk}]$};
\end{scope}
\begin{scope}[shift={(5,2)}]
\node at (  0,0.5) {${A}^{a0}$};
\node at (  -1.25,0) {$a$};
\node at (  1.25,0) {$b$};
\draw[gluon] ( -1,0) -- (1,0);
\node at (0,-1.3) {$\frac{-i\delta_{ab}}{p^2 + i \epsilon}$};
\end{scope}
\begin{scope}[shift={(5,-2)}]
\node at (  0,0.5) {$\vec{A}$};
\node at (  -1.5,0) {$a,m$};
\node at (  1.5,0) {$b,k$};
\draw[gluon] ( -1,0) -- (1,0);
\node at (0,-1.3) {$\frac{i\delta_{ab}~\delta_{mk}}{p^2 + i \epsilon}$};
\end{scope}

\end{scope}
\end{tikzpicture}
\caption[]{\label{fig:FeynmanRulesNRQCD} 
Feynman rules that are used to derive the non-relativistic potential that determines the radiative capture into bound states. In the gauge fields $A^{a,\mu}$, the first index denotes colour, while the second one if the space-time index. The first column depicts the NRQCD Feynman rules which distinguish between temporal (black dot) and spatial (black rectangle) couplings in the quark-gluon vertex. Hereby, $\psi$ represents a fermion field. The Feynman rules for the anti-fermion field $\chi$ are obtained by simply substituting $T^a_{ij} \rightarrow \overline{T}^a_{ij} =-T^a_{ji} $. The second column shows the non-relativistic 3-gluon-vertex, and the third column shows the gluon propagators. We distinguish again between temporal and spatial gauge fields. Note that according to the NRQCD Lagrangian \eqref{eq:NRQCDManoharLagrangian}, the NRQCD Feynman rules are defined with upper indices only, and incorporate the signs arising from the metric, $g_{\mu \nu}={\rm diag}(1,-1,-1,-1)$, in 4-vector contractions done according to the relativistic conventions.
}
\end{figure}
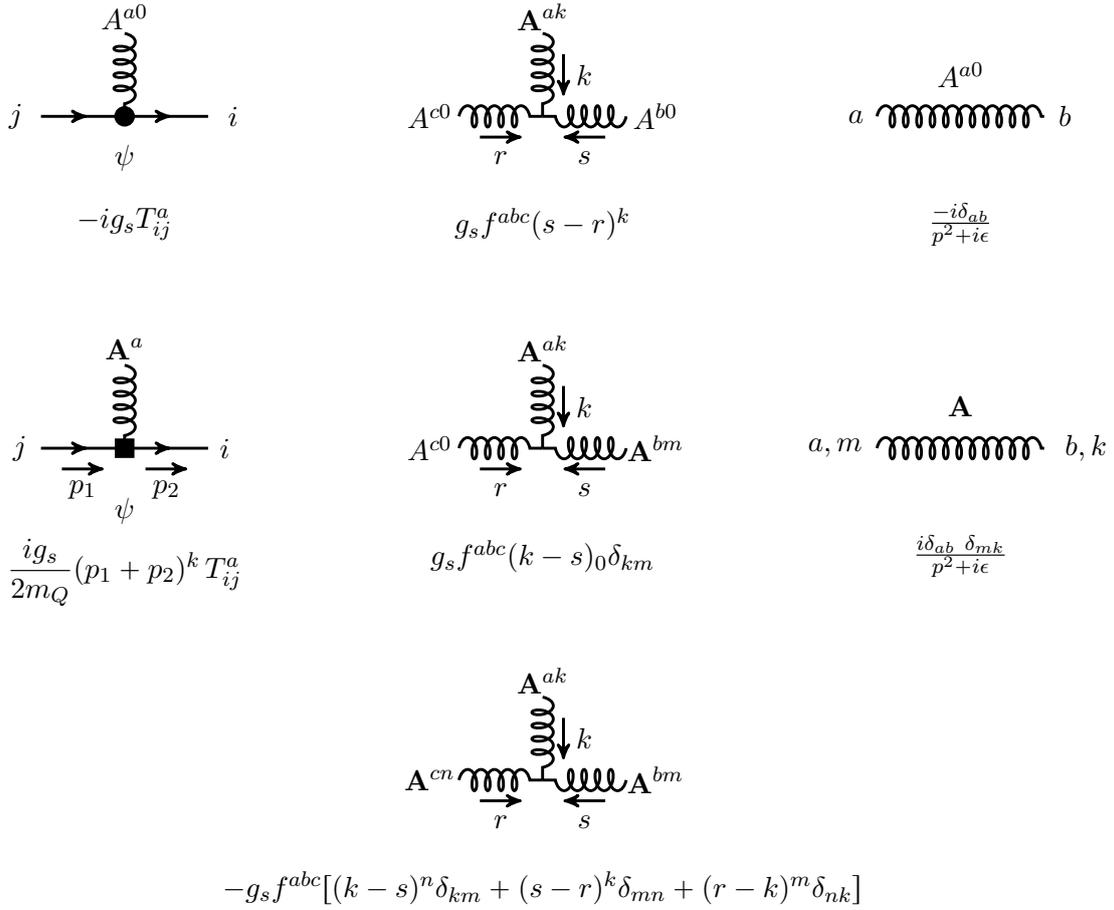

\bigskip
We now move on to NRQCD. At leading order, a heavy quark-antiquark system yields the same potential as a scalar particle-antiparticle system. Thus, we may compare our result of \cref{eq:ourpotential} with that derived for quarkonium in NRQCD, where the ``potential'' gluons are integrated out, leaving four-quark operators in the effective Lagrangian in analogy to the Fermi theory. There are different formulations, e.g.~pNRQCD~\cite{Pineda:1997bj,Beneke:1999zr} or vNRQCD~\cite{Luke:1999kz,Manohar:2000kr,Manohar:1999xd}\footnote
{For a comprehensive comparison, we refer to \cite{ioffe2001frontier}.
}.  
Here, we will follow the conventions of \cite{Manohar:2000kr,Manohar:1999xd,Hoang:2011gy}. We start from the ultrasoft NRQCD Lagrangian given in eq.~(7) of ref.~\cite{Manohar:2000kr}
\begin{equation}
\mathcal{L}_u
\supset \sum_\vec{p} \psi^\dagger_\vec{p}\left\{ i D^0 - \frac{(\vec{p} - i \vec{D})^2}{2m} + \frac{\vec{p}^4}{8m^3} \right\} \psi_\vec{p} + \sum_\vec{p} \chi^\dagger_\vec{p}\left\{ i D^0 - \frac{(\vec{p} - i \vec{D})^2}{2m} + \frac{\vec{p}^4}{8m^3} \right\} \chi_\vec{p}\,,
\label{eq:NRQCDManoharLagrangian}
\end{equation}
where $D^\mu = \partial^\mu + i g_s A^{a\mu} T^a= (D^0,-\vec{D})$ with $D^{0}=\partial^0 + i g_s A^{a0} T^a$ and $\vec{D}={\bf\nabla} - i g_s \vec{A}^a T^a$. The momentum $\vec{p}$ represents momenta of the soft scale. In contrast to the usual relativistic conventions, here $\psi^\dagger$ ($\chi^\dagger$) annihilates (anti)particles, while $\psi$ ($\chi$) creates (anti)particles. Their generators are related via $\overline{T}^a_{ij} =-T^a_{ji}$. $m_Q$ denotes the (anti)quark mass. From \cref{eq:NRQCDManoharLagrangian}, we can derive the Feynman rules for the fermion-antifermion system, treating the temporal and spatial component of the gauge field separately. We present them in  \cref{fig:FeynmanRulesNRQCD}.

We may now compute the contributions to the radiative transition amplitudes, shown in \cref{fig:DiagramsNRQCD}. In the CM frame, the momenta of the incoming/outgoing fermions are
\begin{subequations}
\label{eq:momentadef}
\label[pluralequation]{eqs:momentadef}
\begin{align}
&q_1^\mu =  (q_1^0, \  \vec{q}) \,, & 
&q_2^\mu =  (q_2^0, \ -\vec{q}) \,, & 
\\
&p_1^\mu = (p_1^0, \ - \vec{P}_g / 2 + \vec{p}) \,, & 
&p_2^\mu = (p_2^0, \ - \vec{P}_g / 2 - \vec{p}) \,, &
\end{align}
where $\vec{P}_g$ is the momentum of the emitted gluon. The energies of the scattering and bound states are
\begin{align}
(q_1+q_2)^0 &= 2m_Q + \vec{k}^2/m_Q \,, 
\\
(p_1 + p_2)^0 &= M_B + \vec{P}_g^2/(2M_B) \,,
\end{align}
\end{subequations}
where $\vec{k}$ is the expectation value of $\vec{q}$ and $M_B =2 m_Q - E_B$ is the mass of the bound state, with $E_B$ being the binding energy (cf.~\cref{App:Wavefunctions}). 
The energy of the radiated gluon, $\omega = |\vec{P}_g|$, is found from the conservation of energy to be  [cf.~\cref{eq:omega}],
\beq
\omega = (q_1+q_2 - p_1 -p_2)^0 \simeq E_B + {\bf k}^2/m_Q \,.
\label{eq:energyconservation}
\eeq
In the following, we shall extract the leading order contributions to the radiative BSF amplitude taking into account that $|\vec{q}| \sim |\vec{k}| = (m_Q/2) \vrel$, $|\vec{p}| \sim \kappa = (m_Q/2) \as$ and $\omega = (m_Q/4) (\as^2 + \vrel^2)$.

From the Lagrangian of \cref{eq:NRQCDManoharLagrangian} and the Feynman rules of \cref{fig:FeynmanRulesNRQCD}, it is straightforward to obtain the Abelian contributions to the transition amplitude, shown in \cref{fig:DiagramsNRQCD_dipole},
\begin{align}
&i{\cal W}_{\rm Abelian} =
\nn \\
&\!\! = \!\frac{i \gs}{2 m_Q}
\!\left[\!
(\vec{q}_1+\vec{p}_1) 
T_{i'i}^a \delta_{jj'}
(2\pi)^3 \delta^3(\vec{q}_1-\vec{p}_1 - \vec{P}_g)
\!+\! (\vec{q}_2 +\vec{p}_2) 
\overline{T}_{j'j}^a \delta_{ii'}
(2\pi)^3 \delta^3(\vec{q}_2-\vec{p}_2 - \vec{P}_g) \right] 
\! \boldsymbol{\epsilon}^a
\nn \\
&\!\!\simeq \frac{i \gs}{m_Q} 
\[ T_{i'i}^a \delta_{jj'}  \, (2\pi)^3 \delta^3(\vec{q}-\vec{p} - \vec{P}_g/2) 
-  \overline{T}_{j'j}^a \delta_{i'i}  \, (2\pi)^3 \delta^3(\vec{q}-\vec{p} + \vec{P}_g/2)\]
(\vec{q} \cdot \boldsymbol{\epsilon}^a)
\,.
\label{eq:Wdipole}
\end{align}
\begin{subfigures}
\label{fig:DiagramsNRQCD}
\label{figs:DiagramsNRQCD}
\begin{figure}
\centering
\begin{tikzpicture}[line width=1.1pt, scale=1.8]
\begin{scope}[shift={(0,0)}]
\begin{scope}[shift={(-2,0)}]
\draw[->] (-0.25,-0.25) -- ( 0.25,-0.25);
\draw[->] (-0.9,  1.25) -- (-0.4,  1.25);
\draw[->] ( 0.4,  1.25) -- ( 0.9,  1.25);
\draw[->] ( 0.2,  1.1)  -- ( 0.2,  1.4);
\node at ( 0  ,-0.5)  {${q_2}$};
\node at (-0.7, 1.5)  {${q_1}$};
\node at ( 0.7, 1.5)  {${p_1}$};
\node at ( 0.25, 1.5) {${P_g}$};

\node at (-1.25, 1) {$i$};
\draw[fermion]    (-1,1) -- ( 0,1);
\draw[fermion]    ( 0,1) -- ( 1,1);
\node at ( 1.25, 1) {$i'$};
\node at (-1.25, 0) {$j$};
\draw[anti]  ( -1,0) -- (1,0);
\node at ( 1.25, 0) {$j'$};
\filldraw (-0.05, 0.95) rectangle (0.05,1.05);
\draw[gluon] (0,1.5) -- (0,1);
\node at 	 (0,1.75) {$A^{ak}$};
\end{scope}
\begin{scope}[shift={(2,0)}]
\draw[->] (-0.25, 1.25) -- ( 0.25, 1.25);
\draw[->] (-0.9, -0.25) -- (-0.4, -0.25);
\draw[->] ( 0.4, -0.25) -- ( 0.9, -0.25);
\draw[->] ( 0.2, -0.1)  -- ( 0.2, -0.4);
\node at ( 0  , 1.5)  {${q_1}$};
\node at (-0.7,-0.5)  {${q_2}$};
\node at ( 0.7, -0.5)  {${p_2}$};
\node at ( 0.25,-0.55) {${P_g}$};

\node at (-1.25, 1) {$i$};
\draw[fermion]    (-1,1) -- ( 1,1);
\node at ( 1.25, 1) {$i'$};
\node at (-1.25, 0) {$j$};
\draw[anti]       ( -1,0) -- (0,0);
\draw[anti]       ( 0,0) -- ( 1,0);
\node at ( 1.25, 0) {$j'$};
\filldraw (-0.05,-0.05) rectangle (0.05,0.05);
\draw[gluon] (0,-0.5) -- (0,0);
\node at 	 (0,-0.75) {$A^{ak}$};
\end{scope}
\end{scope}
\end{tikzpicture}
\caption[]{\label{fig:DiagramsNRQCD_dipole} 
Abelian contributions to the capture of a fermion-antifermion pair into a bound state via gluon emission. The fermion field $\psi$ is depicted as a single solid line, the antifermion field $\chi$ as a double line.}
\begin{tikzpicture}[line width=1.1pt, scale=1.8]
\begin{scope}[shift={(0,0)}]
\begin{scope}[shift={(-2,1.3)}]
\draw[->] (-0.75,-0.25) -- (-0.25,-0.25);
\draw[->] (-0.75, 1.25) -- (-0.25, 1.25);
\draw[->] ( 0.25,-0.25) -- ( 0.75,-0.25);
\draw[->] ( 0.25, 1.25) -- ( 0.75, 1.25);
\draw[->] ( 0.9,  0.5)  -- ( 1.4,  0.5);
\draw[->] (-0.25, 0.85) -- (-0.25, 0.55);
\draw[->] (-0.25, 0.15) -- (-0.25, 0.45);
\node at (-0.5,-0.5) {${ q_2}$};
\node at (-0.5, 1.5) {$ {q_1}$};
\node at ( 0.7,-0.5) {${ p_2}$};
\node at ( 0.7, 1.5) {${ p_1}$};
\node at ( 1.1, 0.65) {${P_g}$};
\node at (-0.6, 0.75) {$q_1-p_1$};
\node at (-0.6, 0.25) {$q_2-p_2$};
\node at ( 0.0, 1.20) {$b,0$};
\node at ( 0.0,-0.20) {$c,0$};

\node at (-1.25, 1) {$i$};
\draw[fermion]    (-1,1) -- ( 0,1);
\draw[fermion]    ( 0,1) -- ( 1,1);
\node at ( 1.25, 1) {$i'$};
\node at (-1.25, 0) {$j$};
\draw[anti]       ( -1,0) -- (0,0);
\draw[anti]       ( 0,0) -- ( 1,0);
\node at ( 1.25, 0) {$j'$};
\filldraw (0.0,0.0) circle (2pt);
\filldraw (0.0,1.0) circle (2pt);
\draw[gluon] (0,0)   -- (0,0.5);
\draw[gluon] (0,1)   -- (0,0.5);
\draw[gluon] (0.8,0.5) -- (0,0.5);
\node at 	 ( 1.75 ,0.5) {$A^{ak}$};
\end{scope}
\begin{scope}[shift={(2,1.3)}]
\node at (-1.25, 1) {$i$};
\draw[fermion]    (-1,1) -- ( 0,1);
\draw[fermion]    ( 0,1) -- ( 1,1);
\node at ( 1.25, 1) {$i'$};
\node at (-1.25, 0) {$j$};
\draw[anti]       ( -1,0) -- (0,0);
\draw[anti]       ( 0,0) -- ( 1,0);
\node at ( 1.25, 0) {$j'$};
\filldraw (-0.05,-0.05) rectangle (0.05,0.05);
\filldraw (-0.05, 0.95) rectangle (0.05,1.05);
\draw[gluon] (0,0)   -- (0,0.5);
\draw[gluon] (0,1)   -- (0,0.5);
\draw[gluon] (0.8,0.5) -- (0,0.5);
\node at 	 ( 1.1 ,0.5) {$A^{ak}$};
\node at ( 0.0, 1.20) {$b,n$};
\node at ( 0.0,-0.20) {$c,m$};
\end{scope}
\begin{scope}[shift={(-2,-1)}]
\node at (-1.25, 1) {$i$};
\draw[fermion]    (-1,1) -- ( 0,1);
\draw[fermion]    ( 0,1) -- ( 1,1);
\node at ( 1.25, 1) {$i'$};
\node at (-1.25, 0) {$j$};
\draw[anti]       ( -1,0) -- (0,0);
\draw[anti]       ( 0,0) -- ( 1,0);
\node at ( 1.25, 0) {$j'$};
\filldraw ( 0,   0) circle (2pt);
\filldraw (-0.05,0.95) rectangle (0.05,1.05);
\draw[gluon] (0,0)   -- (0,0.5);
\draw[gluon] (0,1)   -- (0,0.5);
\draw[gluon] (0.8,0.5) -- (0,0.5);
\node at 	 ( 1.1 ,0.5) {$A^{ak}$};
\node at ( 0.0, 1.20) {$b,n$};
\node at ( 0.0,-0.20) {$c,0$};
\end{scope}
\begin{scope}[shift={(2,-1)}]
\node at (-1.25, 1) {$i$};
\draw[fermion]    (-1,1) -- ( 0,1);
\draw[fermion]    ( 0,1) -- ( 1,1);
\node at ( 1.25, 1) {$i'$};
\node at (-1.25, 0) {$j$};
\draw[anti]       ( -1,0) -- (0,0);
\draw[anti]       ( 0,0) -- ( 1,0);
\node at ( 1.25, 0) {$j'$};
\filldraw (-0.05,-0.05) rectangle (0.05,0.05);
\filldraw ( 0.0,  1.0) circle (2pt);
\draw[gluon] (0,0)   -- (0,0.5);
\draw[gluon] (0,1)   -- (0,0.5);
\draw[gluon] (0.8,0.5) -- (0,0.5);
\node at 	 ( 1.1 ,0.5) {$A^{ak}$};
\node at ( 0.0, 1.20) {$b,0$};
\node at ( 0.0,-0.20) {$c,m$};
\end{scope}
\end{scope}
\end{tikzpicture}
\caption[]{\label{fig:DiagramsNRQCD_NonAbelian} 
Non-Abelian contributions to the capture of a fermion-antifermion pair into a bound state, via gluon emission. The fermion field $\psi$ is depicted as a single solid line, the antifermion field $\chi$ as a double line. As in \cref{fig:FeynmanRulesNRQCD}, we distinguish between the temporal (circle) and spatial (rectangle) components of $A^{a \mu}$. 
}
\end{figure}
\end{subfigures}

We shall now derive the potential NRQCD Lagrangian term that describes the non-Abelian contribution to the radiative transition amplitudes. We will demonstrate that (a) we recover the same sign as in \cite{Manohar:2000kr,Manohar:1999xd,Hoang:2011gy} using our conventions \cref{eq:potdef,eq:TransAmpl_QM}, and that (b) this sign disagrees with the result of \cite{Mitridate:2017izz}. From the NRQCD Feynman rules of \cref{fig:FeynmanRulesNRQCD}, it is immediately evident that the first of the four diagrams shown in \cref{fig:DiagramsNRQCD_NonAbelian} yields the dominant contribution (see also comment below \cref{eqs:BSF_offshellamplitudes}). The diagrams involving the spatial components of the gluon propagators are suppressed by higher orders in the momenta $\vec{q}$ and $\vec{p}$, as shown explicitly below. Allowing for a non-zero gluon mass $m_g$, we obtain the following contributions for
$i{\cal W}_{\rm \mathsmaller{NA}}$ respectively:
\begin{subequations}
\label{eq:amplitudesNRQCD}
\label[pluralequation]{eqs:amplitudesNRQCD}	
\begin{align}
i \mathcal{W}_{\bullet\bullet} 
=&  \, (-i \gs T^b_{i^\prime i}) (-i\gs \overline{T}^c_{ j^\prime j}) 
\[\frac{-i}{(q_1-p_1)^2 - m_g^2}\] 
\[\frac{-i}{(q_2-p_2)^2 - m_g^2}\]  
\nn \\
&\times g_s f^{abc} (q_1-p_1 -q_2 + p_2)^k  \epsilon^{a,k}
\nn \\
\simeq&~
~2 g_s^2 g_s \ f^{abc} T^b_{i^\prime i} \overline{T}^c_{ j^\prime j} 
\ \frac{(\vec{q} - \vec{p})}{[({\bf q-p})^2+m_g^2]^2}  \cdot \boldsymbol{\epsilon}^a\,,
\label{eq:W1}
\\
i \mathcal{W}_{\sqbullet\sqbullet} = 
& \, 
\[ \frac{i\gs}{2 m_Q}  ( q_1 + p_1)^m \, T^b_{i'i} \]
\[\frac{i\gs}{2 m_Q}  (q_2 + p_2)^n \, \overline{T}^c_{ j^\prime j} \]
\[\frac{i \delta_{mm'}}{(q_1-p_1)^2 - m_g^2}\] 
\[\frac{i \delta_{nn'}}{(q_2-p_2)^2 - m_g^2}\]
\nn \\
&\times 
(- \gs) f^{abc} \, \epsilon^{a,k}
\nn \\
&\times \left[ 
(q_1-p_1 -q_2 + p_2)^k  \delta_{m'n'} +
(q_2 - p_2 + P_g)^{m'} \delta_{n'k} + 
(-P_g - q_1 + p_1)^{n'} \delta_{km'} 
\right]  
\nn \\
\simeq& 
~g_s^2 g_s f^{abc} T^b_{i^\prime i} \overline{T}^c_{ j^\prime j}  \ 
\frac{[
(\vec{q}+\vec{p})^2 (\vec{q} - \vec{p}) +
(\vec{p}^2 - \vec{q}^2)(\vec{q} + \vec{p})
]}
{2 m^2_Q \, [({\bf q-p})^2 + m_g^2]^2} \cdot \boldsymbol{\epsilon}^a \,,
\label{eq:W2}
\\
i \mathcal{W}_{\sqbullet\bullet} = 
& \, 
\[\frac{i\gs}{2 m_Q} T^b_{i^\prime i}  (q_1 + p_1)^m \] 
(- i \gs\overline{T}^c_{ j^\prime j} )
\[\frac{i \delta_{mm'}}{(q_1-p_1)^2 - m_g^2}\] 
\[\frac{-i}{(q_2-p_2)^2 - m_g^2}\]
\nonumber \\
& \times \gs f^{abc} 
(-P_g - q_1 + p_1)^0 \delta_{km'} \ \epsilon^{a,k}
\nonumber \\
\simeq& 
\, - g_s^2 g_s f^{abc} T^b_{i^\prime i} \overline{T}^c_{ j^\prime j}  
\[\frac{(q_1-p_1+P_g)^0}{2 m_Q} \]
\frac{(\vec{q} + \vec{p})}{[({\bf q-p})^2 + m_g^2]^2}\cdot \boldsymbol{\epsilon}^a \,,
\label{eq:W3}
\\
i \mathcal{W}_{\sqbullet\bullet}
=& \, 
(- i \gs T^b_{i' i})
\[\frac{i\gs}{2 m_Q} \overline{T}^c_{ j^\prime j} (q_2 + p_2)^n \]
\[\frac{-i}{(q_1-p_1)^2 - m_g^2}\] 
\[\frac{i \delta_{nn'}}{(q_2-p_2)^2 - m_g^2}\]
\nonumber \\
&\times \gs f^{abc} (q_2 - p_2 + P_g)^0 \delta_{n'k} \ \epsilon^{a,k}
\nonumber \\
\simeq& 
\,- g_s^2 g_s f^{abc} T^b_{i^\prime i}\overline{T}^c_{ j^\prime j}  
\[ \frac{(q_2-p_2+P_g)^0}{2 m_Q} \]
\frac{(\vec{q} + \vec{p})}{[({\bf q-p})^2 + m_g^2]^2}\cdot \boldsymbol{\epsilon}^a \,,
\label{eq:W4}
\nn
\end{align}
where we have used \cref{eqs:momentadef}. Using the energy conservation \eqref{eq:energyconservation}, the last two terms add up to
\beq
i (\mathcal{W}_{\bullet\sqbullet} + \mathcal{W}_{\sqbullet\bullet}) \simeq
- g_s^2 g_s f^{abc} T^b_{i^\prime i}\overline{T}^c_{ j^\prime j}  
\( \frac{3\omega}{2 m_Q} \)
\frac{(\vec{q} + \vec{p})}{[({\bf q-p})^2 + m_g^2]^2}\cdot \boldsymbol{\epsilon}^a \,.
\label{eq:W3plusW4}
\eeq
\end{subequations}
Since $|\vec{q}|^2$, $|\vec{p}|^2$, $\omega \ll m_Q$, the terms
${\cal W}_{\sqbullet\sqbullet}$, 
${\cal W}_{\bullet\sqbullet}$ and 
${\cal W}_{\sqbullet\bullet}$ 
are subdominant with respect to 
${\cal W}_{\bullet\bullet}$. 
Thus, the leading order non-Abelian contribution is 
$\mathcal{W}_{\rm NA} \simeq \mathcal{W}_{\bullet\bullet}$, 
which agrees with our findings in \cref{sec:BSF_Amplitude}.

The total transition amplitude is 
$i {{\cal W}}_{\rm trans} \simeq  i {{\cal W}}_{\rm Abelian} +  i {{\cal W}}_{\mathsmaller{\rm NA}}$. 
Collecting \cref{eq:Wdipole,eq:W1}, we find it to be
\begin{align}
i {{\cal W}}_{\rm trans}  \simeq  \frac{i 2\gs}{m_Q} 
&\left\{
\[\frac12 T_{i'i}^a \delta_{jj'}  \, (2\pi)^3 \delta^3(\vec{q}-\vec{p} - \vec{P}_g/2) 
- \frac12 \overline{T}_{j'j}^a \delta_{i'i}  \, (2\pi)^3 \delta^3(\vec{q}-\vec{p} + \vec{P}_g/2)\]
\vec{q}
\right. \nn \\
&\left.
-i 8 \pi f^{abc} T^b_{i^\prime i} \overline{T}^c_{ j^\prime j}  
\(\frac{m_Q \alpha_s}{2} \)
\ \frac{(\vec{q} - \vec{p})}{[({\bf q-p})^2+m_g^2]^2}  
\right\} \cdot \boldsymbol{\epsilon}^a
\,.
\label{eq:EFTamplitude}
\end{align}
This can be compared with 
$i{\cal M}_{\rm trans} / (4M\mu) = -i\boldsymbol{{\cal M}}_{\rm trans}^a 
\cdot \boldsymbol{\epsilon}^a / (4M\mu)$ 
with 
$\boldsymbol{{\cal M}}_{\rm trans}$ given by \cref{eq:MBSF_trans_computed}. As earlier, the factor $4M\mu$ accounts for the relativistic normalisation of states. We see that the two results are in agreement. 

In order to compare \cref{eq:EFTamplitude} with \cite{Manohar:2000kr,Hoang:2011gy}, we construct the effective action that recovers the amplitude,
\begin{align}
\langle p_1, p_2|i S_{\mathrm{eff}} | q_1, q_2\rangle = 
(2 \pi)^4 \delta^{(4)}(q_1 + q_2 - p_1 - p_2) 
\, i\mathcal{W}_{\rm trans}(q_1,q_2,p_1,p_2) \,,
\end{align}
such that
\begin{align}
S_{\mathrm{eff}} &=  
\int \frac{d^4 q_1 d^4 q_2 d^4 p_1 d^4 p_2}{(2 \pi)^{16}} 
\ \psi^\dagger(p_1)   \chi^\dagger(p_2)
\, \mathcal{W}_{\rm trans}(q_1,q_2,p_1,p_2) \, 
\psi(q_1) \chi(q_2)  
\nn \\
&\times (2 \pi)^4 \delta^{(4)}(q_1 + q_2 - p_1 - p_2) \,.
\label{eq:effectiveaction}
\end{align}
Exchanging the polarisation vector $\boldsymbol{\epsilon}^a$ for the background field $\vec{A}^a$, we arrive at the Lagrangian
\begin{align}
\mathcal{L}_u + \mathcal{L}_p
\supset &\sum_\vec{p} \psi^\dagger_\vec{p}\left\{ i D^0 - \frac{(\vec{p} - i \vec{D})^2}{2m} + \frac{\vec{p}^4}{8m^3} \right\} \psi_\vec{p} + \sum_\vec{p} \chi^\dagger_\vec{p}\left\{ i D^0 - \frac{(\vec{p} - i \vec{D})^2}{2m} + \frac{\vec{p}^4}{8m^3} \right\} \chi_\vec{p}\,
\nonumber \\
+&\sum_{\vec{p},\vec{q}} (-  2 i g_s^2 g_s \ f^{abc} )
\( \psi^\dagger_\vec{p} T^b \psi_\vec{q} \)
\( \chi^\dagger_{-\vec{p}} \overline{T}^c \chi_{-\vec{q}} \)
\ \frac{(\vec{q} - \vec{p})}{({\bf q-p})^4}  \cdot \vec{A}^a 
\,.
\label{eq:lagrangianManohar}
\end{align}
This Lagrangian can be compared with eqs.~(6), (7), (12) and the first term of eq.~(16) of ref.~\cite{Manohar:2000kr}.\footnote{Note that $\vec{k}$ in eq.~(16) of ref.~\cite{Manohar:2000kr} is defined as $\vec{k}=-(\vec{q}-\vec{p})$.} It agrees perfectly, including the sign of the non-Abelian term.\footnote{
The effective  Lagrangian is invariant under ultrasoft gauge transformations~\cite{Beneke:1999zr,Hoang:2011gy}.}

Similarly to before, we may now derive the full interaction potential in position space. Starting from \cref{eq:EFTamplitude} and following similar steps as in the beginning of this appendix, we arrive again at \cref{eq:ourpotential}, with the identification 
$T_{i'i}^a = (T_1)_{i^\prime i}^a$, 
$\overline{T}_{j'j}^a = (T_2)_{j^\prime j}^a =  - (T_1)_{j j^\prime}^a$ and 
$m_Q = \mX$, 
thereby confirming the validity of the computation of \cref{Sec:BSF} and the disagreement with ref.~\cite{Mitridate:2017izz}. We note that we have also compared our results to those of ref.~\cite{Beneke:1999zr} and found them in agreement.\footnote{
For comparison with eq.~(1.8) in ref.~\cite{Beneke:1999zr}, it is important to note the different definition of fields with respect to ref.~\cite{Manohar:2000kr}. In ref.~\cite{Beneke:1999zr}, $\psi$ ($\chi^\dagger$) annihilates (anti)particles, while $\psi^\dagger$ ($\chi$) creates (anti)particles. In that notation, one has to account for the odd Wick permutations, such that the corresponding non-Abelian amplitude $\mathcal{M}^\prime_{\mathrm{NA}}$ is recovered by
\begin{align}
S_{\mathrm{eff}} =  
\int \frac{d^4 q_1 d^4 q_2 d^4 p_1 d^4 p_2}{(2 \pi)^{16}} 
\psi^\dagger(p_1)  \chi(-p_2)~(-1)~\mathcal{M}^\prime_{\rm NA}(q_1,q_2,p_1,p_2)\psi(q_1)  \chi^\dagger(-q_2)\,,
\end{align}
in contrast to \cref{eq:effectiveaction}. 
From $\mathcal{M}^\prime_{\rm NA}$, we may derive the non-relativistic potential, as in the beginning of this appendix, and compare it with \cref{eq:ourpotential}. A global sign difference is expected due to the different definition for the covariant derivative: 
while here we have used $D_\mu = \partial_\mu + i g_s A^a_\mu T^a$, in ref.~\cite{Beneke:1999zr} the covariant derivative is defined as $D_\mu = \partial_\mu - i g_s A^a_\mu T^a$.
}
%

\section{The Milne relation \label{App:MilneRelation}}

For a 2-to-2 process $1 + 2 \leftrightarrow A + B$, the cross-section, averaged over the degrees of freedom of the initial state, is
\beq
\sigma_{1+2 \to A+B} = \frac{1}{g_1 g_2}
\ \frac{1}{2\sqrt{(s-m_1^2-m_2^2)^2-4m_1^2m_2^2}}
\ \frac{|\vec{P}_{A}^{\mathsmaller{\rm CM}}|}{16 \pi^2 \sqrt{s}}
\ \int d\Omega \: |{\cal M}_{1+2 \to A+B}|^2 \,,
\label{eq:sigma_2to2}
\eeq
where $s$ is the first Mandelstam variable and $\vec{P}_A^{\mathsmaller{\rm CM}}$ is the momentum of $A$ (or $B$) in the CM frame. Since 
$|{\cal M}_{1+2 \to A+B}|^2 = |{\cal M}_{A+B \to 1+2}|^2$, the cross-sections of two 2-to-2 inverse processes are related via 
\beq
\frac{\sigma_{A+B \to 1+2}}{\sigma_{1+2 \to A+B}} = 
\frac{g_1 g_2}{g_A g_B} \[ 
\frac{(s-m_1^2-m_2^2)^2 - 4 m_1^2 m_2^2}{(s-m_A^2-m_B^2)^2 - 4 m_A^2 m_B^2} 
\]^{1/2}
\frac{|\vec{P}_1|}{|\vec{P}_A|} \,.
\label{eq:sigmaratio_inverseprocesses}
\eeq

We now consider the radiative BSF and ionisation processes $X_1 + X_2 \leftrightarrow {\cal B}_{n\ell} + g$, where $g$ stands generally for the massless field radiated in a BSF process with energy $\omega$. In the non-relativistic regime,
\beq
s 
\simeq (m_1 + m_2 + {\cal E}_{\vec{k}} )^2 
\simeq (m_1 + m_2 + {\cal E}_{n\ell} + \omega )^2 \,,
\label{eq:s}
\eeq
where ${\cal E}_{\vec{k}} = \mu \vrel^2/2$ is the kinetic energy of the initial state in the CM frame, and ${\cal E}_{n\ell} < 0$ is the binding energy of the bound state (cf.~\cref{App:Wavefunctions}), with ${\cal E}_{\vec{k}}, \ |{\cal E}_{n\ell}| \ll m_1+m_2$. Then, $|\vec{P}_1| = |\vec{P}_2| = \mu \vrel$ and 
$|\vec{P}_g| = |\vec{P}_{\cal B}| = \omega = {\cal E}_{\vec k} - {\cal E}_{n\ell}$, and from \cref{eq:sigmaratio_inverseprocesses} we find the Milne relation
\beq
\sigma_{\ion} = 
\frac{g_1 g_2}{g_g g_{\cal B}^{}} 
\( \frac{\mu^2 \vrel^2}{\omega^2} \)
\ \sigma_{\BSF}
\,.
\label{eq:MilneRelation}
\eeq

\section*{Acknowledgements}

We thank Adam Falkowski, Carlos Tamarit and Simone Biondini for useful discussions. J.H. was supported by the Labex ILP (reference ANR-10-LABX-63) part of the Idex SUPER, and received financial state aid managed by the Agence Nationale de la Recherche, as part of the programme Investissements d'avenir under the reference ANR-11-IDEX-0004-02. K.P. was supported by the ANR ACHN 2015 grant (``TheIntricateDark" project), and by the NWO Vidi grant ``Self-interacting asymmetric dark matter".

\bibliography{Bibliography.bib}

\end{document}